\documentclass[preprint,notoc,11pt]{JHEP}

\usepackage{axodraw}

\def\PlusBreak#1{+ \nonumber \\          % splits line with plus sign
          &&  \hphantom{#1} \! \null   +}
\def\MinusBreak#1{- \nonumber \\         % splits line with minus sign
          &&  \hphantom{#1} \!  \null  -}

\def\qb{{\bar{q}}}
\def\Qb{{\bar{Q}}}

\def\Tr{\mathop{\rm tr}\nolimits}

\def\bigglB{\biggl[}
\def\biggrB{\biggr]}
\def\bigglP{\biggl(}

\def\BigglP{\Biggl\{ }

\def\BigrB{\Bigr]}

\def\NeqOne{{\cal N} = 1}
\def\ksl{\not{\hbox{\kern-2.3pt $k$}}}

\def\e{\epsilon}

\def\Ord{{\cal O}}
\def\cm{{\cal M}}

\def\Nf{{N_{\! f}}}

\def\ra{\rangle}
\def\RS{{\scriptscriptstyle\rm R\!.S\!.}}

\def\bom#1{{\mbox{\boldmath $#1$}}}
\def\MSbar{\overline{\rm MS}}
\def\DRbar{\overline{\rm DR}}

\def\FDH{{\rm FDH}}

\def\li#1{{\mathop{\rm Li}\nolimits}_#1}
\def\Li{\mathop{\rm Li}\nolimits}
\def\ggtogg{{gg \to gg}}
\def\qqtogg{{q\bar{q} \to gg}}

\def\qgtogq{{qg \to gq}}
\def\qqtoqq{{q\bar{q} \to \bar{Q}Q}}

\def\tg{\tilde{g}}

\def\alphas{\alpha_s}

\def\Susy{{\scriptscriptstyle\rm SYM}}

\def\eqn#1{eq.~(\ref{#1})}
\def\eqns#1#2{eqs.~(\ref{#1}) and~(\ref{#2})}

\def\Boxsix{{\rm Box}^{(6)}}

\def\Trifour{{\rm Tri}^{(4)}}

\def\spa#1.#2{\left\langle#1\,#2\right\rangle}
\def\spb#1.#2{\left[#1\,#2\right]}
\def\lor#1.#2{\left(#1\,#2\right)}
\def\trc{{\rm Tr}}

\def\tS{{\tt S}}
\def\tT{{\tt T}}
\def\tU{{\tt U}}
\def\ibar{\bar{\imath}}

\def\fig#1{figure~{\ref{#1}}}

\def\sect#1{section~\ref{#1}}
\def\Sect#1{Section~\ref{#1}}
\def\app#1{appendix~\ref{#1}}

\newbox\charbox
\newbox\slabox
\def\s#1{{      % Feynman slash
        \setbox\charbox=\hbox{$#1$}
        \setbox\slabox=\hbox{$/$}
        \dimen\charbox=\ht\slabox
        \advance\dimen\charbox by -\dp\slabox
        \advance\dimen\charbox by -\ht\charbox
        \advance\dimen\charbox by \dp\charbox
        \divide\dimen\charbox by 2
        \raise-\dimen\charbox\hbox to \wd\charbox{\hss/\hss}
        \llap{$#1$}
}}
\preprint{
  DESY 04-061 \\
  UCLA/04/TEP/32\\
  August, 2004}

\title{
Two-Loop Helicity Amplitudes for Quark-Quark Scattering in QCD and
Gluino-Gluino Scattering in Supersymmetric Yang-Mills Theory}

\author{Abilio De Freitas\thanks{Alexander von Humboldt Fellow } \\
	Deutsches Elektronen Synchrotron \\
	DESY, D-15738 Zeuthen, Germany\\
	E-mail: \email{dfreitas@ifh.de}}

\author{Zvi Bern\thanks{Research supported by the US Department of
Energy under grant DE-FG03-91ER40662.} \\
	Department of Physics and Astronomy\\
	UCLA, Los Angeles, CA 90095-1547, USA\\
	E-mail: \email{bern@physics.ucla.edu}}

\abstract{We present the two-loop QCD helicity amplitudes for
quark-quark and quark-antiquark scattering. These amplitudes are
relevant for next-to-next-to-leading order corrections to (polarized)
jet production at hadron colliders.  We give the results in the
`t~Hooft-Veltman and four-dimensional helicity (FDH) variants of
dimensional regularization and present the scheme dependence of the
results.  We verify that the finite remainder, after subtracting the
divergences using Catani's formula, are in agreement with previous
results.  We also provide the amplitudes for gluino-gluino scattering
in pure $\NeqOne$ supersymmetric Yang-Mills theory.  We describe
ambiguities in continuing the Dirac algebra to $D$ dimensions,
including ones which violate fermion helicity conservation. The finite
remainders after subtracting the divergences using Catani's formula,
which enter into physical quantities, are free of these ambiguities.
We show that in the FDH scheme, for gluino-gluino scattering, the
finite remainders satisfy the expected supersymmetry Ward identities.}


\begin{document}

\section{Introduction}
\label{IntroSection}

In the coming years, our field is looking forward to the ongoing
experiments at the Tevatron at Fermilab and the future ones at the
Large Hadron Collider at CERN for unlocking the physics of the
electroweak symmetry breaking scale.  At hadron colliders, for large
momentum transfer, the most copious events are hadronic jets.  To
explore the validity of the Standard Model at the shortest possible
distances, therefore, it would be helpful to determine jet production
cross sections with high precision.  Calculations of jet production at
next-to-leading order (NLO) in the strong coupling constant
$\alpha_s$~\cite{Aversa,EKS,JETRAD} agree well with the data over a
broad range of transverse momentum.  Still, the NLO predictions have
an uncertainty from higher-order corrections, which is traditionally
estimated using dependence on the renormalization and factorization
scales, of order 10\% or more.  For very large momentum transfer the
predictions can be improved by resumming threshold
logarithms~\cite{Kidonakis}.  There are also sizable uncertainties
associated with the experimental input to the parton distribution
functions~\cite{PDFuncertainty}.  Nevertheless, an exact
next-to-next-to-leading order (NNLO) computation would be desirable.
An important step has recently been accomplished with the computation
of the three-loop splitting function by Moch, Vermaseren and
Vogt~\cite{MVV}.  There has also been some earlier work on global fits
to the data~\cite{MRSTNNLO} within an approximate
next-to-next-to-leading order (NNLO) framework~\cite{NNLOPDFApprox}.
Once combined with the matrix elements in complete programs, this
should considerably reduce the renormalization and factorization scale
uncertainties in production rates. For a summary of the various
expected improvements see, for example, ref.~\cite{GloverReview}.

Recent years have seen rapid progress in our ability to compute
two-loop matrix elements, especially when there is dependence on more
than a single kinematic
variable~\cite{BRY,AllPlusTwo,BhabhaTwoLoop,GOTYqqqq,GOTYqqgg,GOTYgggg,
gggamgamPaper,PhotonPaper,BDDgggg,TwoLoopee3Jets,TwoLoopee3JetsHel,BDDqqgg,
Gloverqqgg,Gloverqqqq}.  Much of this progress has relied on new
developments in loop
integration~\cite{IBP,PBScalar,NPBScalar,Lorentz,NPBReduction,
PBReduction,IntegralsAGO} and in understanding the infrared
divergences of the theory~\cite{Catani}.

An NNLO calculation of jet production requires requires six-point
tree-level, one-loop five-point amplitudes and two-loop four-point
amplitudes.  The tree amplitudes for six external
partons~\cite{TreeSixPoint,MPReview} and the one-loop amplitudes for
five external partons~\cite{OneloopFivePoint} have been determined
some time ago. Anastasiou, Glover, Oleari, and Tejeda-Yeomans have
provided the NNLO interferences of the two-loop amplitudes with the
tree amplitudes, for all QCD four-parton processes, summed over all
external helicities and colors~\cite{GOTYgggg}. The helicity
amplitudes for $gg \to gg$ were presented in ref.~\cite{BDDgggg}.  The
$\bar q q \rightarrow gg$ and $q g \rightarrow qg$ helicity amplitudes
were presented in refs.~\cite{BDDqqgg,Gloverqqgg}.  Recently, while
preparing this paper, Glover presented the four-quark helicity
amplitudes~\cite{Gloverqqqq}.  Here we present the same amplitudes using
somewhat different methods, as well as the $\NeqOne$ supersymmetric
version of these amplitudes.  We also describe ambiguities in the 
amplitudes that first arise in four-fermion amplitudes at NNLO.

For jet production in collisions of unpolarized hadrons, which is the
main phenomenological application of the amplitudes, the additional
helicity and color information contained in the helicity amplitudes is
not necessary.  However, for the case of polarized proton scattering
at the relativistic heavy ion collider (RHIC) at Brookhaven, the
helicity amplitudes are of direct relevance for improving predictions
to NNLO accuracy.  This may help with the determination of the
poorly-known polarized gluon distribution in the
proton~\cite{SofferVirey}, which is currently available through
NLO~\cite{dFFSV}.

Many formal properties of scattering amplitudes are simpler in a
helicity basis.  As a striking recent example, Witten has linked
tree-level helicity amplitudes to a twistor space topological string
theory~\cite{Witten} leading to simple~\cite{CSW} and
efficient~\cite{BBK} rules for dealing with general~\cite{Khoze} tree
amplitudes in massless gauge theories. It may also lead to new
insights into loop calculations~\cite{LoopsCSW}.  Other examples are
supersymmetry Ward identities~\cite{SWI,TwoLoopSUSY}, the behavior
of amplitudes as momenta become
collinear~\cite{MPReview,Neq4,LoopReview}, and high-energy
behavior~\cite{BFKL,TwoloopBFKL} which all become more evident in a
helicity basis.  The full color dependence is also helpful for
exploring the general structure of infrared
singularities~\cite{Catani,BDDgggg,StermanIR,BDKtwoloopsplit}.

In general, scattering amplitudes with massless gluon exchange possess
infrared (soft and collinear) divergences.  In dimensional
regularization two-loop amplitudes generically contain poles in the
dimensional regularization parameter $\e = (4-D)/2$ up to $1/\e^4$.
These singularities have been predicted and organized into a compact
form by Catani~\cite{Catani} and cancel from final physical results.
As is now standard, we use Catani's formula and color space notation
to organize the helicity amplitudes into singular terms (which do
contain finite terms in their series expansion in $\e$), plus finite
remainders.  The precise form of the $1/\e$ poles was not predicted in
ref.~\cite{Catani} for general processes at two loops. However, it is
now apparent that these terms also have a universal structure
depending only on the external legs, based on explicit
calculation~\cite{BhabhaTwoLoop,GOTYqqqq,GOTYqqgg,GOTYgggg,
gggamgamPaper,PhotonPaper,BDDgggg,TwoLoopee3Jets,TwoLoopee3JetsHel,BDDqqgg,
Gloverqqgg,Gloverqqqq}, matching to resummations~\cite{StermanIR} and
constraints on the functional form as momenta become
collinear~\cite{BDKtwoloopsplit}.

A number of dimensional regularization variants are commonly employed
for QCD loop calculations.  The conventional dimensional
regularization (CDR) scheme~\cite{CDR} is usually applied in
calculations of amplitude interferences, such as in
refs.~\cite{EllisSexton,BhabhaTwoLoop,GOTYqqqq,GOTYqqgg,GOTYgggg}.  In
the helicity approach, the two commonly used schemes are the
't~Hooft-Veltman (HV) scheme~\cite{HV} and the four-dimensional
helicity (FDH) scheme~\cite{BKgggg,TwoLoopSUSY}.  These schemes differ
in the number of polarization components for unobserved gluons.  The
't~Hooft-Veltman (HV) scheme~\cite{HV} contains $2 - 2\e$ virtual
gluon states (as does the CDR scheme), whereas the four-dimensional
helicity (FDH) scheme~\cite{BKgggg,TwoLoopSUSY} assigns $2$ states.
The FDH scheme is related to dimensional reduction (DR)~\cite{DR}, but
is more compatible with the helicity method, because it allows two
transverse dimensions in which to define helicity.  A more detailed
description of the differences between schemes, as well as a
definition of the FDH scheme beyond one loop, may be found in
ref.~\cite{TwoLoopSUSY}.  As in ref.~\cite{BDDqqgg} we find
non-trivial scheme dependence in the finite remainders.

The continuation of the Dirac algebra from four to $D$ dimensions
suffers from a variety of ambiguities.  One well known ambiguity is in
the continuation of $\gamma_5$~\cite{HV,Chanowitz}.  Another lesser
known ambiguity, which we also investigate in this paper, is tied to
charge conjugation.  The appearance of an ambiguity is perhaps not
surprising in hindsight given that like $\gamma_5$ in $D=4$ the charge
conjugation matrix is a discrete product of $\gamma$ matrices, whose
continuation to $D$ dimensions is inherently delicate. In QCD it is a
simple matter to sidestep this ambiguity by avoiding use of any charge
conjugation identities that are valid only in four dimensions.  For
$\NeqOne$ super-Yang-Mills theory, the situation is more complicated,
because of the Majorana nature of the gluino implying it is its own
antiparticle.  At tree level and one loop it turns out to affect the
amplitudes only at $\Ord(\e)$ and is therefore irrelevant through NLO
in the coupling constant.  Moreover, it is does not appear even at two loops in
the previously computed four-gluon or two-fermion two-gluon helicity
amplitudes.  This ambiguity first becomes relevant at NNLO when there
are at least four fermions present.  From Catani's infrared divergence
formula, an ambiguity in $\Ord(\e)$ parts of a one-loop amplitude
necessarily feeds into an ambiguity in the two-loop amplitude starting
at $\Ord(1/\e)$. Nevertheless, it is reassuring that this ambiguity
turns out not to affect the finite remainder of the two-loop amplitudes as
long as the different loop orders are computed consistently: We find
that the entire ambiguity, including associated finite parts, may be
absorbed into the Catani subtraction.  Moreover, as noted in
ref.~\cite{BDDqqgg}, when computing a physical process, at NNLO one
does not need $\Ord(\e)$ contributions to the one-loop amplitudes
because such terms always cancel.  Thus the ambiguity completely
cancels from physical quantities, as one may have anticipated from
general considerations~\cite{Einhorn}.

For theories without infrared divergences, one can straightforwardly
add local counterterms to undo violations of supersymmetry by the
regulator. For theories with infrared divergences the situation is
more subtle. In any case, it is simpler and more elegant to use a
regularization scheme which automatically preserves
supersymmetry~\cite{DR}.  The supersymmetry preserving properties of
the FDH scheme~\cite{BKgggg,TwoLoopSUSY} have been verified explicitly
at two loops, for gluon-gluon and gluon-gluino scattering
amplitudes~\cite{TwoLoopSUSY,BDDgggg,BDDqqgg}. In this paper we study
the supersymmetry identities satisfied by gluino-gluino scattering
amplitudes.  The ambiguities in the gluino-gluino amplitudes, however,
causes a difficulty, since the precise value depends on the details of
how the calculation was performed. However, after subtracting the
singularities using the Catani formula, in the FDH scheme we
explicitly show that the finite remainders, which are free of the
ambiguities, do satisfy the expected supersymmetry identities.

This paper is arranged as follows. In \sect{GeneralitiesSection} we
present the helicity and color structure of the four-quark QCD
amplitudes.  The color space structure of the divergent part of the
amplitudes is reviewed in \sect{IRSection}.  \Sect{OneloopSection}
contains the one-loop QCD amplitudes, which appear in Catani's formula
for the divergent parts.  The one-loop amplitudes are presented in a
form valid through $\Ord(\e^2)$ since these are needed in Catani's
formula for the two-loop divergences. The structure of the finite
remainders after subtracting out the divergences using Catani's
formula is given in \sect{TwoLoopFiniteQCDSection}. These finite
remainders are tabulated in \app{QCDRemainderAppendix}. The $\NeqOne$
supersymmetric amplitudes, the ambiguity in their value and the
supersymmetry Ward identities, are discussed in
\sect{N=1AmplitudesSection}.  Some auxiliary functions needed for
describing the scheme shifts are given in
\app{OrdepsdeltaRemainderAppendix}.

%%%%%%%%%%%%%%%%%%%%%%%%%%%%%%%%%%%%%%%%%%%%%%%%%%%%%%%
\section{Helicity and color structure}
\label{GeneralitiesSection}

The three QCD processes considered in this paper are
\begin{eqnarray}
    q(p_1,\lambda_1) + \bar{q}(p_2,\lambda_2)
&\to& \bar{Q}(p_3,\lambda_3) + Q(p_4,\lambda_4)\,,
\label{qbQBlabel} \\
    q(p_1,\lambda_1) + \bar{Q}(p_2,\lambda_2)
&\to& q(p_3,\lambda_3) + \bar{Q}(p_4,\lambda_4)\,,
\label{qBqBlabel} \\
    q(p_1,\lambda_1) + Q(p_2,\lambda_2)
&\to& q(p_3,\lambda_3) + Q(p_4,\lambda_4)\,,
\label{qQqQlabel}
\end{eqnarray}
where we use a ``standard'' convention for the external momentum ($p_i$)
and helicity labeling ($\lambda_i$), {\it i.e}., particles 1 and 2 are taken
to be incoming, while particles 3 and 4 are assumed outgoing. The
identical quarks cases are easily obtained from these, as discussed below.

We handle ultraviolet and infrared singularities using dimensional
regularization. We consider a continuous set of schemes, labeled by a
parameter $\delta_R$ characterizing the number of virtual gluon
degrees of freedom circulating in loops.  Specifically, when the trace
of the Minkowski metric is encountered, we set
\begin{equation}
\eta^{\mu}{}_{\mu} \equiv D_s \equiv 4 - 2 \e \, \delta_R  \,,
\label{EtaTrace}
\end{equation}
corresponding to $2(1-\e\,\delta_R)$ gluon states in the loop.
Setting $\delta_R = 1$ corresponds to the HV scheme~\cite{HV}, while setting
$\delta_R = 0$ corresponds to the FDH scheme~\cite{BKgggg,TwoLoopSUSY}.

The CDR and HV schemes have the same standard $\MSbar$ coupling
constant, $\bar{\alpha}_s(\mu)$.  The coupling in a general $\delta_R$
scheme is related to this coupling at NNLO by~\cite{TwoLoopSUSY}
\begin{eqnarray}
\alpha_s^{\delta_R}(\mu) &=& \bar{\alpha}_s(\mu) \biggl[ 1
  + {C_A \over 6} (1-\delta_R) {\bar{\alpha}_s(\mu) \over 2\pi}
  \PlusBreak{ \bar{\alpha}_s(\mu) \bigglB }
    \biggl( {C_A^2 \over 36} (1-\delta_R)^2
        + {7 C_A^2 - 6 C_F T_R \Nf \over 12 } (1-\delta_R) \biggr)
          \biggl( {\bar{\alpha}_s(\mu) \over 2\pi} \biggr)^2
  \PlusBreak{ \bar{\alpha}_s(\mu) \bigglB }
    \Ord([\bar{\alpha}_s(\mu)]^3) \biggr] \,.
\label{deltaRMSconv}
\end{eqnarray}
Henceforth, for simplicity, we suppress the $\delta_R$ index on
$\alpha_s(\mu)$.

We work with ultraviolet renormalized amplitudes.  The relation
between the bare coupling $\alphas^u$ and renormalized coupling
$\alphas(\mu)$, through two-loop order, is~\cite{Catani},
\begin{equation}
\alphas^u \;\mu_0^{2\e} \;S_{\e} = \alphas(\mu) \;\mu^{2\e}
\left[ 1 - { \alphas(\mu) \over 2\pi } \; { b_0 \over \e }
         + \biggl( { \alphas(\mu) \over 2\pi} \biggr)^2
           \left( { b_0^2 \over \e^2 } - { b_1 \over 2\e } \right)
+ {\cal O}(\alphas^3(\mu)) \right]\,,
\label{TwoloopCoupling}
\end{equation}
where $\mu$ is the renormalization scale,
$S_\e = \exp[\e (\ln4\pi + \psi(1))]$, and
$\gamma = -\psi(1) = 0.5772\ldots$ is Euler's constant.
The first two coefficients appearing in the beta function for QCD,
or more generally $SU(N)$ gauge theory with $\Nf$ flavors of
massless fundamental representation quarks, are scheme-independent,
\begin{equation}
b_0 = {11 C_A - 4 T_R \Nf \over 6} \,, \hskip 2 cm
b_1 = {17 C_A^2 - ( 10 C_A + 6 C_F ) T_R \Nf \over 6} \,,
\label{QCDBetaCoeffs}
\end{equation}
where $C_A = N$, $C_F = (N^2-1)/(2N)$, and $T_R = 1/2$.
(Note that ref.~\cite{Catani} uses the notation
$\beta_0 = b_0/(2\pi)$, $\beta_1 = b_1/(2\pi)^2$.)

The perturbative expansion of the $\qqtoqq$ amplitude is
\begin{eqnarray}
\cm_\qqtoqq(\alphas(\mu), \mu;\{p\}) &=&
4\pi\alphas(\mu) \, \bigglB \cm_\qqtoqq^{(0)}(\mu;\{p\}) +
\label{RenExpand} \\
&& \hskip 1.5 cm
+ { \alphas(\mu) \over 2\pi } \cm_\qqtoqq^{(1)}(\mu;\{p\}) +
\nonumber \\
&& \hskip 1.5 cm
+ \biggl( { \alphas(\mu) \over 2\pi } \biggr)^2
\cm_\qqtoqq^{(2)}(\mu;\{p\}) + \Ord(\alphas^3(\mu)) \biggrB \,,
\nonumber
\end{eqnarray}
where $\cm_\qqtoqq^{(L)}(\mu;\{p\})$ is the $L^{\rm th}$ loop
contribution.  The same type of expansion holds, of course, for the
$q\bar{Q} \to q\bar{Q}$ and $qQ \to qQ$ amplitudes.
Equation~(\ref{TwoloopCoupling}) is equivalent to the following
$\MSbar$ renormalization prescriptions at one and two loops,
\begin{eqnarray}
\cm_\qqtoqq^{(1)}
&=&  S_\e^{-1} \, \cm_\qqtoqq^{(1){\rm unren}}
    - {b_0\over\e} \, \cm_\qqtoqq^{(0)} \,,
\label{OneloopCounterterm} \\
\cm_\qqtoqq^{(2)}
&=&  S_\e^{-2} \, \cm_\qqtoqq^{(2){\rm unren}}
  -  2 {b_0\over\e} \, S_\e^{-1}
                     \, \cm_\qqtoqq^{(1){\rm unren}}
  + \biggl( { b_0^2\over \e^2 } - {b_1 \over 2 \e} \biggr)
             \, \cm_\qqtoqq^{(0)} \,.
\label{TwoloopCounterterm}
\end{eqnarray}

We consider the following set of independent helicity configurations $h$
\begin{eqnarray}
h=1: &\;\;\;& q(p_1,+)+\bar{q}(p_2,-) \to \bar{Q}(p_3,-)+Q(p_4,+) \,,
\label{h1} \\
h=2: &\;\;\;& q(p_1,+)+\bar{q}(p_2,-) \to \bar{Q}(p_3,+)+Q(p_4,-) \,,
\label{h2} \\
h=3: &\;\;\;& q(p_1,+)+\bar{Q}(p_2,+) \to q(p_3,+)+\bar{Q}(p_4,+) \,,
\label{h3} \\
h=4: &\;\;\;& q(p_1,+)+\bar{Q}(p_2,-) \to q(p_3,+)+\bar{Q}(p_4,-) \,,
\label{h4} \\
h=5: &\;\;\;& q(p_1,+)+Q(p_2,+) \to q(p_3,+)+Q(p_4,+) \,,
\label{h5} \\
h=6: &\;\;\;& q(p_1,+)+Q(p_2,-) \to q(p_3,+)+Q(p_4,-) \,.
\label{h6}
\end{eqnarray}
Other configurations are simply related to these by symmetries.  For
example, the $q(p_1,-)$ amplitudes are obtained by parity (P), while
the $\bar{q}Q \to \bar{q}Q$ and $\bar{q}\bar{Q} \to \bar{q}\bar{Q}$
amplitudes are related to $q\bar{Q} \to q\bar{Q}$ and $qQ \to qQ$,
respectively, by an overall charge conjugation (C).  (Overall charge
conjugation is unaffected by the ambiguity to be discussed in
\sect{Ambiguities}.)  In defining these helicity configurations we
impose helicity conservation on the quark lines.

The cases where both quark lines are identical, {\it i.e.} $q\bar{q} \to
q\bar{q}$, $qq \to qq$ and $\bar{q} \bar{q} \to \bar{q} \bar{q}$, can
also be obtained from configurations (\ref{h1})-(\ref{h6}).  For
example, $q(p_1,+)+\bar{q}(p_2,-) \to \bar{q}(p_3,-)+q(p_4,+)$ can be
obtained by taking process (\ref{h1}) and subtracting the
process (\ref{h4}) with $p_3$ and $p_4$ interchanged.
The relative minus sign is due to the Fermi statistics.
More generally we have for the $qq \rightarrow qq$ process,
\begin{eqnarray}
{\cal M}_{q^+_1 {\bar q}^-_2 \rightarrow\, {\bar q}^-_3 q^+_4} &=&
{\cal M}_{q^+_1 {\bar q}^-_2 \rightarrow\, {\bar Q}^-_3 Q^+_4} -
{\cal M}_{q^+_1 {\bar Q}^-_2 \rightarrow \,q^+_4 {\bar Q}^-_3} \,,
\nonumber \\
{\cal M}_{q^+_1 {\bar q}^-_2 \rightarrow\, {\bar q}^+_3 q^-_4} &=&
{\cal M}_{q^+_1 {\bar q}^-_2 \rightarrow\, {\bar Q}^+_3 Q^-_4} \,,
\nonumber \\
{\cal M}_{q^+_1 q^+_2 \rightarrow\, q^+_3 q^+_4} &=&
{\cal M}_{q^+_1 Q^+_2 \rightarrow\, q^+_3 Q^+_4} -
{\cal M}_{q^+_1 Q^+_2 \rightarrow\, q^+_4 Q^+_3}\,,
\nonumber \\
{\cal M}_{q^+_1 q^-_2 \rightarrow\, q^+_3 q^-_4} &=&
{\cal M}_{q^+_1 Q^-_2 \rightarrow\, q^+_3 Q^-_4}\,,
\end{eqnarray}
where we use $q^+_1$ as a shorthand for $q(p_1,+)$ and so forth.

The color decomposition of the amplitudes is given by
\begin{equation}
\cm^{(L)}_{h} = S_{h} \times
\sum_{c=1}^2 \trc^{[c]} \times M^{(L),[c]}_{h} \,,
\qquad h=1,\dots 6,
\label{RemoveColorPhase}
\end{equation}
where the elements in the color bases are
\begin{eqnarray}
\trc^{[1]} = \delta^{i_4}_{~\ibar_1} \delta^{i_2}_{~\ibar_3} \,, \hskip0.5cm
\trc^{[2]} = \delta^{i_2}_{~\ibar_1} \delta^{i_4}_{~\ibar_3} \,, \qquad 
h=1,2,
\label{basis12}
\\
\trc^{[1]} = \delta^{i_2}_{~\ibar_1} \delta^{i_3}_{~\ibar_4} \,, \hskip0.5cm
\trc^{[2]} = \delta^{i_3}_{~\ibar_1} \delta^{i_2}_{~\ibar_4} \,, \qquad 
h=3,4,
\label{basis34}
\\
\trc^{[1]} = \delta^{i_4}_{~\ibar_1} \delta^{i_3}_{~\ibar_2} \,, \hskip0.5cm
\trc^{[2]} = \delta^{i_3}_{~\ibar_1} \delta^{i_4}_{~\ibar_2} \,, \qquad 
h=5,6.
\label{basis56}
\end{eqnarray}
In our previous calculations of the $gg \to gg$ and $q\bar{q} \to gg$
processes \cite{BDDgggg,BDDqqgg}, traces of products of color matrices
appeared. For consistency of notation, we have maintained here the ``{\rm 
Tr}''
notation for the color bases, even though there are no traces in
the present case.

The helicity-dependent, phase-containing factors $S_h$ arise
from evaluating the amplitudes in the spinor helicity
formalism~\cite{SpinorHelicity}. They are,
\begin{eqnarray}
&& S_{1} = - i {\spa3.1 \over \spa4.2} \,, \hskip1cm
S_{2} = - i {\spa4.1 \over \spa3.2} \,, \hskip1cm
S_{3} = - i {\spa2.1 \over \spa4.3} \,,
   \nonumber \\
&& S_{4} = - i {\spa4.1 \over \spa3.2} \,, \hskip1cm
S_{5} = - i {\spa2.1 \over \spa4.3} \,, \hskip1cm
S_{6} = - i {\spa4.1 \over \spa2.3} \,.
\label{SpinorPhases}
\end{eqnarray}
The spinor inner products~\cite{SpinorHelicity,MPReview} are
$\spa{i}.j = \langle i^- | j^+\rangle$ and
$\spb{i}.j = \langle i^+| j^-\rangle$,
where $|i^{\pm}\rangle$ are massless Weyl spinors of momentum $k_i$,
labeled with the sign of the helicity.  They are anti-symmetric, with
norm $|\spa{i}.j| = |\spb{i}.j| = \sqrt{s_{ij}}$, where
$s_{ij} = 2k_i\cdot k_j$. Notice that the  prefactors
$S_h$ are all pure phases, {\it i.e., }
\begin{equation}
|S_h|^2 = 1 \,, \qquad h = 1 \ldots 6 \,.
\end{equation}
The quantities $M^{(0),[c]}_h$ depend only on the Mandelstam variables
$s=(p_1+p_2)^2$, $t=(p_1-p_4)^2$ and $u=(p_1-p_3)^2$.
At tree level in the color bases~(\ref{basis12}),
(\ref{basis34}) and (\ref{basis56}) they are given by,
\begin{eqnarray}
&&
M^{(0),[1]}_1 = {u \over s} \,, \qquad
M^{(0),[1]}_2 = {t \over s} \,, \qquad
M^{(0),[1]}_3 = {s \over u} \,,
\nonumber \\
&&
M^{(0),[1]}_4 = {t \over u} \,, \qquad
M^{(0),[1]}_5 = {s \over u} \,, \qquad
M^{(0),[1]}_6 = {t \over u} \,, \nonumber \\
&& M^{(0),[2]}_h = -{1 \over N} M^{(0),[1]}_h \,, \qquad h=1 \ldots 6 \,.
\label{TreeAmps}
\end{eqnarray}
%

%%%%%%%%%%%%%%%%%%%%%%%%%%%%%%%%%%%%%%%%%%%%%%%%%%%%%%%%%%%%%%%%%%%%%%%

\section{Infrared singularities}
\label{IRSection}

We now briefly review the structure of the soft and collinear singularities
of dimensionally regularized one- and two-loop QCD
amplitudes, using Catani's color space notation~\cite{Catani}.
The finite remainders are given in \sect{OneloopRemainderSubsection},
\sect{TwoLoopFinRemSubsection} and \app{QCDRemainderAppendix}.

The infrared divergences of renormalized one- and two-loop $n$-point
amplitudes are given by~\cite{KunsztIR,Catani},
\begin{eqnarray}
| \cm_n^{(1)}(\mu; \{p\}) \ra_{\RS} &=& {\bom I}^{(1)}(\e, \mu; \{p\})
  \; | \cm_n^{(0)}(\mu; \{p\}) \ra_{\RS}
  + |\cm_n^{(1){\rm fin}}(\mu; \{p\}) \ra_{\RS} \,,
	\label{OneloopCatani} \\
| \cm_n^{(2)}(\mu; \{p\}) \ra_{\RS} &=& {\bom I}^{(1)}(\e, \mu; \{p\})
  \; | \cm_n^{(1)}(\mu; \{p\}) \ra_{\RS}
\nonumber \\ && \null
+ {\bom I}^{(2)}_{\RS}(\e, \mu; \{p\}) \;
         | \cm_n^{(0)}(\mu; \{p\}) \ra_{\RS}
+ |\cm_n^{(2){\rm fin}}(\mu; \{p\}) \ra_{\RS} \,, \hskip .5 cm
\label{TwoloopCatani}
\end{eqnarray}
where the ``ket'' notation $|\cm_n^{(L)}(\mu; \{p\}) \ra_{\RS}$
indicates that the $L$-loop amplitude is treated as a vector in
color space.
The components of this vector are given by the $M^{(L),[c]}_h$ in
\eqn{RemoveColorPhase}.
The subscript $\RS$ indicates that a quantity depends on the choice of
regularization and renormalization scheme. The divergences of
$\cm_n^{(1)}$ are encoded in the color operator ${\bom I}^{(1)}$,
while those of $\cm_n^{(2)}$ also involve the scheme-dependent operator
${\bom I}^{(2)}_{\RS}$.

The operator ${\bom I}^{(1)}$, which controls the one-loop
singularity structure, is given by
\begin{equation}
{\bom I}^{(1)}(\e,\mu;\{p\}) = \frac{1}{2} {e^{-\e \psi(1)} \over
\Gamma(1-\e)} \sum_{i=1}^n \, \sum_{j \neq i}^n \, {\bom T}_i \cdot
{\bom T}_j \Biggl[ {1 \over \e^2} + {\gamma_i \over {\bom T}_i^2 } \,
{1 \over \e} \Biggr] \Biggl( \frac{\mu^2 e^{-i\lambda_{ij} \pi}}{2
p_i\cdot p_j} \Biggr)^{\e} \,,
\label{CataniGeneral}
\end{equation}
where $\lambda_{ij}=+1$ if $i$ and $j$ are both incoming or outgoing
partons, and $\lambda_{ij}=0$ otherwise. The color charge ${\bom T}_i =
\{T^a_i\}$ is a vector with respect to the generator label $a$, and an
$SU(N)$ matrix with respect to the color indices of the outgoing
parton $i$. For external fermions, the ratio
\begin{equation}
{ \gamma_q \over {\bom T}_i^2}  = {3\over2} \,,
\label{QCDfermionValues}
\end{equation}
is independent of the representation.   For quarks,
${\bom T}_i^2 = C_F = (N^2-1)/(2N)$; for gluinos ${\bom T}_i^2 = C_A = N$.
The two-loop operator ${\bom I}^{(2)}_{\RS}$ is~\cite{Catani}
\begin{eqnarray}
{\bom I}^{(2)}_{\RS}(\e,\mu;\{p\})
& =& - \frac{1}{2} {\bom I}^{(1)}(\e,\mu;\{p\})
\left( {\bom I}^{(1)}(\e,\mu;\{p\}) + {2 b_0 \over \e} \right)
  \PlusBreak{}
{e^{+\e \psi(1)} \Gamma(1-2\e) \over \Gamma(1-\e)}
\left( {b_0 \over \e} + K_\RS \right) {\bom I}^{(1)}(2\e,\mu;\{p\})
  \PlusBreak{}
  {\bom H}^{(2)}_{\RS}(\e,\mu;\{p\}) \,,
\label{CataniGeneralI2}
\end{eqnarray}
where the coefficient $K_\RS$ depends on $\delta_R$ and
is given by~\cite{Catani,BDDgggg}
\begin{equation}
K_\RS = \left[ \frac{67}{18} - \frac{\pi^2}{6}
    - \biggl( {1\over6} + {4\over9} \e \biggr) (1-\delta_R) \right] C_A
- \frac{10}{9} T_R \Nf \,.    \label{CataniK}
\end{equation}

The function ${\bom H}^{(2)}_{\RS}$ contains only {\em single} poles,
and splits into two types of terms,
\begin{equation}
{\bom H}^{(2)}(\e) =
{ e^{-\e\psi(1)} \over 4\e \, \Gamma(1-\e) }
\biggl( { \mu^2 \over -s } \biggr)^{2\e}
  \Bigl( 4 H_q^{(2)} \, {\bom 1}
       + \hat{\bom H}^{(2)} \Bigr) \,,
\label{OurH}
\end{equation}
where a standard analytic continuation is needed to bring $s$ to the
physical region.  We find that the term proportional to the identity
matrix in color space ${\bom 1}$ is given by
\begin{eqnarray}
H_q^{(2)} &=&
\biggl( {13\over2} \zeta_3 - {23\over48} \pi^2 + {245\over216} \biggr)
C_A C_F
+ \biggl( - 6 \zeta_3 + {\pi^2\over2} - {3\over8} \biggr) C_F^2
+ \biggl( {\pi^2\over12} - {25\over54} \biggr) C_F T_R \Nf
\nonumber \\
\hskip0.5cm &&
+ \biggl( - {4\over3} C_A C_F + {1\over2} C_F^2
           + {1\over6} C_F T_R \Nf \biggr) (1-\delta_R) \,.
\label{Hquark}
\end{eqnarray}
This term survives the sum over colors, and the expression for
$H_q^{(2)}$ in the HV scheme ($\delta_R=1$) agrees, as expected, with
previous color-summed results in the CDR
scheme~\cite{GOTYqqqq,GOTYqqgg,GOTYgggg,AGTYphotons}.

The second term in ${\bom H}^{(2)}(\e)$ has exactly the same type of
nontrivial color and kinematic dependence found in the $\ggtogg$ and 
$\qqtogg$
helicity amplitudes~\cite{BDDgggg,BDDqqgg}, namely
\begin{equation}
\hat{\bom H}^{(2)} = - 4 \, \ln\biggl( {-s\over-t} \biggr)
                            \ln\biggl( {-t\over-u} \biggr)
                            \ln\biggl( {-u\over-s} \biggr)
      \times \Bigl[ {\bom T}_1 \cdot {\bom T}_2 \,,
                    {\bom T}_2 \cdot {\bom T}_3 \Bigr] \,,
\label{HExtra}
\end{equation}
where again appropriate analytic continuation are required.  For
example, in the $s$-channel, $\ln((-s)/(-t)) \to \ln s - \ln(-t) -
i\pi$.  In refs. \cite{BDDgggg,BDDqqgg,Gloverqqgg} it was shown that
for the $\ggtogg$ and $q\bar{q} \to gg$ amplitudes the structure of
this term is independent of the helicity configuration, and whether
the external legs are quarks or gluons.  Here once again, we find this
to be the case.  An ansatz generalizing $\hat{\bom H}^{(2)}$ for an
arbitrary number of external legs has been presented recently in
\cite{BDKtwoloopsplit}.  Because of the commutator structure,
$\hat{\bom H}^{(2)}$ vanishes when sandwiched between tree amplitudes,
after performing a sum over colors; hence it drops out of the
color-summed interference of the two-loop amplitudes with the tree
amplitudes~\cite{BDDgggg,BDDqqgg}.

For each color basis we will have a different ${\bom I}^{(1)}$ matrix.
For the basis (\ref{basis12}) we have,
\begin{equation}
{\bom I}^{(1)}(\epsilon) =
{e^{-\epsilon \psi (1)} \over \Gamma (1-\epsilon )} \, \xi_q
\left(
\begin{array}{c c}
2 C_F \tT - {1 \over N} (\tS-\tU) &            \tT - \tU              \\
          \tS - \tU               & 2 C_F \tS - {1 \over N} (\tT - \tU)
\end{array}
\right) \,,
\label{I1Matrix}
\end{equation}
where
\begin{equation}
\tS = \left({\mu^2\over -s}\right)^\e \,, \qquad
  \tT = \left({\mu^2\over -t}\right)^\e \,, \qquad
  \tU = \left({\mu^2\over -u}\right)^\e \,, \qquad
\xi_q = {1\over\e^2} + {3\over 2\e}\,.
\label{STUedef}
\end{equation}
The corresponding operator for $q\bar{Q} \to q\bar{Q}$ in the basis
(\ref{basis34}) is obtained by changing $\tS \to \tU $, $\tT \to \tS $
and $\tU \to \tT$ in \eqn{I1Matrix}. Similarly, the operator for $qQ \to qQ$
in the basis (\ref{basis56}) is obtained by exchanging $\tS$ and $\tU$
in (\ref{I1Matrix}).

A typical partonic cross section requires an amplitude interference,
summed over all external colors.  Such interferences are evaluated in
the color bases~(\ref{basis12}), (\ref{basis34}), (\ref{basis56}) as
\begin{equation}
{\cal I}^{(L,L')}_{\lambda_1\lambda_2\lambda_3\lambda_4}
\equiv
\langle \cm_{\lambda_1\lambda_2\lambda_3\lambda_4}^{(L)}
      | \cm_{\lambda_1\lambda_2\lambda_3\lambda_4}^{(L')} \rangle
= \sum_{c,c'=1}^2 M^{(L),[c] \, *}_{\lambda_1\lambda_2\lambda_3\lambda_4}
                 {\cal C\!C}_{cc'}
                  M^{(L'),[c']}_{\lambda_1\lambda_2\lambda_3\lambda_4} \,,
\label{ColorSumGen}
\end{equation}
where the symmetric matrix
${\cal C\!C}_{cc'} \equiv \sum_{\rm colors} \trc^{[c]\,*} \trc^{[c']}$
is
\begin{equation}
{\cal C\!C} = \left(
\begin{array}{cc}
N^2 & N \\
N   & N^2
\end{array}
\right) \,.
\label{ColorSumMatrix}
\end{equation}

\section{One-loop QCD amplitudes}
\label{OneloopSection}

The one-loop amplitudes for $\qqtoqq$ were first evaluated through
$\Ord(\e^0)$ as an interference with the tree amplitude in the
CDR scheme~\cite{EllisSexton}.  Later they were evaluated as helicity
amplitudes in the HV, FDH and $\DRbar$ schemes~\cite{KSTfourparton}.

Because ${\bom I}^{(1)}$ contains terms of order $1/\e^2$, the ${\bom
I}^{(1)} | \cm^{(1)} \rangle_\RS$ term in the infrared
decomposition~(\ref{TwoloopCatani}) of the two-loop $\qqtoqq$
amplitudes requires the series expansion of the one-loop amplitudes
through $\Ord(\e^2)$.  In section~\ref{AllOrdersSubsection} we present
the all-order results in the color bases~(\ref{basis12}),
(\ref{basis34}), (\ref{basis56}), with the normalizations implicit in
\eqn{RenExpand}, in terms of integral functions whose series
expansions have been evaluated to the required
order~\cite{AllPlusTwo,BhabhaTwoLoop}.

In ref.~\cite{BDDgggg} it was shown that the $\Ord(\e)$ terms in
one-loop amplitudes such as $\cm_\qqtoqq^{(1)}$ are not required for
the construction of a numerical NNLO program, once the divergences
have been subtracted from $\cm_\qqtoqq^{(2)}$ using Catani's formula.
Thus we need only present explicit {formul\ae} for the $\Ord(\e^0)$
finite remainders $\cm_\qqtoqq^{(1){\rm fin}}$ of the one-loop
amplitudes, after ultraviolet
renormalization~(\ref{OneloopCounterterm}) and subtraction of infrared
divergences~(\ref{OneloopCatani}).  We do this in
section~\ref{OneloopRemainderSubsection}, for a general $\delta_R$
scheme.  The formula~(\ref{Twoloopc12}) for converting the two-loop
finite remainders $\cm_\qqtoqq^{(2){\rm fin}}$ from one scheme to
another are most compactly presented in terms of the
$\delta_R$-dependent parts of the one-loop amplitudes at order $\e$;
the explicit values of these quantities are collected in
appendix~\ref{OrdepsdeltaRemainderAppendix}.

%%%%%%%%%%%%%%%%%%%%%%%%%%%%%%%%%%%%%%%%%%%%%%%%%%%%%%%%%%

\subsection{One-loop all orders in $\epsilon$ QCD amplitudes}
\label{AllOrdersSubsection}

We now present the one-loop $\qqtoqq$ amplitudes in the color
bases~(\ref{basis12}), (\ref{basis34}), (\ref{basis56}), with the
normalizations implicit in \eqn{RenExpand}, in a form valid to all
orders in $\e$.

At one loop the crossing properties of the amplitudes are relatively
simple, so we present the explicit values of the helicity amplitudes
for the process $\qqtoqq$. The process $q \bar{Q} \to q \bar{Q}$ may
be obtained from these by crossing the initial antiquark and final
quark into the final and initial states respectively, and the process
$qQ \to qQ$ may be obtained by crossing antiquark 2 into the final
state, and antiquark 3 into the initial state,
\begin{eqnarray}
M^{(1),[c]}_3(s,t,u) &=& M^{(1),[c]}_2(u,s,t) \,, \qquad c=1,2, \\
M^{(1),[c]}_4(s,t,u) &=& M^{(1),[c]}_1(u,s,t) \,, \qquad c=1,2, \\
M^{(1),[c]}_5(s,t,u) &=& M^{(1),[c]}_1(u,t,s) \,, \qquad c=1,2, \\
M^{(1),[c]}_6(s,t,u) &=& M^{(1),[c]}_2(u,t,s) \,, \qquad c=1,2,
\end{eqnarray}
where $M^{(L),[c]}_h$ is defined in \eqn{RemoveColorPhase} with
the color bases (\ref{basis12}), (\ref{basis34}) and (\ref{basis56})
using the helicity configurations $h$ defined in
eqs. (2.11)-(2.16). After crossing, appropriate analytic continuations
are required to bring each function into the physical region.

A compact representation for the unrenormalized amplitudes is,
\begin{eqnarray}
M^{(1),[1]}_1(s,t,u) &=&
%%%%% begin : M111EpsPaper
A_+(s,u,t) + {2 \over N}A_1(s,t,u)
+ \biggl( N - {2 \over N} \biggl)A_2(s,t,u)
%%%%% end : M111EpsPaper
\,, \label{M111Oneloop} \\
%-{b_0 \over \e} M^{(0)[1]}_1
M^{(1),[2]}_1(s,t,u) &=&
%%%%% begin : M121EpsPaper
-{1 \over N}A_+(s,u,t)-\biggl( 1 + {1 \over N^2} \biggr)A_1(s,t,u)
+{1 \over N^2}A_2(s,t,u)
%%%%% end : M121EpsPaper
\,, \hskip1cm  \label{M121Oneloop}\\
%{} \MinusBreak{}
% {b_0 \over \e} M^{(0)[2]}_1 \,, \hskip1cm \\
M^{(1),[1]}_2(s,t,u) &=&
%%%%% begin : M112EpsPaper
-A_-(s,t,u) + {2 \over N}A_3(s,t,u)
+ \biggl( N - {2 \over N} \biggr)A_4(s,t,u)
%-{b_0 \over \e} M^{(0)[1]}_2
%%%%% end : M112EpsPaper
\,,  \label{M112Oneloop}\\
M^{(1),[2]}_2(s,t,u) &=&
%%%%% begin : M122EpsPaper
{1 \over N}A_-(s,t,u)
- \biggl( 1 + {1 \over N^2} \biggr)A_3(s,t,u)
         + {1 \over N^2}A_4(s,t,u)
%%%%% end : M122EpsPaper
\,, \label{M122Oneloop}
%{} \MinusBreak{}
% {b_0 \over \e} M^{(0)[2]}_2 \,,
\end{eqnarray}
where,
\begin{eqnarray}
A_{\pm}(s,t,u) &=&
%%%%% begin : ApmEpsPaper
{1 \over 2 \e-1} \biggl[ {N \over 2} \e^3 \delta_R^2 s
    \pm \biggl( {\e {} (1-\e \delta_R) \over 2 (\e-1)}
            \biggl( {\e-2 \over 2 \e-3} N
                 + {2 \e-1 \over N} \biggr)
{} \PlusBreak{ {1 \over 2 \e-1}
               \bigglB {N \over 2} \e^3 \delta_R^2 s \pm \bigglP \null }
                   { \e {} (\e-1) \over 2 \e-3} N_f
                 - {1 \over N} \biggr) t  \biggr] \Trifour{(s)}
%%%%% end : ApmEpsPaper
\,, \\
A_1(s,t,u) &=&
%%%%% begin : A1EpsPaper
{\e^2 \over 2 (2 \e-1)} \delta_R s \Trifour{(s)}
-\biggl[ {\e^2 \over 2 (2 \e-1)} \delta_R {} (1+\e \delta_R) u
        + {u^2 \over s} \biggr]  \Trifour{(u)}
{} \MinusBreak{}
\biggl[ {\e^2 \over 2} \delta_R {} (1+\e \delta_R) s
        -\biggl(\e^2 \delta_R - (2 \e-1) {t \over s} \biggr) u
\biggr] \Boxsix(s,u)
%%%%% end : A1EpsPaper
\,, \\
A_2(s,t,u) &=&
%%%%% begin : A2EpsPaper
{\e \over 2 (2 \e-1)} s \Trifour{(s)}
-\biggl[ {\e \over 2 \e-1} (1+\e^2 \delta_R^2)
        +2 {u \over s} \biggr] {t \over 2} \Trifour{(t)}
{} \PlusBreak{}
\biggl[ {1-\e \over 2} (1+\e^2 \delta_R^2) s
       - \biggl( \e^2 \delta_R+(1-2 \e) {t \over s} \biggr) u
\biggr] \Boxsix(s,t)
%%%%% end : A2EpsPaper
\,, \\
A_3(s,t,u) &=&
%%%%% begin : A3EpsPaper
\biggl[ {\e \over 2} (\e-1) \delta_R {} (1+\e \delta_R) s
  + \e^2 \delta_R t - \e {} {t^2 \over s}
+ (1-\e) {(u^2+t^2) \over 2 s} \biggr] \Boxsix(s,u)
{} \PlusBreak{}
\biggl[ {\e^2 \over 2} \delta_R {} (1+\e \delta_R)
+ (1-2 \e) {t \over s} - {\e \over 2}
\biggr] {u \over 2 \e-1} \Trifour{(u)}
{} \PlusBreak{}
{\e {} (1-\e \delta_R) \over 2 (2 \e-1)} s \Trifour{(s)}
%%%%% end : A3EpsPaper
\,, \\
A_4(s,t,u) &=&
%%%%% begin : A4EpsPaper
\biggl[ {\e^3 \over 2 (2 \e-1)} \delta_R^2 t
       - {t^2 \over s} \biggr] \Trifour{(t)}
{} \PlusBreak{}
\biggl[ {\e^3 \over 2} \delta_R^2 s - \e^2 \delta_R t
  +((1-\e) u + \e t) {t \over s} \biggr] \Boxsix(s,t)
%%%%% end : A4EpsPaper
\,.
\end{eqnarray}
Here $\Trifour(s)$ is the scalar triangle integral in $4-2\e$
dimensions with one external massive leg, and $\Boxsix(s,t)$ is the
all-massless scalar box integral in $6-2\e$ dimensions.  The expansion
of these integrals to $\Ord(\e^2)$ in the various kinematic channels is
given, for example, in refs.~\cite{BhabhaTwoLoop,gggamgamPaper}. The
renormalized amplitudes are obtained by subtracting $b_0
M^{(0),[c]}_h/\e$ from each of
eqs.~(\ref{M111Oneloop})-(\ref{M122Oneloop}), where $c,h$ correspond
to the amplitude under consideration.

\subsection{Finite remainders}
\label{OneloopRemainderSubsection}

We now give the finite remainders of the one-loop $\qqtoqq$, $q\bar{Q}
\to q\bar{Q}$ and $qQ \to qQ$ amplitudes at $\Ord(\e^0)$, defined by
$\cm_\qqtogg^{(1){\rm fin}}$ and $\cm_\qgtogq^{(1){\rm fin}}$
in~\eqn{OneloopCatani} and color decomposed into $M^{(1),[c]{\rm
fin}}_{h}$ in \eqn{RemoveColorPhase}.  We write,
\begin{eqnarray}
M^{(1),[1]{\rm fin}}_h &=&
\left[ -b_0 \left( \ln \left( {s \over \mu^2} \right) -i\pi \right) +
\left( {N \over 3} - {1 \over 2 N} \right) (1-\delta_R)\right] M^{(0),[1]}_h
\PlusBreak{}
N a^{[1]}_h + {1 \over N} b^{[1]}_h + N_f d^{[1]}_h \,, 
\label{Mh11fin} \\
M^{(1),[2]{\rm fin}}_h &=&
\left[ -b_0 \left( \ln \left( {s \over \mu^2} \right) -i\pi \right) +
\left( {N \over 3} - {1 \over 2 N} \right) (1-\delta_R)\right] M^{(0),[2]}_h
\PlusBreak{}
h^{[2]}_h + {N_f \over N} j^{[2]}_h + {1 \over N^2} k^{[2]}_h \,.
\label{Mh12fin} 
\end{eqnarray}

For the $h=1$ helicity amplitude, the independent remainder functions $a$, 
$b$,
$d$, $h$, $j$ and $k$ are
\begin{eqnarray}
a^{[1]}_1 &=&
%%%%% begin : aa11Paper
{13 \over 18} y + \biggl( {x \over 2} - y \biggr) X
+{1 \over 4} \biggl( y + {x^2 \over y} \biggr) X^2
+i \pi {} \biggl[ {x \over 2} - y
              + {1 \over 2} \biggl( {x^2 \over y} + y \biggr) X \biggr]
%%%%% end : aa11Paper
\,, \\
b^{[1]}_1 &=&
%%%%% begin : bb11Paper
(2 y - x) X - {1 \over 2} \biggl( {x^2 \over y} + y \biggr) X^2
+ y {} (Y^2-3 Y+4)
{} \PlusBreak{}
i \pi {}\biggl[ 1 - \biggl( {x^2 \over y} + y \biggr) X + 2 y Y \biggr]
%%%%% end : bb11Paper
\,, \\
d^{[1]}_1 &=&
%%%%% begin : dd11Paper
-{5 \over 9} y
%%%%% end : dd11Paper
\,, \\
h^{[2]}_1 &=&
%%%%% begin : hh12Paper
-{y \over 2} \biggl( Y^2 - 3 Y + {13 \over 9} \biggr)
-i \pi y {}\biggl( Y - {3 \over 2} \biggr)
%%%%% end : hh12Paper
\,, \\
j^{[2]}_1 &=&
%%%%% begin : jj12Paper
{5 \over 9} y
%%%%% end : jj12Paper
\,, \\
k^{[2]}_1 &=&
%%%%% begin : kk12Paper
\biggl( {x \over 2} - y \biggr) X
+ {1 \over 4} \biggl( {x^2 \over y} + y \biggr) X^2
+ y {} \biggl( {3 \over 2} Y - {1 \over 2} Y^2 - 4 \biggr)
{} \MinusBreak{}
i \pi {}\biggl[ {1 \over 2}
              - {1 \over 2} \biggl( {x^2 \over y} + y \biggr) X + y Y 
\biggr]
%%%%% end : kk12Paper
\,,
\end{eqnarray}
where
\begin{equation}
x={t \over s} \,,\qquad y={u \over s} \,,\qquad X=\ln (-x) \,,\qquad Y=\ln 
(-y)
\,.
\end{equation}

For $h=2$ the functions are
\begin{eqnarray}
a^{[1]}_2 &=&
%%%%% begin : aa21Paper
{x \over 2} \biggl( X^2 - 3 X + {13 \over 9} \biggr)
+i \pi x {}\biggl( X - {3 \over 2} \biggr)
%%%%% end : aa21Paper
\,, \\
b^{[1]}_2 &=&
%%%%% begin : bb21Paper
x {} (4 + 3 X - X^2) + {1 \over 2} \biggl( {y^2 \over x} + x \biggr) Y^2
+(y-2 x) Y
{} \MinusBreak{}
i \pi {}\biggl[ 1 + 2 x X - \biggl( x + {y^2 \over x} \biggr) Y \biggr]
%%%%% end : bb21Paper
\,, \\
d^{[1]}_2 &=&
%%%%% begin : dd21Paper
-{5 \over 9} x
%%%%% end : dd21Paper
\,, \\
h^{[2]}_2 &=&
%%%%% begin : hh22Paper
-{13 \over 18} x - {1 \over 4} \biggl( {y^2 \over x} + x \biggr) Y^2
-\biggl( {y \over 2} - x \biggr) Y
-i \pi {}\biggl[ {y \over 2} - x
              + {1 \over 2} \biggl( x + {y^2 \over x} \biggr) Y \biggr]
%%%%% end : hh22Paper
\,, \\
j^{[2]}_2 &=&
%%%%% begin : jj22Paper
{5 \over 9} x
%%%%% end : jj22Paper
\,, \\
k^{[2]}_2 &=&
%%%%% begin : kk22Paper
-{1 \over 4} \biggl( {y^2 \over x} + x \biggr) Y^2
- \biggl( {y \over 2} - x \biggr) Y
+ x {} \biggl( {1 \over 2} X^2 - 4 - {3 \over 2} X \biggr)
{} \PlusBreak{}
i \pi {} \biggl[ {1 \over 2} + x X
             - {1 \over 2} \biggl( x + {y^2 \over x} \biggr) Y \biggr]
%%%%% end : kk22Paper
\,.
\end{eqnarray}

For $h=3$ the functions are
\begin{eqnarray}
a^{[1]}_3 &=&
%%%%% begin : aa31Paper
\biggl( {1 \over 2} Y^2 - {1 \over 3} Y + {13 \over 18} \biggr) {1 \over y}
+i \pi {}\biggl( Y - {1 \over 3} \biggr) {1 \over y}
%%%%% end : aa31Paper
\,, \\
b^{[1]}_3 &=&
%%%%% begin : bb31Paper
\biggl[ (1 + x^2) \biggl( {\pi^2 \over 2} - X Y + {1 \over 2} X^2 \biggr)
+ 4 - (2 - x) X + {1 \over 2} (x^2 - 1) Y^2 \biggr] {1 \over y}
+ Y
{} \MinusBreak{}
i \pi {}(2 Y + 3) {1 \over y}
%%%%% end : bb31Paper
\,, \\
d^{[1]}_3 &=&
%%%%% begin : dd31Paper
\biggl( {1 \over 3} Y - {5 \over 9} \biggr) {1 \over y}
+ i {\pi \over 3 y}
%%%%% end : dd31Paper
\,, \\
h^{[2]}_3 &=&
%%%%% begin : hh32Paper
\biggl[ (1 + x^2) \biggl( {1 \over 2} X Y
                           - {1 \over 4} (\pi^2 + Y^2 + X^2) \biggr)
        + \biggl( 1 - {x \over 2} \biggr) X + {1 \over 3} Y
        -{13 \over 18} \biggr] {1 \over y}
{} \MinusBreak{}
{1 \over 2} Y + i \pi {} {11 \over 6 y}
%%%%% end : hh32Paper
\,, \\
j^{[2]}_3 &=&
%%%%% begin : jj32Paper
\biggl( {5 \over 9} - {1 \over 3} Y \biggr) {1 \over y} - i {\pi \over 3 y}
%%%%% end : jj32Paper
\,, \\
k^{[2]}_3 &=&
%%%%% begin : kk32Paper
\biggl[ {1 \over 4} (1 + x^2) (2 X Y - \pi^2 - X^2)
       -4 + \biggl( 1 - {x \over 2} \biggr) X
       + {1 \over 4} (1 - x^2) Y^2 \biggr] {1 \over y}
- {1 \over 2} Y
{} \PlusBreak{}
i \pi {}\biggl( Y + {3 \over 2} \biggr) {1 \over y}
%%%%% end : kk32Paper
\,.
\end{eqnarray}

For $h=4$ the functions are
\begin{eqnarray}
a^{[1]}_4 &=&
%%%%% begin : aa41Paper
\biggl[ x {} \biggl( {13 \over 18} - {1 \over 3} Y \biggr)
+ {1 \over 4} \biggl( x + {1 \over x} \biggr) Y^2 \biggr] {1 \over y}
+ {1 \over 2} Y
{} \MinusBreak{}
  i \pi {}\biggl[ {5 \over 3} x + 1
              - \biggl( {1 \over x} + x \biggr) Y \biggr] {1 \over 2 y}
%%%%% end : aa41Paper
\,, \\
b^{[1]}_4 &=&
%%%%% begin : bb41Paper
\biggl[ x {} (X^2 + \pi^2 + 4 - 3 X - 2 X Y)
- {1 \over 2} \biggl( {1 \over x} - x \biggr) Y^2 \biggr] {1 \over y} - Y
{} \MinusBreak{}
i \pi {}\biggl[ 2 x - 1
                 +\biggl( x + {1 \over x} \biggr) Y \biggr] {1 \over y}
%%%%% end : bb41Paper
\,, \\
d^{[1]}_4 &=&
%%%%% begin : dd41Paper
\biggl( {1 \over 3} Y - {5 \over 9} \biggr) {x \over y} + i \pi {}{x \over 3 
y}
%%%%% end : dd41Paper
\,, \\
h^{[2]}_4 &=&
%%%%% begin : hh42Paper
\biggl[ -{\pi^2 \over 2} + {3 \over 2} X + {1 \over 3} Y
-{1 \over 2} (X^2 + Y^2) + X Y - {13 \over 18} \biggr] {x \over y}
+ i \pi {} {11 x \over 6 y}
%%%%% end : hh42Paper
\,, \\
j^{[2]}_4 &=&
%%%%% begin : jj42Paper
-\biggl( {1 \over 3} Y
- {5 \over 9} \biggr) {x \over y} - i \pi {}{x \over 3 y}
%%%%% end : jj42Paper
\,, \\
k^{[2]}_4 &=&
%%%%% begin : kk42Paper
-\biggl[ x {} \biggl( {\pi^2 \over 2} + 4 - {3 \over 2} X
                  + {1 \over 2} X^2 - X Y \biggr)
-{1 \over 4} \biggl( {1 \over x} - x \biggr) Y^2 \biggr]
   {1 \over y} + {1 \over 2} Y
{} \PlusBreak{}
i \pi {} \biggl[ x - {1 \over 2}
      + {1 \over 2} \biggl( {1 \over x} + x \biggr) Y \biggr] {1 \over y}
%%%%% end : kk42Paper
\,.
\end{eqnarray}

For $h=5$ the functions are
\begin{eqnarray}
a^{[1]}_5 &=&
%%%%% begin : aa51Paper
\biggl[ {1 \over 4} (1 + x^2) (\pi^2 - 2 X Y + Y^2 + X^2)
       +{13 \over 18} - \biggl( 1 - {x \over 2} \biggr) X
       - {1 \over 3} Y \biggr] {1 \over y}
{} \PlusBreak{}
{1 \over 2} Y - i \pi {} {11 \over 6 y}
%%%%% end : aa51Paper
\,, \\
b^{[1]}_5 &=&
%%%%% begin : bb51Paper
\biggl[ {1 \over 2} (1 + x^2) (2 X Y - X^2 - \pi^2)
+ 4 - (x - 2) X + {1 \over 2} (1 - x^2) Y^2 \biggr] {1 \over y}
{} \MinusBreak{}
  Y + i \pi {}(2 Y+3) {1 \over y}
%%%%% end : bb51Paper
\,, \\
d^{[1]}_5 &=&
%%%%% begin : dd51Paper
\biggl( {1 \over 3} Y - {5 \over 9} \biggr) {1 \over y}
      + i {\pi \over 3 y}
%%%%% end : dd51Paper
\,, \\
h^{[2]}_5 &=&
%%%%% begin : hh52Paper
-\biggl( {1 \over 2} Y^2 - {1 \over 3} Y + {13 \over 18} \biggr) {1 \over y}
- i \pi {} \biggl( Y - {1 \over 3} \biggr) {1 \over y}
%%%%% end : hh52Paper
\,, \\
j^{[2]}_5 &=&
%%%%% begin : jj52Paper
-\biggl( {1 \over 3} Y - {5 \over 9} \biggr)
       {1 \over y} - i {\pi \over 3 y}
%%%%% end : jj52Paper
\,, \\
k ^{[2]}_5&=&
%%%%% begin : kk52Paper
\biggl[ {1 \over 4} (1 + x^2) (X^2 + \pi^2 - 2 X Y)
       - 4 + \biggl( {x \over 2} - 1 \biggr) X
       - {1 \over 4} (1 - x^2) Y^2 \biggr] {1 \over y}
{} \PlusBreak{}
{1 \over 2} Y - i \pi {}\biggl( Y + {3 \over 2} \biggr) {1 \over y}
%%%%% end : kk52Paper
\,.
\end{eqnarray}

For $h=6$ the functions are
\begin{eqnarray}
a^{[1]}_6 &=&
%%%%% begin : aa61Paper
\biggl[ {\pi^2 \over 2} - {3 \over 2} X - {1 \over 3} Y
+{1 \over 2} (X^2 + Y^2) - X Y + {13 \over 18} \biggr] {x \over y}
- i \pi {}{11 x \over 6 y}
%%%%% end : aa61Paper
\,, \\
b^{[1]}_6 &=&
%%%%% begin : bb61Paper
\biggl[ x {} (4 - \pi^2 + 3 X - X^2 + 2 X Y)
+ {1 \over 2} \biggl( {1 \over x} - x \biggr) Y^2 \biggr] {1 \over y} + Y
{} \PlusBreak{}
i \pi {}\biggl[ 2 x - 1
                 + \biggl( x + {1 \over x} \biggr) Y \biggr] {1 \over y}
%%%%% end : bb61Paper
\,, \\
d^{[1]}_6 &=&
%%%%% begin : dd61Paper
\biggl( {1 \over 3} Y - {5 \over 9} \biggr) {x \over y}
+ i \pi {} {x \over 3 y}
%%%%% end : dd61Paper
\,, \\
h ^{[2]}_6&=&
%%%%% begin : hh62Paper
\biggl[ x {} \biggl( {1 \over 3} Y - {13 \over 18} \biggr)
       - {1 \over 4} \biggl( x + {1 \over x} \biggr) Y^2 \biggr] {1 \over y}
- {1 \over 2} Y
{} \PlusBreak{}
i \pi {} \biggl[ {5 \over 6} x + {1 \over 2}
  - {1 \over 2} \biggl( {1 \over x} + x \biggr) Y \biggr] {1 \over y}
%%%%% end : hh62Paper
\,, \\
j ^{[2]}_6&=&
%%%%% begin : jj62Paper
\biggl( {5 \over 9} - {1 \over 3} Y \biggr) {x \over y}
    - i \pi {} {x \over 3 y}
%%%%% end : jj62Paper
\,, \\
k^{[2]}_6 &=&
%%%%% begin : kk62Paper
\biggl[ x {} \biggl( {\pi^2 \over 2} - 4 - {3 \over 2} X
      + {1 \over 2} X^2 - X Y \biggr)
      - {1 \over 4} \biggl( {1 \over x} - x \biggr) Y^2 \biggr] {1 \over y}
- {1 \over 2} Y
{}\MinusBreak{}
i \pi {} \biggl[ x - {1 \over 2}
      + {1 \over 2} \biggl( {1 \over x} + x \biggr) Y \biggr] {1 \over y}
%%%%% end : kk62Paper
\,.
\end{eqnarray}
For the HV scheme ($\delta_R=1$), the results (4.14)-(4.52) for the finite
remainders of the one-loop helicity amplitudes are in
agreement with those of ref.~\cite{Gloverqqqq}.

%%%%%%%%%%%%%%%%%%%%%%%%%%%%%%%%%%%%%%%%%%%%%%%%%%%%%%%%%%

\section{Two-loop QCD amplitudes and finite remainders}
\label{TwoLoopFiniteQCDSection}

\subsection{Construction of amplitudes}
%%%%%%%%%%%%%%%%%%%%%%%%%%%%%%%%%%%%%%%%%%%%%%%%%%%%%%%%%%

We generated the Feynman graphs for $\qqtoqq$ using {\tt
QGRAF}~\cite{QGRAF}, from which a {\tt MAPLE} program was constructed
to evaluate each graph.  We employed the integral reduction
algorithms developed for the all-massless four-point
topologies~\cite{PBReduction,IntegralsAGO,NPBReduction,Lorentz,BKgggg},
in order to reduce the loop integrals to a basis of master
integrals.  To put the integrands into a form suitable for applying
the general reduction algorithms, spinor strings were converted to
traces over $\gamma$ matrices, by multiplying and dividing by
appropriate spinor inner products constructed from the external
momenta. To illustrate the method consider the diagram
depicted in
fig. \ref{TwoLoopDiagram}.  The numerator of the integrand is
\begin{equation}
\langle 2^- |
\gamma^{\mu} {\s k_3} \gamma^{\nu} {\s k_2}
\gamma^{\rho} {\s k_1} \gamma_{\mu}
| 1^- \rangle
\langle 4^+ |
\gamma_{\rho} {\s k_4} \gamma_{\nu}
| 3^+ \rangle \,.
\end{equation}
We multiply and divide this by
\begin{equation}
\langle 1^- | 4^+ \rangle \langle 3^+ | 2^- \rangle =
\langle 14 \rangle [32] =
{\langle 14 \rangle \over \langle 23 \rangle} t \,,
\end{equation}
so that the numerator can be rewritten as,
\begin{eqnarray}
&& { \langle 23 \rangle \over \langle 14 \rangle }  {1 \over t}
\langle 2^- |
\gamma^{\mu} {\s k_3} \gamma^{\nu} {\s k_2}
\gamma^{\rho} {\s k_1} \gamma_{\mu}
| 1^- \rangle
\langle 1^- | 4^+ \rangle
\langle 4^+ |
\gamma_{\rho} {\s k_4} \gamma_{\nu}
| 3^+ \rangle
\langle 3^+ | 2^- \rangle \nonumber \\
&& \hskip1.3cm
= { \langle 23 \rangle \over \langle 14 \rangle } {1 \over t} \sum_{\rm 
spins}
\langle 2 | P_+
\gamma^{\mu} {\s k_3} \gamma^{\nu} {\s k_2}
\gamma^{\rho} {\s k_1} \gamma_{\mu}
P_- | 1 \rangle
\langle 1 | 4 \rangle
\langle 4 | P_-
\gamma_{\rho} {\s k_4} \gamma_{\nu}
P_+ | 3 \rangle
\langle 3 | 2 \rangle  \nonumber \\
&& \hskip1.3cm
= { \langle 23 \rangle \over \langle 14 \rangle } {1 \over t}
{\rm Tr}\Bigl({\s p_2} P_+ \gamma^{\mu} {\s k_3} \gamma^{\nu} {\s k_2}
\gamma^{\rho} {\s k_1} \gamma_{\mu} P_- {\s p_1} {\s p_4} P_- \gamma_{\rho}
{\s k_4} \gamma_{\nu} P_+ {\s p_3}\Bigr)  \nonumber \\
&& \hskip1.3cm
= { \langle 23 \rangle \over \langle 14 \rangle } {1 \over t}
{\rm Tr}\Bigl({\s p_2} \gamma^{\mu} {\s k_3} \gamma^{\nu} {\s k_2}
\gamma^{\rho} {\s k_1} \gamma_{\mu} {\s p_1} {\s p_4} P_- \gamma_{\rho}
{\s k_4} \gamma_{\nu} P_+ {\s p_3}\Bigr) \,,
\label{Numer}
\end{eqnarray}
where $p_i$ is the momentum of external leg $i$ and
we have introduced the helicity projectors $P_+$ and $P_-$, given by
\begin{equation}
P_+ = {1 \over 2} (1+\gamma_5) \,, \qquad
P_- = {1 \over 2} (1-\gamma_5) \,,
\end{equation}
in order to reduce the sum over spins to the original helicity 
configuration.

%%%%%%%%%%%%%%%%%%%%%%%%%%%%%%%%%%%%%%%%%%%%%%%%%%%%%%%%%
%FIGURE
%
\vskip .2 cm
\FIGURE[t]{
%\begin{figure}[ht]
%\begin{center}
\begin{picture}(410,152)(0,0)
%
%%%%%%%% First diagram %%%%%%%%%%%%
%
% First arrow line
\ArrowLine(30,110)(10,130)
\ArrowLine(80,110)(30,110)
\ArrowLine(80,60)(80,110)
\ArrowLine(30,60)(80,60)
\ArrowLine(10,40)(30,60)
%
% Second arrow line
\ArrowLine(130,60)(150,40)
\ArrowLine(130,110)(130,60)
\ArrowLine(150,130)(130,110)
%
% Gluons
\Gluon(30,60)(30,110){3}{4}
\Gluon(130,60)(80,60){3}{4}
\Gluon(80,110)(130,110){3}{4}
%
% Closers
\DashCArc(30,40)(20,180,270){3}
\DashCArc(130,40)(20,270,360){3}
\DashCArc(30,130)(20,90,180){3}
\DashCArc(130,130)(20,0,90){3}
\DashLine(30,20)(60,20){3} \DashLine(100,20)(130,20){3}
\DashLine(30,150)(60,150){3} \DashLine(100,150)(130,150){3}
%
% Text
\Text(7,40)[r]{$1^-$}
\Text(7,130)[r]{$2^+$}
\Text(153,40)[l]{$4^-$}
\Text(153,130)[l]{$3^+$}
\Text(24,85)[r]{$\mu$}
\Text(85,85)[l]{$k_2$}
\Text(135,85)[l]{$k_4$}
\Text(55,50)[c]{$k_1$}
\Text(55,120)[c]{$k_3$}
\Text(105,50)[c]{$\rho$}
\Text(105,120)[c]{$\nu$}
\Text(80,20)[c]{\scriptsize $\langle 1^- | 4^+ \rangle$}
\Text(80,150)[c]{\scriptsize $\langle 3^+ | 2^- \rangle$}
\Text(80,0)[c]{(a)}
%
%%%%%%%% Second diagram %%%%%%%%%%%
%
% First arrow line
\ArrowLine(240,110)(220,130)
\ArrowLine(290,110)(240,110)
\ArrowLine(290,60)(290,110)
\ArrowLine(240,60)(290,60)
\ArrowLine(220,40)(240,60)
%
% Second arrow line
\ArrowLine(340,60)(360,40)
\ArrowLine(340,110)(340,60)
\ArrowLine(360,130)(340,110)
%
% Gluons
\Gluon(240,60)(240,110){3}{4}
\Gluon(340,60)(290,60){3}{4}
\Gluon(290,110)(340,110){3}{4}
%
% Closers
\DashCArc(220,60)(20,180,270){3}
\DashCArc(360,60)(20,270,360){3}
\DashCArc(220,110)(20,90,180){3}
\DashCArc(360,110)(20,0,90){3}
\DashLine(200,60)(200,78){3} \DashLine(200,92)(200,110){3}
\DashLine(380,60)(380,78){3} \DashLine(380,92)(380,110){3}
%
% Text
\Text(220,30)[c]{$1^-$}
\Text(220,140)[c]{$2^+$}
\Text(360,30)[c]{$4^-$}
\Text(360,140)[c]{$3^+$}
\Text(234,85)[r]{$\mu$}
\Text(295,85)[l]{$k_2$}
\Text(345,85)[l]{$k_4$}
\Text(265,50)[c]{$k_1$}
\Text(265,120)[c]{$k_3$}
\Text(315,50)[c]{$\rho$}
\Text(315,120)[c]{$\nu$}
\Text(195,85)[c]{\scriptsize $\langle 1^- | {\s p_3} | 2^- \rangle$}
\Text(389,85)[c]{\scriptsize $\langle 3^+ | {\s p_1} | 4^+ \rangle$}
\Text(290,0)[c]{(b)}
\end{picture}
\caption{
\label{TwoLoopDiagram}
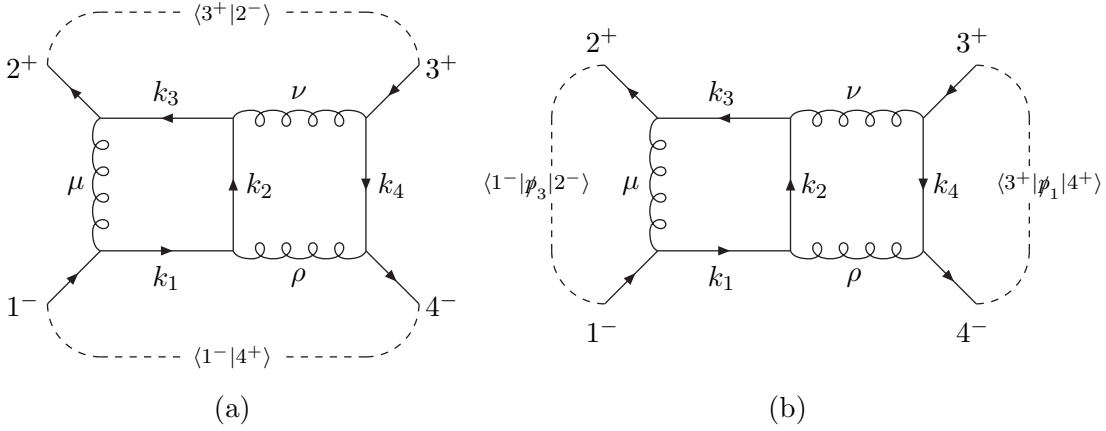
An example of a two-loop diagram showing two different ways
of ``closing'' the fermion lines. The dashed lines represent the
spinor inner products inserted in order to form a trace.}
%\end{center}
%\end{figure}
}

To evaluate the projectors we use the 't~Hooft-Veltman
prescription~\cite{HV} for $\gamma_5$,
\begin{equation}
\{ \gamma_5 , \gamma_{\mu}^{(4)} \} = 0 \,, \qquad
[ \gamma_5 , \gamma_{\mu}^{(-2\e)} ] = 0 \,,
\label{gamma5commute}
\end{equation}
where the notation ``$(4)$'' and ``$(-2\e)$'' is used to indicate
whether the Lorentz index $\mu$ lies the four-dimensional or
$(-2\e)$-dimensional subspaces, respectively. (Another prescription is
to take $\gamma_5$ to anti-commute~\cite{Chanowitz} with all
components, but this has the unwanted side-effect of ruining the
smooth connection of the 't~Hooft-Veltman and FDH scheme as a function
$\delta_R$.)  Using \eqn{gamma5commute} we move one of the projectors
in the last line of \eqn{Numer} until it hits the other, giving,
\begin{eqnarray}
&& {\langle 23 \rangle \over \langle 14 \rangle} {1 \over t}
\Bigl[
  {\rm Tr}\Bigl({\s p_2} \gamma^{\mu} {\s k_3} \gamma^{\nu} {\s k_2}
   \gamma^{\rho} {\s k_1} \gamma_{\mu} {\s p_1} {\s p_4}  
\gamma^{(4)}_{\rho}
   {\s k^{(4)}_4} \gamma^{(4)}_{\nu} P_+ {\s p_3}\Bigr)
{} \PlusBreak{\null { \langle 23 \rangle \over \langle 14 \rangle \BigrB }
              {1 \over t}  \null}
   {\rm Tr}\Bigl({\s p_2} \gamma^{\mu} {\s k_3} \gamma^{\nu} {\s k_2}
   \gamma^{\rho} {\s k_1} \gamma_{\mu} {\s p_1} {\s p_4}  
\gamma^{(-2\e)}_{\rho}
   {\s k^{(-2\e)}_4} \gamma^{(4)}_{\nu} P_+ {\s p_3}\Bigr)
{} \PlusBreak{\null { \langle 23 \rangle \over \langle 14 \rangle \BigrB }
              {1 \over t}  \null}
   {\rm Tr}\Bigl({\s p_2} \gamma^{\mu} {\s k_3} \gamma^{\nu} {\s k_2}
   \gamma^{\rho} {\s k_1} \gamma_{\mu} {\s p_1} {\s p_4}  
\gamma^{(-2\e)}_{\rho}
   {\s k^{(4)}_4} \gamma^{(-2\e)}_{\nu} P_+ {\s p_3}\Bigr)
{} \PlusBreak{\null { \langle 23 \rangle \over \langle 14 \rangle \BigrB}
               {1 \over t} \null}
    {\rm Tr}\Bigl({\s p_2} \gamma^{\mu} {\s k_3} \gamma^{\nu} {\s k_2}
   \gamma^{\rho} {\s k_1} \gamma_{\mu} {\s p_1} {\s p_4}  
\gamma^{(4)}_{\rho}
   {\s k^{(-2\e)}_4} \gamma^{(-2\e)}_{\nu} P_+ {\s p_3}\Bigr) \Bigr] \,.
\label{SampleEvaluate}
\end{eqnarray}
In the last equation, the $\gamma_5$ part of $P_+$ will produce terms
containing a Levi-Civita tensor. Upon integration this tensor can 
appear contracted only with external momenta, and since we have only
three independent ones, these terms must vanish. Therefore the
$\gamma_5$ term from the $P_+$ in \eqn{SampleEvaluate} can be dropped here.

After expanding the traces over Dirac matrices using standard
formul\ae, we obtain dot products of momenta (external and/or internal)
in our expressions, making possible the application of the reduction
algorithms previously mentioned. In evaluating the trace we take the
four-dimensional and $(-2 \e)$-dimensional subspaces to be orthogonal
following the discussion of ref.~\cite{TwoLoopSUSY}. The rules for
evaluating integrands containing dot products of $(-2\e)$ dimensional
components of momenta may be found in ref.~\cite{BDDgggg}.

Alternatively, we could have also chosen to ``close'' the fermion lines
in a different way, multiplying and dividing, for instance, by
\begin{equation}
\langle 1^- | {\s p_3} | 2^- \rangle
\langle 3^+ | {\s p_1} | 4^+ \rangle =
\langle 14 \rangle [32] u \,.
\end{equation}
This is depicted in \fig{TwoLoopDiagram}(b).  However, this would have
given a product of two traces instead of just one, which would produce
some terms containing a product of two Levi-Civita tensors, which
would then need to be evaluated since they would not vanish under
integration.

After all the tensor loop integrals in the amplitudes have been
reduced to a linear combination of master integrals, the next step is
to expand the master integrals in a Laurent series in $\e$, beginning
at order $1/\e^4$, using results from
refs.~\cite{PBScalar,NPBScalar,PBReduction,NPBReduction,IntegralsAGO}.
The results can be expressed solely in terms of ordinary
polylogarithms~\cite{Lewin,NielsenIds},
\begin{eqnarray}
\Li_n(x) &=& \sum_{i=1}^\infty {x^i \over i^n}
= \int_0^x {dt \over t} \Li_{n-1}(t)\,,
\\
\Li_2(x) &=& -\int_0^x {dt \over t} \ln(1-t) \,,
\label{PolyLogDef}
\end{eqnarray}
with $n=2,3,4$.  The analytic properties of the non-planar double box
integrals appearing in the amplitudes are somewhat
intricate~\cite{AllPlusTwo,NPBScalar}; there is no Euclidean region in
any of the three kinematic channels, $s$, $t$ or $u$.  We do not
attempt to give a crossing-symmetric representation, but instead quote
all our results in the physical $s$-channel $(s > 0; \; t, \, u < 0)$
for the $\qqtoqq$, $q\bar{Q} \to q\bar{Q}$ and $qQ \to qQ$ kinematics,
eqs.~(\ref{qbQBlabel}), (\ref{qBqBlabel}) and (\ref{qQqQlabel}).

%%%%%%%%%%%%%%%%%%%%%%%%%%%%%%%%%%%%%

\subsection{Finite remainders}
\label{TwoLoopFinRemSubsection}

The two-loop finite remainders are defined in~\eqn{TwoloopCatani} and
are color decomposed into $M^{(2),[c]{\rm fin}}_h$ in
\eqn{RemoveColorPhase}.  Their dependence on the renormalization scale
$\mu$, the number of colors $N$, the number of fermion flavors $\Nf$,
and scheme label $\delta_R$ may be extracted as
\begin{eqnarray}
M^{(2),[c]{\rm fin}}_h
&=&
\Bigl[- b_0^2 \, (\ln(s/\mu^2) - i\pi)^2
        - b_1 \, (\ln(s/\mu^2) - i\pi)
        - {1\over36} ( C_A + 6 C_F )
             C_A (1-\delta_R)^2
  \PlusBreak{ ~~~~~ }
           \Bigl( 4 R_q + {1\over9} ( C_A + 9 C_F ) Q_0
            + b_0 Q_1^{(qq)} (\ln(s/\mu^2) - i\pi)
  \MinusBreak{ ~~~~~~~~~~~~ }
              {1\over3} ( C_A - 6 C_F ) Q_0 i \pi \Bigr)
                (1-\delta_R) \Bigr]
    M^{(0),[c]}_h
  \PlusBreak{ }
\Bigl[ - 2 b_0 \, (\ln(s/\mu^2) - i\pi) \,
      + Q_1^{(qq)} (1-\delta_R) \Bigr]
    M^{(1),[c]{\rm fin}}_h
  \PlusBreak{ }
   Q_0  M^{(1),[c]\,\e,\delta_R}_h (1-\delta_R) + P^{[c]}_h \,,
\label{Twoloopc12}
\end{eqnarray}
where,
\begin{eqnarray}
P^{[1]}_h &=& N^2 A^{[1]}_h + B^{[1]}_h +{1 \over N^2}C^{[1]}_h
+ N N_f D^{[1]}_h + {N_f \over N} E^{[1]}_h + N_f^2 F^{[1]}_h \,, 
\nonumber\\
P^{[2]}_h &=& N G^{[2]}_h + {1 \over N} H^{[2]}_h + {1 \over N^3} I^{[2]}_h
+ N_f J^{[2]}_h + {N_f \over N^2} K^{[2]}_h + {N_f^2 \over N} L^{[2]}_h \,.
\label{FiniteRemainder}
\end{eqnarray}
The $\mu$-dependence is a consequence of renormalization group invariance.

The tree and one-loop functions,
$M^{(0),[c]}_h$
and $M^{(1),[c]{\rm fin}}_h$,
are given in~\eqn{TreeAmps} and
\eqns{Mh11fin}{Mh12fin},
while $b_0$ and $b_1$ are given
in~\eqn{QCDBetaCoeffs}.
The following combinations of color constants also appear in
\eqn{Twoloopc12},
\begin{eqnarray}
  Q_0 &=& {5\over6} C_A - C_F + {1\over3} T_R \Nf \,,
\label{Q0} \\
  Q_1^{(qq)} &=&  - {1 \over 3} C_A + C_F \,,
\label{Q1qq} \\
  R_q &=&  - {7\over48} C_A^2
  + \Bigl( {\pi^2\over 192} + {617\over864} \Bigr) C_A C_F
  - \Bigl( {\pi^2\over24} + {1\over4} \Bigr) C_F^2
  - {1\over16} C_F T_R \Nf \,,
\label{Rq}
\end{eqnarray}
The constants $Q_0$ and $R_q$ also appeared in the two-loop finite 
remainders
for the $q\bar{q} \to gg$ process, and $Q_1^{(qq)}$ is just twice the
constant $Q_1^{(qg)}$ appearing in that case.
The quantities $M^{(1),[c]\,\e,\delta_R}_h$ are the $\delta_R$-dependent
parts of the $\Ord(\e^1)$ coefficients of the one-loop amplitude
remainders, after subtracting the poles using \eqn{OneloopCatani}.
The explicit values of $M^{(1),[c]\,\e,\delta_R}_h$ are tabulated
in appendix~\ref{OrdepsdeltaRemainderAppendix}.
The coefficient functions $A,B,C,D,E,F,G,H,I,J,K$ depend only on
the Mandelstam variables.
In appendix~\ref{QCDRemainderAppendix},
we give the explicit forms for the independent finite remainder
functions appearing in \eqn{FiniteRemainder}.

We have compared our results for the independent two-loop finite
remainder functions $M^{(2),[c]{\rm fin}}_h$ with corresponding
results obtained recently by Glover ref.~\cite{Gloverqqqq}.
Our results agree with the corrected version of ref.~\cite{Gloverqqqq}, 
once a slightly different definition of ${\bom H}^{(2)}(\e)$ in \eqn{OurH} 
is accounted for. (Instead of dressing the
$1/\e$ pole with $(\mu^2/(-s))^{2\e}$, as we do in \eqn{OurH},
in ref.~\cite{Gloverqqqq} it is dressed with
$(\mu^2/(-s))^{2\e}+(\mu^2/(-t))^{2\e}-(\mu^2/(-u))^{2\e}$).  We have
also checked that the interference of the tree and two-loop helicity
amplitudes, summed over helicities and colors, reproduces the
results given in ref.~\cite{GOTYqqqq}.

%%%%%%%%%%%%%%%%%%%%%%%%%%%%%%%%%%%%%%%%%%%%%%%%%%

\section{Amplitudes in pure $\NeqOne$ super-Yang-Mills theory}
\label{N=1AmplitudesSection}

Supersymmetric gauge theories have a wide range of applications both
for phenomenological and theoretical purposes.  Here we present the
amplitudes in $\NeqOne$ pure super-Yang-Mills theory, obtained from
the QCD ones by modifying the fermions to be in the adjoint
representation and by altering their multiplicity to correspond to a
single Majorana fermion circulating in the loops. (A Majorana
fermion counts as half of a Dirac fermion.) The fermion in the
$\NeqOne$ pure super-Yang-Mills theory is the gluino superpartner of
the gluon.  Besides the inherent interest in supersymmetric theories,
a useful consequence is that supersymmetry imposes a set of powerful
constraints on the amplitudes which can be used to verify
their correctness. The supersymmetry identities have been applied
previously to the same one-loop amplitudes discussed
here~\cite{KSTfourparton}, but only through $\Ord(\e^0)$, as needed in
an NLO calculation.  In this section we discuss the supersymmetry
identities for gluino-gluino scattering up to two loops.

%%%%%%%%%%%%%%%%%%%%

\subsection{Supersymmetry Ward Identities}
\label{SWISubsection}

In refs.~\cite{BDDgggg,BDDqqgg,TwoLoopSUSY} it was shown that
using the FDH scheme ($\delta_R=0$) the following identities are satisfied
through $\Ord(\e^0)$ at two loops,
\begin{eqnarray}
\cm^{\rm SUSY}(g_1^\pm,g_2^-,g_3^+,g_4^+) &=& 0 \,,
\label{SUSYVanish} \\
\cm^{\rm SUSY}(\tg_1^+,\tg_2^-,g_3^+,g_4^+) &=& 0 \,,
\label{SUSYVanishGluinos} \\
\cm^{\rm SUSY}(\tg_1^+,\tg_2^-,g_3^-,g_4^+) &=&
{ \spa2.3 \over \spa1.3 } \, \cm^{\rm SUSY}(g_1^+,g_2^-,g_3^-,g_4^+) \,,
\label{SUSYNonVanish}
\end{eqnarray}
where $g$ and $\tg$ denote a gluon and gluino, and the superscripts
denote helicities. There is also an identity relating the four-gluino
amplitudes to the four-gluon ones,
\begin{eqnarray}
\cm^{\rm SUSY}(\tg_1^+,\tg_2^-,\tg_3^-,\tg_4^+) &=&
- { \spa2.4 \over \spa1.3 } \, \cm^{\rm SUSY}(g_1^+,g_2^-,g_3^-,g_4^+) 
\nonumber \\
&=& 
- { \spa2.4 \over \spa2.3 } \cm^{\rm SUSY}(\tg_1^+,\tg_2^-,g_3^-,g_4^+) 
\label{SUSYFourGluinos}
\end{eqnarray}
which we discuss here.  In ref.~\cite{KSTfourparton} this was shown to
be valid at one loop through $\Ord(\e^0)$, as required in an NLO
calculation.

%%%%%%%%%%%%%%%%%%%%%%%%%%%%%%%%%%%%%%%%

\subsection{Ambiguities in $D$-dimensional Dirac algebra}
\label{Ambiguities}

Before presenting the results, we first describe
ambiguities that affect the amplitudes.  The methods
we use rely on continuing the Dirac algebra appearing in the
spinor inner products away from four dimensions.  This continuation is
of course not unique, with a variety of schemes such as the
't~Hooft-Veltman or FDH scheme for doing so.  However, even within each
of these schemes there can be further ambiguities.  In particular, the
charge conjugation properties of inner products of helicity states are
not well defined as one moves away from four dimensions.  In four
dimensions we have the identity,
\begin{equation}
\langle p_1^\pm |
\gamma_{\mu_1} \cdots \gamma_{\mu_{2n+1}}
| p_2^\pm \rangle =
\langle p_2 ^\mp |
\gamma_{\mu_{2n+1}} \cdots \gamma_{\mu_1}
| p_2^\mp \rangle \,,
\label{ChargeConjugation}
\end{equation}
which is useful when performing Fierz rearrangements of the spinor
products.  It can also be used to construct simpler Dirac traces to
evaluate.  But can we use this identity when we continue the Lorentz
indices of the Dirac matrices away from four dimensions?

%%%%%%%%%%%%%%%%%%%%%%%%%%%%%%%%%%%%%%%%%%%%%%%%%%%%%%%%%
%FIGURE
%
\vskip .2 cm
\FIGURE[t]{
%\begin{figure}[ht]
%\begin{center}
\begin{picture}(360,115)(0,0)
%
%%%%%%%% First diagram %%%%%%%%%%%%
%
% First arrow line
\ArrowLine(30,90)(10,110)
\ArrowLine(30,40)(30,90)
\ArrowLine(10,20)(30,40)
%
% Second arrow line
\ArrowLine(80,40)(100,20)
\ArrowLine(80,90)(80,40)
\ArrowLine(100,110)(80,90)
%
% Gluons
\Gluon(30,90)(80,90){3}{4}
\Gluon(80,40)(30,40){3}{4}
%
% Text
\Text(7,20)[r]{$1$}
\Text(7,110)[r]{$2$}
\Text(103,20)[l]{$4$}
\Text(103,110)[l]{$3$}
%\Text(27,65)[r]{$k_1$}
%\Text(84,65)[l]{$k_2$}
%\Text(55,30)[c]{$\mu$}
%\Text(55,100)[c]{$\nu$}
\Text(55,0)[c]{(a)}
%
%%%%%%%% Second diagram %%%%%%%%%%%
%
% First arrow line
\SetOffset(-12,0)
\ArrowLine(180,90)(160,110)
\Gluon(180,40)(180,90){3}{4}
\ArrowLine(160,20)(180,40)
%
% Second arrow line
\ArrowLine(230,40)(250,20)
\Gluon(230,90)(230,40){3}{4}
\ArrowLine(250,110)(230,90)
%
% Gluons
\ArrowLine(230,90)(180,90)
\ArrowLine(180,40)(230,40)
%
% Text
\Text(157,20)[r]{$1$}
\Text(157,110)[r]{$2$}
\Text(253,20)[l]{$4$}
\Text(253,110)[l]{$3$}
%\Text(177,65)[r]{$k_1$}
%\Text(234,65)[l]{$k_2$}
%\Text(205,30)[c]{$\mu$}
%\Text(205,100)[c]{$\nu$}
\Text(205,0)[c]{(b)}
%
%%%%%%%% Third diagram %%%%%%%%%%%
%
\SetOffset(120,0)
\ArrowLine(180,90)(160,110)
\Gluon(180,40)(180,90){3}{4}
\ArrowLine(160,20)(180,40)
%
% Second arrow line
\ArrowLine(250,20)(230,40)
\Gluon(230,90)(230,40){3}{4}
\ArrowLine(230,90)(250,110)
%
% Gluons
\Line(180,90)(230,90)
\Line(230,40)(180,40)
%
% Text
\Text(157,20)[r]{$1$}
\Text(157,110)[r]{$2$}
\Text(253,20)[l]{$3$}
\Text(253,110)[l]{$4$}
%\Text(177,65)[r]{$k_1$}
%\Text(234,65)[l]{$k_2$}
%\Text(205,30)[c]{$\mu$}
%\Text(205,100)[c]{$\nu$}
\Text(205,0)[c]{(c)}

\end{picture}
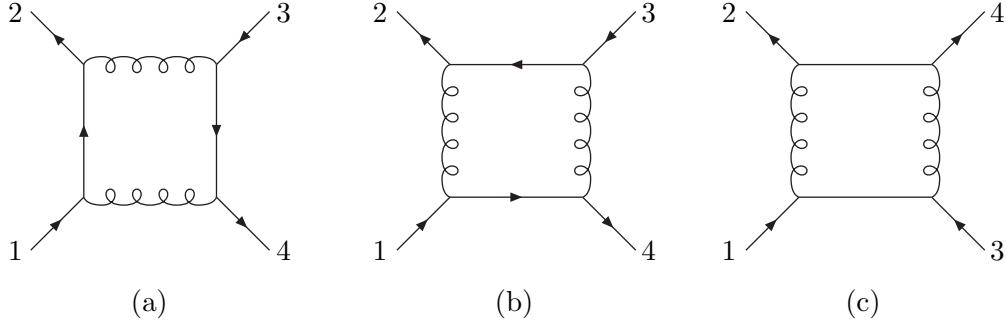
\caption{Three contributions for Majorana fermions.  The assignment
of fermion arrows is inconsistent since there is no distinction between
particles and antiparticles.
\label{FermionArrowFigure}
}
%\end{center}
%\end{figure}
}
%%%%%%%%%%%%%%%%%%%%%%%%%%%%%%%%%%%%%%%%%%%%%%%%%%%%%%%%%%

For the case of Dirac fermions such as the quarks of QCD, we can avoid
answering this question by imposing the rule that
\eqn{ChargeConjugation} should not be used in any manipulations.
Indeed, in performing the QCD calculations described in the
previous section, we did not use this equation.  However, with
Majorana fermions, as appear in the $\NeqOne$ theory, there is no
distinction between particles and antiparticles.  In particular, one
cannot consistently enforce a global requirement that a given fermion
be treated as a particle and no global assignment of fermion arrows is
consistent for all diagrams~\cite{Majorana}.  For example, in
\fig{FermionArrowFigure} if we chose legs 1 and 4 to be particles and
legs 2 and 3 to be antiparticles, this would be inconsistent with
diagram \fig{FermionArrowFigure}(c) which is also a perfectly valid
diagram for Majorana fermions.

%%%%%%%%%%%%%%%%%%%%%%%%%%%%%%%%%%%%%%%%%%%%%%%%%%%%%%%%%
%FIGURE
%
\vskip .2 cm
\FIGURE[t]{
%\begin{figure}[ht]
%\begin{center}
\begin{picture}(260,115)(0,0)
%
%%%%%%%% First diagram %%%%%%%%%%%%
%
% First arrow line
\ArrowLine(30,90)(10,110)
\ArrowLine(30,40)(30,90)
\ArrowLine(10,20)(30,40)
%
% Second arrow line
\ArrowLine(80,40)(100,20)
\ArrowLine(80,90)(80,40)
\ArrowLine(100,110)(80,90)
%
% Gluons
\Gluon(30,90)(80,90){3}{4}
\Gluon(80,40)(30,40){3}{4}
%
% Text
\Text(7,20)[r]{$1^-$}
\Text(7,110)[r]{$2^+$}
\Text(103,20)[l]{$4^-$}
\Text(103,110)[l]{$3^+$}
\Text(27,65)[r]{$k_1$}
\Text(84,65)[l]{$k_2$}
\Text(55,30)[c]{$\mu$}
\Text(55,100)[c]{$\nu$}
\Text(55,0)[c]{(a)}
%
%%%%%%%% Second diagram %%%%%%%%%%%
%
% First arrow line
\ArrowLine(180,90)(160,110)
\ArrowLine(180,40)(180,90)
\ArrowLine(160,20)(180,40)
%
% Second arrow line
\ArrowLine(250,20)(230,40)
\ArrowLine(230,40)(230,90)
\ArrowLine(230,90)(250,110)
%
% Gluons
\Gluon(180,90)(230,90){3}{4}
\Gluon(230,40)(180,40){3}{4}
%
% Text
\Text(157,20)[r]{$1^-$}
\Text(157,110)[r]{$2^+$}
\Text(253,20)[l]{$4^-$}
\Text(253,110)[l]{$3^+$}
\Text(177,65)[r]{$k_1$}
\Text(234,65)[l]{$k_2$}
\Text(205,30)[c]{$\mu$}
\Text(205,100)[c]{$\nu$}
\Text(205,0)[c]{(b)}
\end{picture}
\caption{
\label{OneLoopDiagram}
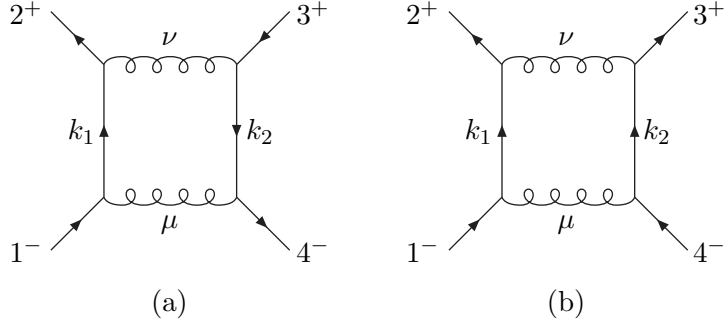
A One-loop box diagram, and its ``equivalent'' where one fermion line has 
been
charge-conjugated.}
%\end{center}
%\end{figure}
}
%%%%%%%%%%%%%%%%%%%%%%%%%%%%%%%%%%%%%%%%%%%%%%%%%%%%%%%%%%

Consider, for example, the box diagrams of \fig{OneLoopDiagram}. With
Majorana fermions it is not clear which of the two diagrams should be
used. The two diagrams differ by an application of the identity
(\ref{ChargeConjugation}) and have identical values
in four dimensions, up to the overall sign.  In order to calculate these
diagrams we can multiply and divide the integrand of diagram (a) by
$\langle 1^- | 4^+ \rangle \langle 3^+ | 2^- \rangle$ and the
integrand of diagram (b) by $\langle 1^- | {\s p_2} | 3^- \rangle
\langle 4^- | {\s p_1} | 2^- \rangle$. After performing the Dirac
algebra, it is straightforward to show that the integrands of these
two diagrams differ by a term proportional to $\lambda^2$, where
$\lambda$ is the $(-2\e)$-dimensional part of the loop momentum.
Since the one-loop box diagram with a single $\lambda^2$ in the
numerator is of $\Ord(\e)$, this ambiguity is irrelevant at
NLO. However, at two loops this ambiguity does enter, as one might
expect by observing that $\Ord(\e)$ terms at one loop contribute to
$\Ord(1/\e)$ terms at two loops in Catani's formula
(\ref{TwoloopCatani}).  With only a single or no fermion pair the results
are unaffected by applying \eqn{ChargeConjugation}. This ambiguity
therefore does not appear in the $gggg$ or $\tilde g \tilde g gg$
amplitudes.

A related issue is the lack of helicity conservation at a fermion
vertex with the 't Hooft-Veltman prescription~\cite{HV} for $\gamma_5$
(\ref{gamma5commute}). Since the $(-2\e)$-dimensional part of
$\gamma$-matrices commute, helicity violating terms of the form
$\langle a^-| \gamma^{(-2\e)} | b^+\rangle = \langle a| P_+
\gamma^{(-2\e)} P_+| b\rangle$ do in general contribute.  At tree
level the helicity violating contributions vanish in the FDH scheme.
However, they do contribute at loop level because $\Ord(\e)$ terms are
generated which interfere with the divergences.  This is a reflection
of the well known violation of the chiral Ward identity when using the
't~Hooft-Veltman $\gamma_5$ prescription.  Because of their connection
to the divergences, it is not surprising that the fermion helicity
violating terms can all be absorbed into Catani's formula for two-loop
divergences, as confirmed by our calculations, and hence do not affect
physical quantities.  It is therefore perfectly consistent to drop
these contributions (as we have done in QCD), if this is done
systematically throughout the calculation.  In $\NeqOne$
super-Yang-Mills theory, although it is again consistent to drop these
contributions, they do affect supersymmetry Ward identities.

One can choose various prescriptions for resolving these ambiguities,
but it is not a priori clear which will allow the amplitudes to
satisfy the supersymmetry identities.  Whether or not one keeps fermion
helicity violating contributions and the choices for assigning fermion
arrows alters the value of the full amplitudes.  Effectively, these
ambiguities mean that the supersymmetry Ward identities will not hold
unless an additional prescription is imposed.  As we discuss 
below it is possible to choose the ambiguous terms so that the supersymmetry 
identities are satisfied. 

However, a more straightforward way to deal with these ambiguities is to
use the fact that they are all tied to the divergences controlled
by Catani's formula.  Our explicit calculations verify that the finite
remainders after subtracting the divergences using Catani's formula
are completely free of these ambiguities, as long as the tree,
one-loop and two-loop amplitudes are treated uniformly with the same
prescriptions.  That is, we obtain the same finite remainders whether
or not we apply \eqn{ChargeConjugation} to one of the fermion lines,
or whether we keep or drop contributions that violate fermion helicity
conservation. Moreover, as we discuss below, in the FDH scheme the
finite remainders satisfy the expected supersymmetry identities
without any additional prescriptions.

%%%%%%%%%%%%%%%%%%%%%%%%%%%%%%%%%%%%%%%%

\subsection{Color and infrared structure}
\label{IRSusySubsection}

Because the four gluinos are identical particles, and are Majorana,
if we enforce helicity conservation, there are
only two independent processes,
\begin{eqnarray}
h=1: \quad   \tg(p_1,+) + \tg(p_2,-) &\to& \tg(p_3,-) + \tg(p_4,+)\,,
\label{susyhel1} \\
h=2: \quad   \tg(p_1,+) + \tg(p_2,+) &\to& \tg(p_3,+) + \tg(p_4,+)\,.
\label{susyhel2}
\end{eqnarray}
We ignore the helicity violating processes, {\it e.g.},
\begin{eqnarray}
h=3: \quad   \tg(p_1,+) + \tg(p_2,-) &\to& \tg(p_3,+) + \tg(p_4,+)\,,
\label{susyhel3} \\
h=4: \quad   \tg(p_1,-) + \tg(p_2,-) &\to& \tg(p_3,+) + \tg(p_4,+)\,,
\label{susyhel4}
\end{eqnarray}
since, as mentioned above, and confirmed by explicit computation,
they can be absorbed entirely into Catani's formula for divergences.

Since the gluinos are in the adjoint representation we use the same
color basis as used for the four-gluon helicity
amplitudes~\cite{BDDgggg}
\begin{equation}
\tilde {\cal M}^{(L)}_h =
  \tilde{S}_h \times
  \sum_{c=1}^9 \trc^{[c]} \times
       \tilde M^{(L),[c]}_h \,,
\label{RemoveColorPhaseAdjoint}
\end{equation}
where
\begin{eqnarray}
&& \trc^{[1]} = \Tr(T^{a_1} T^{a_2} T^{a_3} T^{a_4})\,, \nonumber 
\hskip1.4cm
   \trc^{[2]} = \Tr(T^{a_1} T^{a_2} T^{a_4} T^{a_3})\,, \nonumber \\
&& \trc^{[3]} = \Tr(T^{a_1} T^{a_4} T^{a_2} T^{a_3})\,, \nonumber 
\hskip1.4cm
   \trc^{[4]} = \Tr(T^{a_1} T^{a_3} T^{a_2} T^{a_4})\,, \nonumber \\
&& \trc^{[5]} = \Tr(T^{a_1} T^{a_3} T^{a_4} T^{a_2})\,, \nonumber 
\hskip1.4cm
   \trc^{[6]} = \Tr(T^{a_1} T^{a_4} T^{a_3} T^{a_2})\,, \nonumber \\
&& \trc^{[7]} = \Tr(T^{a_1} T^{a_2}) \Tr(T^{a_3} T^{a_4})\,, \nonumber 
\qquad
   \trc^{[8]} = \Tr(T^{a_1} T^{a_3}) \Tr(T^{a_2} T^{a_4})\,, \nonumber\\
&& \trc^{[9]} = \Tr(T^{a_1} T^{a_4}) \Tr(T^{a_2} T^{a_3})\,.
\label{TraceBasisAdj}
\end{eqnarray}
A reflection identity implies that the $c=4,5,6$ coefficients are
equal to the $c=3,2,1$ coefficients (respectively), so there are
really only six different coefficients for each $h$, namely
$\tilde M^{(L),[c]}_h$, $c=1,2,3,7,8,9$.
The corresponding spinor factors are
\begin{equation}
\tilde{S}_1 = S_1 = - i {\spa3.1 \over \spa4.2} \,, \qquad
\tilde{S}_2 = S_5 = - i {\spa2.1 \over \spa4.3} \,.
\end{equation}
The infrared divergence structure is similar to that of gluon-gluon
and gluino-gluon scattering amplitudes~\cite{BDDgggg,BDDqqgg}.  For
the case of $\NeqOne$ pure super-Yang-Mills theory, in the basis
(\ref{TraceBasisAdj}) the matrix ${\bom I}^{(1)}$
is~\cite{GOTYgggg,BDDgggg,BDDqqgg}
{\small
\begin{eqnarray}
&&{\bom {\tilde I}}^{(1)}(\e) = - {e^{-\e\psi(1)} \over \Gamma(1-\e)}
\biggl( {1\over\e^2} + {\tilde b_0\over N\e} \biggr) \times
\label{explicitI1Susy} \\
&\times&\left( {
\begin{array}{ccccccccc}
N(\tS+\tT) & 0 & 0 & 0 & 0 & 0 & (\tT-\tU) & 0 & (\tS-\tU) \\
0 & N(\tS+\tU) & 0 & 0 & 0 & 0 & (\tU-\tT) & (\tS-\tT) & 0 \\
0 & 0 & N(\tT+\tU) & 0 & 0 & 0 & 0 & (\tT-\tS) & (\tU-\tS) \\
0 & 0 & 0 & N(\tT+\tU) & 0 & 0 & 0 & (\tT-\tS) & (\tU-\tS) \\
0 & 0 & 0 & 0 & N(\tS+\tU) & 0 & (\tU-\tT) & (\tS-\tT) & 0 \\
0 & 0 & 0 & 0 & 0 & N(\tS+\tT) & (\tT-\tU) & 0 & (\tS-\tU) \\
(\tS-\tU) & (\tS-\tT) & 0 & 0 & (\tS-\tT) & (\tS-\tU) & 2N\tS & 0 & 0 \\
0 & (\tU-\tT) & (\tU-\tS) & (\tU-\tS) & (\tU-\tT) & 0 & 0 & 2N\tU & 0 \\
(\tT-\tU) & 0 & (\tT-\tS) & (\tT-\tS) & 0 & (\tT-\tU) & 0 & 0 & 2N\tT
\end{array}
}
\right) \nonumber
\end{eqnarray}
}

\noindent
where $\tS$, $\tT$ and $\tU$ are defined in \eqn{STUedef}.
For $\NeqOne$ super-Yang-Mills theory the first two coefficients of the
$\beta$-function are
\begin{equation}
\tilde b_0 = {3 \over 2 } C_A \,, \hskip 2 cm
\tilde b_1 = {3 \over 2} C_A^2  \,.
\label{N=1SYBetaCoeffs}
\end{equation}
The ${\bom I}^{(2)}$ operator for super-Yang-Mills theory is
\begin{eqnarray}
{ \bom {\tilde I}}^{(2)}_{\FDH}(\e,\mu;\{p\})
& =& -\frac{1}{2} {\bom {\tilde I}}^{(1)}(\e,\mu;\{p\})
\left({\bom {\tilde I}}^{(1)}(\e,\mu;\{p\}) + {2 \tilde b_0 \over \e} 
\right)
  \PlusBreak{}
{e^{+\e \psi(1)} \Gamma(1-2\e) \over \Gamma(1-\e)}
\left( {\tilde b_0 \over \e} + K^{\Susy}_{\FDH} \right)
   {\bom {\tilde I}}^{(1)}(2\e,\mu;\{p\})
  \PlusBreak{}
  {\bom {\tilde H}}^{(2)}_{\FDH}(\e,\mu;\{p\}) \,,
\label{CataniI2Susy}
\end{eqnarray}
where
\begin{equation}
K_\FDH^\Susy = \left( 3 - {\pi^2 \over 6} - {4 \over 9} \e \right) C_A ,
\label{SUSYK}
\end{equation}
\begin{equation}
{\bom {\tilde H}}^{(2)}_\FDH(\e,\mu;\{p\}) =
{e^{-\e\psi(1)} \over 4\e \, \Gamma(1-\e) }
\biggl( { \mu^2 \over -s } \biggr)^{2\e}
   \Bigl(4 (H_g^{(2)})^\Susy_\FDH  \, {\bom 1}
       + \hat{\bom H}^{(2)} \Bigr) \,,
\label{OurSusyH}
\end{equation}
and
\begin{equation}
(H_g^{(2)})_\FDH^\Susy =
(H_{\tilde g}^{(2)})_\FDH^\Susy =
\biggl( {\zeta_3 \over 2} + {\pi^2 \over 16} - {2 \over 9} \biggr) C_A^2 \,.
\label{SUSYH}
\end{equation}
The equality of $(H_g^{(2)})_\FDH^\Susy$ and $(H_{\tilde
g}^{(2)})_\FDH^\Susy$ is a consequence of supersymmetry.
Equations~(\ref{SUSYK})--(\ref{SUSYH})
are obtained from the QCD formulas,
eqs. (3.7)-(3.8), by the replacements
$\delta_R \rightarrow 0$, $C_F \rightarrow C_A$ and $T_R N_F
\rightarrow C_A/2$ for converting to a single adjoint fermion in
the FDH scheme.   The operator
$\hat{\bom H}^{(2)}$ defined in \eqn{HExtra} does not explicitly
depend on the fermion representation.

The tree amplitudes in this color basis in the FDH scheme are, 
\begin{eqnarray}
&& \tilde M^{(0),[1]}_1 = {u \over s} + {u \over t}
   \,, \qquad
   \tilde M^{(0),[2]}_1 = -{u \over s} 
   \,, \qquad
   \tilde M^{(0),[3]}_1 = -{u \over t} 
   \,, \nonumber \\
&& \tilde M^{(0),[1]}_2 = {s \over t} 
   \,, \phantom{+ {u \over t}} \qquad \hskip0.2cm
   \tilde M^{(0),[2]}_2 = {s \over u} 
   \,, \phantom{-} \qquad 
   \tilde M^{(0),[3]}_2 = -{s \over u} - {s \over t}
   \,, \nonumber \\
&& \tilde M^{(0),[1]}_3 = 0
   \,, \phantom{+ {u \over t}} \qquad \hskip0.25cm
   \tilde M^{(0),[2]}_3 = 0
   \,, \qquad \hskip0.45cm
   \tilde M^{(0),[3]}_3 = 0  \,, \nonumber \\
&& \tilde M^{(0),[3-i]}_h = \tilde M^{(0),[4+i]}_h \,,
   \qquad h=1,2,3, \qquad i=0,1,2, \phantom{{A \over B}} \nonumber \\
&& \tilde M^{(0),[c]}_h = 0 \,, \qquad c=7,8,9, \qquad h=1,2,3 \,,
\label{TreeAmpsAdj}
\end{eqnarray}
and are free of the ambiguities described above.

%%%%%%%%%%%%%%%

\subsection{One-loop amplitudes in pure $\NeqOne$ super-Yang-Mills theory}
\label{OneLoopN=1SubSection}

We now present the results for one-loop four-gluino scattering in a
format valid to all orders in $\e$.  However, because of the ambiguity
discussed above the answer depends on the precise steps used to
calculate the expression. 

\iffalse
a1:  closer mmpp qbQQbq [12]*<34> closer
a2:  closer mmpp qbQqQb [12]*<43> closer
a3:  closer mmpp qbqQbQ [14]*[23] closer
a4:  closer mmpp qbqQQb [13]*[42] closer
\fi

We can, for example, form the $h=1$ gluino amplitude from the $h = 3$
(with traces formed by multiplying and dividing by $\spb1.2 \spa4.3$)
amplitude (\ref{h3}) with adjoint representation quarks and $\Nf =
1/2$ summed with its $3\leftrightarrow 4$ interchange.  This gives, for
the unrenormalized coefficients of the three color structures ${\rm
Tr}^{[1]}$, ${\rm Tr}^{[2]}$ and ${\rm Tr}^{[3]}$,
\begin{eqnarray}
M^{(1),[1]}_1 &=&
%%%%% begin : M111SusyEps
{N \over 2} u {} \biggl[
-{ \e {} (4 s + t) + 2 u \over (2 \e - 1) t} \Trifour{(s)}
-{ \e {} (4 t + s) + 2 u \over (2 \e - 1) s} \Trifour{(t)}
{} \PlusBreak{\null {N \over 2} u \bigglP \null}
\biggl(
(4 \e-2) \biggl( {s \over t} + {t \over s} \biggr) + (5 \e-1)
\biggr) \Boxsix(s,t)
\biggr]
%%%%% end : M111SusyEps
\,,
%{} \MinusBreak{}
%-{\tilde b_0 \over \e} \tilde M^{(0),[1]}_1 \,,
\label{OneloopSUSY11} \\
M^{(1),[2]}_1 &=&
%%%%% begin : M121SusyEps
N u {}\biggl[
{\e-2 \over 2 (2 \e-1)} \Trifour{(s)}
+{u \over s} \Trifour{(u)}
-\biggl( \e {} {s + 2 u \over  s} + {t \over s} \biggr) \Boxsix(u,s)
%-\biggl( \e {} {s + 4 u \over 2 s} + {t \over s} \biggr) \Boxsix(u,s)
\biggr]
%%%%% end : M121SusyEps
\,,
%{} \MinusBreak{}
%{\tilde b_0 \over \e} \tilde M^{(0),[2]}_1 \,,
\label{OneloopSUSY12} \\
M^{(1),[3]}_1 &=&
%%%%% begin : M131SusyEps
N u {} \biggl[
{u \over t} \Trifour{(u)}
+{\e - 2 \over 2 (2 \e - 1)} \Trifour{(t)}
-\biggl( \e {} {t+ 2u \over t} + {s \over t} \biggr) \Boxsix(u,t)
%-\biggl( \e {} {3 t + 4 u \over 2 t} + {s \over t} \biggr) \Boxsix(u,t)
\biggr]
%%%%% end : M131SusyEps
\,,
%{} \MinusBreak{}
%{\tilde b_0 \over \e} \tilde M^{(0),[3]}_1 \,,
\label{OneloopSUSY13}
\end{eqnarray}
where we have taken the FDH scheme ($\delta_R=0$). The renormalization
is trivially performed by subtracting the quantity
${\tilde b_0} \tilde M^{(0),[c]}_h /\e$ from each,
where $\tilde b_0$ is given in \eqn{N=1SYBetaCoeffs}.
The ambiguities are all proportional to
$\e\, \Boxsix$ and hence are of $\Ord(\e)$.

Other choices are also possible.  For example, if include a fermion
helicity violating term $q^+ \qb^+ \rightarrow Q^+ \Qb^+$ (forming
traces by multiplying and dividing by $\langle4^-|{\s k_3}|1^-\rangle
\langle3^-| {\s k_1}|2^- \rangle$) in constructing the $h=2$ gluino
amplitude, this shifts \eqn{OneloopSUSY12} by
\begin{equation}
\delta M^{(1),[2]}_1 = {1\over 2} N u\, \e \,\Boxsix(u,s)\,,
\label{Shifta3}
\end{equation}
leaving the coefficients of the first and third color structures
unshifted.  Similarly, adding in the fermion helicity violating
contribution $q^+ \qb^+ \rightarrow \Qb^+ Q^+$ shifts
\eqn{OneloopSUSY13} by
\begin{equation}
\delta M^{(1),[3]}_1 = {1\over 2} N u\, \e \,\Boxsix(u, t) \,.
\label{Shifta4}
\end{equation}
For any of the above contributions, we can swap one of the particles
with its antiparticle before converting them to gluinos, which again
alters their values by $\Ord(\e)$ terms proportional to $\e\, \Boxsix$.

The remaining color coefficients, up to $\Ord(\e)$ ambiguous terms,
are given in terms of the previous ones by,
\begin{eqnarray}
\tilde M^{(1),[3-i]}_h(s,t,u) &=& \tilde M^{(1),[4+i]}_h(s,t,u)
\qquad i=0,1,2, \nonumber \,, \\
\tilde M^{(1),[7]}_h(s,t,u) &=& {2 \over N}  \Bigl(
         \tilde M^{(1),[3]}_h(s,t,u) +
         \tilde M^{(1),[2]}_h(s,t,u) +
         \tilde M^{(1),[1]}_h(s,t,u)  \Bigr) \,, \hskip 2 cm \nonumber \\
\tilde M^{(1),[8]}_h(s,t,u) &=& \tilde M^{(1),[9]}_h(s,t,u) =
        \tilde M^{(1),[7]}_h(s,t,u) \,,
\label{OtherSusyHelicities}
\end{eqnarray}
The $h=2$ amplitudes are related to the $h=1$ through the following 
relations,
\begin{eqnarray}
\tilde M^{(1),[1]}_2(s,t,u) &=& - \tilde M^{(1),[2]}_1(t,u,s)
\,, \nonumber \\
\tilde M^{(1),[2]}_2(s,t,u) &=& - \tilde M^{(1),[3]}_1(t,u,s)
\,, \nonumber \\
\tilde M^{(1),[3]}_2(s,t,u) &=& - \tilde M^{(1),[1]}_1(t,u,s)
\,, \nonumber \\
\tilde M^{(1),[7]}_2(s,t,u) &=& - \tilde M^{(1),[7]}_1(t,u,s) \,.
\end{eqnarray}

To check the supersymmetry identity (\ref{SUSYFourGluinos}) we compare
these results to the corresponding ones for gluon-gluon and gluino-gluon
scattering given in refs.~\cite{BDDgggg,BDDqqgg}.  We see that
eqs. (\ref{OneloopSUSY11}), (\ref{OneloopSUSY12}) and
(\ref{OneloopSUSY13}) match eqs.~(5.28), (5.29) and (5.30) in
ref.~\cite{BDDqqgg} respectively, up to an overall normalization
factor of $-st/u^2$ (which is similar to the same factor in eq. (5.39)
of ref.~\cite{BDDqqgg}), except for terms proportional to $\e \,\Boxsix$
in $\tilde{M}_2^{(1),[2]}$ and $\tilde{M}_2^{(1),[3]}$.  (Note that in
ref.~\cite{BDDqqgg} the amplitudes include the renormalization
subtraction, not explicitly included here.)  Similarly it matches the
corresponding equations in section 3.1 of ref.~\cite{BDDgggg} for
gluon-gluon scattering, again up to the terms proportional to $\e
\Boxsix$.  Since $\Boxsix(s,u)$ is finite, the one-loop terms that
violate the supersymmetry Ward identity are of $\Ord(\epsilon)$.
These terms are precisely the ambiguous terms described above.  We can
alter such terms by modifying the prescription.  If, for example, we
include in the shifts (\ref{Shifta3}) and (\ref{Shifta4}) arising from
fermion helicity violating terms we find that the supersymmetry
identities are then satisfied to all orders in $\e$.  Of course, this
is not completely satisfactory because of the {\it ad hoc} nature of
such choices.

%%%%%%%%%

\subsection{Two-loop amplitudes in $\NeqOne$ super-Yang-Mills theory}
\label{TwoLoopFiniteN=1Section}

The dependence of the identity (\ref{SUSYFourGluinos}) beyond
$\Ord(\e^0)$ on ambiguous terms at one loop foretells a similar
dependence at two loops except this time at $\Ord(1/\e)$.  However,
after subtracting these divergences, via Catani's formula, we find
that \eqn{SUSYFourGluinos} is indeed satisfied for the finite terms
which are independent of the ambiguous terms and any prescriptions
chosen for fixing them.  However, some attention is required to ensure
that precisely the same prescriptions are applied to tree level and
one loop as applied to two loops. Otherwise, there would be a mismatch
between the Catani subtraction terms and the two-loop divergences
leaving behind dependence on the ambiguous terms.

Our results in the FDH scheme are
\begin{eqnarray}
M^{(2),\Susy,[c]{\rm fin}}_h
&=&
- \Bigl[ \bigl( \tilde b_0 \bigr)^2 \, (\ln(s/\mu^2) - i\pi)^2
        + \tilde b_1 \, (\ln(s/\mu^2) - i\pi) \Bigr]
    M^{(0),[c]}_h
  \MinusBreak{ }
  2 \tilde b_0 \, (\ln(s/\mu^2) - i\pi) \,
    M^{(1),\Susy,[c]{\rm fin}}_h
+ N^2 \, A^{\Susy,[c]}_h
+        B^{\Susy,[c]}_h \,,
\nonumber \\
&&  \hskip8cm   c = 1,2,3,
\label{SUSYTwoloopSingleTrace} \\
M^{(2),\Susy,[c]{\rm fin}}_h
&=&
- 2 \tilde b_0 \, (\ln(s/\mu^2) - i\pi) \,
    M^{(1),\Susy,[c]{\rm fin}}_h
+ N \, G^{\Susy,[c]}_h \,,
\nonumber \\
&&  \hskip8cm c = 7,8,9.
\label{SUSYTwoloopDoubleTrace}
\end{eqnarray}
Because the adjoint color indices of the gluino fields are identical
to those of gluons, and only structure constants $f^{abc}$ appear
in the two-loop Feynman diagrams, the color coefficients for
two-gluino two-gluon scattering obey the same group theory relations
identified for $\ggtogg$ in ref.~\cite{BDDgggg},
\begin{eqnarray}
G^{\Susy,[7]}_h & = &
2 \Bigl( A^{\Susy,[1]}_h
   + A^{\Susy,[2]}_h
   + A^{\Susy,[3]}_h \Bigr)
   - B^{\Susy,[3]}_h
\,, \label{GSUSY7elim}\\[1pt plus 4pt]
%
%%%%%%%%%%%%%%%%%%%%%%%%%%%%%%%%%%%%%%
%
G^{\Susy,[8]}_h & = &
2 \Bigl( A^{\Susy,[1]}_h
   + A^{\Susy,[2]}_h
   + A^{\Susy,[3]}_h \Bigr)
   - B^{\Susy,[1]}_h
\,, \label{GSUSY8elim}\\[1pt plus 4pt]
%
%%%%%%%%%%%%%%%%%%%%%%%%%%%%%%%%%%%
%
G^{\Susy,[9]}_h & = &
2 \Bigl( A^{\Susy,[1]}_h
   + A^{\Susy,[2]}_h
   + A^{\Susy,[3]}_h \Bigr)
   - B^{\Susy,[2]}_h
\,, \label{GSUSY9elim}
\end{eqnarray}
and
\begin{equation}
B^{\Susy,[3]}_h =
    - B^{\Susy,[1]}_h
    - B^{\Susy,[2]}_h \,.
\label{BSUSY3elim}
\end{equation}

%%%%%%%%%%%%%%%%%%%%%%%%%%%%%%%%%%%%%%%%%%%%%%%%%

We have verified that the finite remainder functions match those of
the pure gluon case, {\it i.e.}
\begin{eqnarray}
X_1^{\Susy,[c]} &=& X_{-+-+}^{\Susy,[c]} \,,
\label{h13SYM} \\
X_2^{\Susy,[c]} &=& - X_{--++}^{\Susy,[c]} \,,
\label{h5SYM}
\end{eqnarray}
where $X \in \{ A, B, G \}$.  The functions $X_{-+-+}^{\Susy,[c]}$ and
$X_{--++}^{\Susy,[c]}$ for $\ggtogg$ scattering in pure $\NeqOne$
super-Yang-Mills theory are given in ref.~\cite{BDDgggg}. (Note that
in that reference an all outgoing definition of helicity is used.)
The relations~(\ref{h13SYM})--(\ref{h5SYM}) are precisely equivalent
to the content of the supersymmetry Ward identities~\cite{SWI}, after
removing overall factors and the divergent terms, and extracting the
$N$ and $\mu$ dependence.  As discussed in ref.~\cite{BDDqqgg} this
also matches the corresponding functions for the $\tilde g \tilde g
\rightarrow gg$ amplitudes.  This demonstrates that the supersymmetry
identity (\ref{SUSYFourGluinos}) holds for the finite remainders at
two loops.  In the HV scheme, as expected, it does not hold because of
the mismatch of fermionic and bosonic states.

\section{Conclusions}
\label{ConclusionsSection}

In this paper we presented the two-loop helicity amplitudes for
quark-quark and anti-quark-quark scattering in QCD and gluino-gluino
scattering in $\NeqOne$ super-Yang-Mills theory.  These amplitudes
retain the full information on external colors and helicities.  We
verified that the interference of our two-loop amplitudes with the
tree-level amplitudes, summed over all external colors and helicities,
and converted to the CDR scheme, are in complete agreement with the
results of ref.~\cite{GOTYqqqq}. We also found complete agreement with
recently published results by Glover~\cite{Gloverqqqq}, after correction
of minor errors in the original version of that article.

We also discuss ambiguities in defining the amplitudes, related to the
continutation of $\gamma_5$ or charge conjugation identities away from
four dimensions.  As confirmed by our explicit calculations these
ambiguities drop out of final physical results, since the ambiguous
contributions can be absorbed into infrared singularities that cancel
from physical quantities. In particular, there is no physical content
to contributions that violate helicity conservation of a given
massless fermion.  However, some attention needs to be paid so that
the different loop orders are computed consistently with the 
same set of prescriptions throughout.  Otherwise, the cancellation
of the ambiguous unphysical terms would not be complete.

In ref.~\cite{BDDqqgg} it was shown that the supersymmetry Ward
identities relating the $\tilde{g}\tilde{g} \to gg$ to the $gg \to gg$
processes are satisfied in the FDH scheme at two loops, for both
infrared divergent and finite parts.  For the gluino-gluino
scattering amplitudes discussed in this paper, unless the ambiguities
entering in the amplitude are carefully adjusted, the supersymmetry
identities will not hold starting at $\Ord(1/\e)$.  In any case, after
subtracting the divergences using Catani's formula~\cite{Catani}, the
finite remainders are free of these ambiguities and satisfy the
expected supersymmetry identities in the FDH scheme.

So far, the new $2 \rightarrow 2$ amplitudes amplitudes have been
implemented in a handful of new phenomenological
studies~\cite{PhotonPaper,Higgs}.  Once general algorithms for dealing
with infrared divergent phase space integrations at
next-to-next-leading-order are completed~\cite{NNLOPS}, many more
phenomenological applications will follow.  These applications would
include the implementation of the two-loop four-quark amplitudes of
this paper, or those of refs.~\cite{GOTYqqqq,Gloverqqqq}, as ingredients
in a numerical program for computing dijet production cross sections
at hadron colliders at NNLO in QCD.  When this task is accomplished,
the intrinsic precision on the QCD predictions should reach the few
percent level.

\acknowledgments We thank Lance Dixon for key contributions at early
stages of this paper and Babis Anastasiou, David Kosower
and Henry Wong for helpful comments. We also thank Nigel Glover for
communications regarding ref.~\cite{Gloverqqqq}.  A.D.F thanks
the Alexander von Humboldt Foundation for its support.

%%%%%%%%%%%%%%%%%%%%%%%%%%%%%%%%%%%%%%%%%%%%%%%%%%
\appendix

\section{Finite remainder functions for QCD}
\label{QCDRemainderAppendix}

In this appendix, we present the explicit forms for the independent
finite remainder functions for the processes $\qqtoqq$, $q\bar{Q}\to
q\bar{Q}$ and $qQ \to qQ$ in QCD, which appear in
\eqn{FiniteRemainder}.

For the helicity $h=1$ configuration in \eqn{h1} and color
factor $\trc^{[1]}$ in \eqn{basis12}, the finite remainder
functions are:
\begin{eqnarray}
A^{[1]}_1 & = &
%%%%% begin : AA11Paper
%%%% x/y %%%%%%%%
\biggl[
(x-1) \li4(-x)
-\biggl( x X + {19 \over 6} + {11 \over 6} x \biggr) \li3(-x)
+\biggl( {1 \over 24} + {1 \over 12} x \biggr) X^4
{} \PlusBreak{\null + \null}
   {1 \over 6} (11 x + 19) X \li2(-x)
+\biggl( \biggl( {85 \over 18} - {5 \over 12} \pi^2 \biggr) x
         -{\pi^2 \over 3} + {109 \over 18} \biggr) X^2
{} \PlusBreak{\null + \null}
  \biggl( \biggl( -{1 \over 2} \zeta_3 - {79 \over 27}
                  +{373 \over 144} \pi^2 \biggr) x
         -{62 \over 27} - 3 \zeta_3
         +{59 \over 24} \pi^2 \biggr) X
{} \PlusBreak{\null + \null}
  \biggl( {197 \over 72} \zeta_3
         +{23213 \over 5184} + {113 \over 1440} \pi^4 \biggr) x
+{233 \over 36} \zeta_3 {}
+{23213 \over 2592}
+{137 \over 720} \pi^4
\biggr] {x \over y}
%%%% 1/y %%%%%%%%%
{} \PlusBreak{}
\biggl[
  -{1 \over 2} \li4(-x)
  -{4 \over 3} \li3(-x)
  +\biggl( {1 \over 4} X^2 + {4 \over 3} X \biggr) \li2(-x)
  +{1 \over 48} X^4 - {37 \over 72} X^3
{} \PlusBreak{\null + + \null}
   \biggl( {28 \over 9} - {\pi^2 \over 6} \biggr) X^2
  +\biggl( {17 \over 27} - {3 \over 2} \zeta_3 + {43 \over 48} \pi^2 \biggr) 
X
  +{269 \over 72} \zeta_3
  +{137 \over 1440} \pi^4
{} \PlusBreak{\null + + \null}
{23213 \over 5184}
\biggr] {1 \over y}
%%%% 1 %%%%%%%%%%
-{1 \over 2} x X^2 \li2(-x) + {49 \over 36} x X^3
-{\pi^2 \over 3} y \li2(-x)
{} \MinusBreak{}
\biggl( {11 \over 12} x + {2 \over 3} \biggr) Y X^2
+{61 \over 36} \pi^2 + {49 \over 18} \pi^2 x
- {\pi^2 \over 3} y Y X
+ {1 \over 6} y X^3 Y
%
%%%% Imaginary part %%%%%%%%
%
{} \PlusBreak{}
i \pi {} \Biggl\{
%%%% x/y %%%%%%%%%%
\biggl[
- x \li3(-x)
+ \biggl( {11 \over 6} x + {19 \over 6} \biggr) \li2(-x)
+ \biggl( {1 \over 6} + {1 \over 3} x \biggr) X^3
{} \PlusBreak{\null + i \pi \BigglP \bigglP \null }
   {1 \over 9} (85 x + 109) X
- \biggl( {1 \over 2} \zeta_3 + {79 \over 27} \biggr) x
- 3 \zeta_3 - {62 \over 27}
\biggr] {x \over y}
-x X \li2(-x)
%%%% 1/y %%%%%%%%%%
{} \PlusBreak{\null + i \pi \BigglP }
\biggl[
  \biggl( {1 \over 2} X + {4 \over 3} \biggr) \li2(-x)
  -{3 \over 2} \zeta_3
  +{1 \over 12} X^3
  +{17 \over 27}
  -{37 \over 24} X^2
  +{56 \over 9} X
\biggr] {1 \over y}
%%%% 1 %%%%%%%%%
{} \MinusBreak{\null + i \pi \BigglP }
\biggl( {11 \over 6} x + {4 \over 3} \biggr) Y X
+ {49 \over 12} x X^2
+ \biggl( {1 \over 2} Y X^2
         -{\pi^2 \over 6} X
         -{19 \over 144} \pi^2 \biggr) y
\Biggr\}
%%%%% end : AA11Paper
\,, \label{AA11} \\[1pt plus 4pt]
%_______________________________________________________________________
%
B^{[1]}_1 & = &
%%%%% begin : BB11Paper
%%%%%%% x/y %%%%%%%%
\biggl[
-10 \li4 \biggl( - {x \over y} \biggr)
-(11+3 x) \li4(-x)
+10 \li4(-y)
{} \PlusBreak{\null \bigglP}
  \biggl( (2+3 x) X + {11 \over 3} x+8 Y
         + {47 \over 6} \biggr) \li3(-x)
- \biggl( {11 \over 3} x + 4 X + {40 \over 3} \biggr) \li3(-y)
{} \MinusBreak{\null \bigglP}
   \biggl( \biggl( {47 \over 6} + 2 Y + {11 \over 3} x \biggr) X
          + {1 \over 3} ( 11 x + 40 ) Y
%{} \MinusBreak{\bigglP \null + \bigglP}
         - {5 \over 3} \pi^2 - \pi^2 x \biggr) \li2(-x)
{} \MinusBreak{\null \bigglP}
   \biggl( {1 \over 8} + {1 \over 4} x \biggr) X^4
- \biggl(  \biggl( {1 \over 2} x + {2 \over 3} \biggr) Y
           - {125 \over 36} x - {119 \over 36} \biggr) X^3
{} \PlusBreak{\null \bigglP}
   \biggl( \biggl( {1 \over 4} x - {3 \over 2} \biggr) Y^2
          - \biggl( {53 \over 12} + {31 \over 12} x \biggr) Y
          + \biggl( {3 \over 2} \pi^2 - {617 \over 72} \biggr) x
          - {227 \over 18} + {5 \over 4} \pi^2 \biggr) X^2
{} \PlusBreak{\null \bigglP}
    {5 \over 3} Y^3 X
   + \biggl( \biggl( {9 \over 4} - {2 \over 3} \pi^2 \biggr) x
            + {\pi^2 \over 3} + {19 \over 4} \biggr) Y X
   - \biggl(  {4 \over 27} + {373 \over 72} \pi^2
             - \zeta_3 \biggr) x X
{} \MinusBreak{\null \bigglP}
   \biggl( {119 \over 27} - 6 \zeta_3 + {14 \over 3} \pi^2 \biggr) X
- \biggl( {1 \over 6} - {1 \over 8} x \biggr) Y^4
- \pi^2 \biggl( {1 \over 3} x + {3 \over 2} \biggr) Y^2
{} \MinusBreak{\null \bigglP}
   \biggl( \biggl( \zeta_3 + {158 \over 27} \biggl) x
          +{316 \over 27} + 10 \zeta_3 \biggr) Y
+ {19 \over 720} \pi^4
- {263 \over 36} \zeta_3
- {167 \over 72} \pi^2
{} \PlusBreak{\null \bigglP}
   \biggl( {30659 \over 1296} - {443 \over 72} \zeta_3
          + {79 \over 1440} \pi^4 - {55 \over 144} \pi^2 \biggr) x
+ {30659 \over 648}
\biggr] {x \over y}
+{3 \over 2} x X^2 \li2(-x)
%%%%%%% 1/y %%%%%%%%%
{} \PlusBreak{}
\biggl[
  - 2 \li4 \biggl( - {x \over y} \biggr)
  - {5 \over 2} \li4(-x)
  + 2 \li4(-y)
  + \biggl( Y + X + {19 \over 6} \biggr) \li3(-x)
{} \MinusBreak{\null + \bigglP }
    \biggl( {29 \over 3} + 2 X \biggr) \li3(-y)
  - \bigg( {3 \over 4} X^2 + \biggl( {19 \over 6} + Y \biggr) X
          - {4 \over 3} \pi^2 + {29 \over 3} Y \biggr) \li2(-x)
{} \MinusBreak{\null + \bigglP }
    {1 \over 16} X^4 - \biggl( {1 \over 3} Y - {89 \over 72} \biggr) X^3
  + \biggl( {5 \over 8} \pi^2 - {403 \over 72}
           - {41 \over 24} Y - {3 \over 4} Y^2 \biggr) X^2
{} \PlusBreak{\null + \bigglP }
    \biggl( {1 \over 3} Y^3
           + \biggl( {\pi^2 \over 6} + {5 \over 2} \biggr) Y
           - {115 \over 27} - {41 \over 24} \pi^2
           + 3 \zeta_3 \biggr) X
  + {1 \over 24} Y^4
  - {\pi^2 \over 2} Y^2
{} \MinusBreak{\null + \bigglP }
    \biggl( 2 \zeta_3 + {158 \over 27} \biggr) Y
  + {30659 \over 1296} - {31 \over 16} \pi^2
  - {13 \over 288} \pi^4 - {11 \over 72} \zeta_3
\biggr] {1 \over y}
%%%%% 1 %%%%%%%
{} \PlusBreak{}
  {1 \over 12} \biggl( 67 + 31 x \biggr) Y^2 X
- \pi^2 \biggl( {45 \over 8} + {139 \over 24} x \biggr) Y
%%%%% y/x %%%%%
- \biggl[  {71 \over 36} x Y^3
    + \biggl( {5 \over 2} - {599 \over 72} x \biggr) Y^2
\biggr] {y \over x}
%
%%%%%%% Imaginary part %%%%%%%%
%
{} \PlusBreak{}
i \pi {} \Biggl\{
\biggl[
%%%%%%% x/y %%%%%%%%
  ( 3 x + 10 ) \li3(-x)
- 4 \li3(-y)
- \biggl( {1 \over 2} + x \biggr) X^3
+ \biggl( {113 \over 12} + {29 \over 3} x \biggr) X^2
{} \MinusBreak{\null + i \pi \BigglP \,\, + \null }
   \biggl(   ( 5 + 3 x ) X + {22 \over 3} x
           + {127 \over 6} + 2 Y \biggr) \li2(-x)
- \biggl( 1 - {1 \over 2} x \biggr) Y^2 X
{} \MinusBreak{\null + i \pi \BigglP \,\, + \null }
   \biggl(  \biggl( {71 \over 6} + {20 \over 3} x \biggr) Y
          - \biggl( {\pi^2 \over 3} - {134 \over 9} \biggr) x
          + {737 \over 36} - {\pi^2 \over 2} \biggr) X
+ \biggl( {11 \over 18} \pi^2 - 6 \biggr) x
{} \PlusBreak{\null + i \pi \BigglP \,\, + \null}
   {53 \over 36} \pi^2
- \biggl(   \biggl( {\pi^2 \over 3} - {170 \over 9} \biggr) x
           + {\pi^2 \over 3} - {1189 \over 36} \biggr) Y
-4 \zeta_3 - {145 \over 9}
\biggr] {x \over y}
%%%%%%% 1/y %%%%%%%%
{} \PlusBreak{ \null + i \pi \,\, \BigglP \null }
\biggl[
   2 \li3(-x)
  -2 \li3(-y)
  - \biggl( {77 \over 6} + {5 \over 2} X + Y \biggr) \li2(-x)
  - {1 \over 4} X^3 + {43 \over 12} X^2
{} \MinusBreak{\null + i \pi \BigglP + + \null}
    \biggl(  {59 \over 12} Y - {\pi^2 \over 4}
            +{313 \over 36} + {1 \over 2} Y^2 \biggr) X
  - \biggl( {\pi^2 \over 6} - {329 \over 36} \biggr) Y
{} \PlusBreak{ \null + i \pi \BigglP + \null }
    \zeta_3 - {91 \over 9}
  + {25 \over 36} \pi^2
\biggr] {1 \over y}
%%%%%%%% others %%%%%%%%%
-\biggl(
    Y X^2
  - {1 \over 2} Y^3
  + {20 \over 3} Y^2
\biggr) y
  -{5 \over x y} Y
\Biggr\}
%%%%% end : BB11Paper
\,, \label{BB11}\\[1pt plus 4pt]
%__________________________________________________________________________
%
C^{[1]}_1 & = &
%%%%% begin : CC11Paper
%%%%% x/y %%%%%%%
\biggl[
12 \li4 \biggl( - {x \over y} \biggr)
+(4 x+20) \li4(-x)
+(4+8 x) \li4(-y)
{} \MinusBreak{\null \bigglP}
  ((4 x+6) X+4+6 Y) \li3(-x)
+(12 X-(16+8 x) Y) \li3(-y)
{} \PlusBreak{\null \bigglP}
  \biggl( (6 Y+4) X
         +{4 \over 3} \pi^2 x + {2 \over 3} \pi^2 \biggr) \li2(-x)
+\biggl( {7 \over 12} + {1 \over 24} x \biggr) Y^4
{} \PlusBreak{\null \bigglP}
  \biggl( {1 \over 12} + {5 \over 24} x \biggr) X^4
+\biggl( {1 \over 3} (1 + 2 x) Y
         - {9 \over 4}x - {11 \over 6} \biggr) X^3
- {\pi^2 \over 2} x Y^2
{} \MinusBreak{\null \bigglP}
  \biggl( \biggl( {3 \over 4} x - {9 \over 2} \biggr) Y^2
          - \biggl( {7 \over 2} + {9 \over 4} x \biggr) Y
          + \biggl( {5 \over 8} + {11 \over 6} \pi^2 \biggr) x
          + {7 \over 4} \pi^2 - 4 \biggr) X^2
{} \MinusBreak{\null \bigglP}
  {1 \over 3} (8 x+22) Y^3 X
-\biggl( \biggl( {27 \over 4} - {11 \over 3} \pi^2 \biggr) x
         -{7 \over 3} \pi^2 + {57 \over 4} \biggr) Y X
{} \PlusBreak{\null \bigglP}
  \biggl( 18 - 2 \zeta_3
         +12 x - {5 \over 12} \pi^2 \biggr) X
+(6 \zeta_3 - 12 x -24) Y
-{19 \over 2} \zeta_3 - {23 \over 60} \pi^4
{} \PlusBreak{\null \bigglP}
  \biggl( {511 \over 64} - {15 \over 4} \zeta_3
         + {29 \over 48} \pi^2 - {49 \over 120} \pi^4 \biggr) x
+{19 \over 8} \pi^2 + {511 \over 32}
\biggr] {x \over y}
%%%%%% 1/y %%%%%%%%%%
{} \PlusBreak{}
\biggl[
   6 \li4 \biggl( - {x \over y} \biggr)
  +10 \li4(-x)
  +2 \li4(-y)
  - (1+3 Y+3 X) \li3(-x)
{} \MinusBreak{\null \bigglP \null + \null}
    (8 Y-6 X) \li3(-y)
  + \biggl( X^2 + (3 Y+1) X + {\pi^2 \over 3} \biggr) \li2(-x)
  + {1 \over 24} X^4
{} \PlusBreak{\null \bigglP \null + \null}
   {7 \over 24} Y^4
  -\biggl( {13 \over 24} - {1 \over 6} Y \biggr) X^3
  +\biggl( {9 \over 4} Y^2 + {7 \over 8} Y
          - {7 \over 8} \pi^2 + {1 \over 8} \biggr) X^2
{} \MinusBreak{\null \bigglP \null + \null}
   \biggl(  {11 \over 3} Y^3
           +\biggl( {15 \over 2}-{7 \over 6} \pi^2 \biggr) Y
           +\zeta_3 + {\pi^2 \over 4} - 6 \biggr) X
{} \MinusBreak{\null \bigglP \null + \null}
   (12-3 \zeta_3) Y
  -{35 \over 4} \zeta_3
  +{85 \over 48} \pi^2 - {23 \over 120} \pi^4
  +{511 \over 64}
\biggr] {1 \over y}
%%%%%%%%% 1 %%%%%%%%%%%
-{\pi^2 \over 2} Y
{} \MinusBreak{}
  {3 y \over 2 x} Y^2
-2 x X^2 \li2(-x)
-\biggl[
    4 Y^2 \li2(-x)
  - {9 \over 4} Y^2 X
  + {9 \over 4} Y^3
  - {59 \over 8} Y^2
\biggr] y
%
%%%%%%% Imaginary part %%%%%%%%%%
%
{} \PlusBreak{}
i \pi {} \Biggl\{
%%%%%%%%%% x/y %%%%%%%%%%
\biggl[
-(12+4 x) \li3(-x)
-(4+8 x) \li3(-y)
+ {1 \over 6} (5 x + 2) X^3
{} \MinusBreak{\null + i \pi \BigglP \,\, + \null }
  \biggl( {9 \over 2} x + 4 \biggr) X^2
+((10+4 x) X + 4 - (10+8 x) Y) \li2(-x)
{} \MinusBreak{\null + i \pi \BigglP \,\, + \null }
  \biggl(  \biggl( {11 \over 2} x + 5 \biggr) Y^2
         - (16 + 9 x) Y + 8 x
         + {25 \over 4} - {\pi^2 \over 6} \biggr) X
{} \PlusBreak{\null + i \pi \BigglP \,\, + \null }
  \biggl( \biggl( 8 + {4 \over 3} \pi^2 \biggr) x
         + {5 \over 3} \pi^2 + {49 \over 4} \biggr) Y
- 6 - {7 \over 12} \pi^2
+ 4 \zeta_3
\biggr] {x \over y}
%%%%%%%%%% 1/y %%%%%%%%%%
{} \PlusBreak{\null + i \pi \BigglP }
\biggl[
  -6 \li3(-x)
  -2 \li3(-y)
  +(1 + 5 X - 5 Y) \li2(-x)
  + {1 \over 6} X^3 - {5 \over 4} X^2
{} \PlusBreak{\null + i \pi \BigglP \,\,\,\, + \null }
    \biggl( {25 \over 4} Y + {\pi^2 \over 12}
           - {29 \over 4} - {5 \over 2} Y^2 \biggr) X
  +\biggl( {5 \over 6} \pi^2 + {5 \over 4} \biggr) Y
  +2 \zeta_3 - 6
  - {\pi^2 \over 12}
\biggr] {1 \over y}
%%%%%%%%%% 1 %%%%%%%%%%%%%
{} \PlusBreak{\null + i \pi \BigglP }
\biggl(
   {1 \over 2} Y X^2
  +{1 \over 6}  Y^3
  -{9 \over 2}  Y^2
\biggr) y
-{3 \over x y} Y
\Biggr\}
%%%%% end : CC11Paper
\,, \label{CC11}\\[1pt plus 4pt]
%__________________________________________________________________________
%
D^{[1]}_1 & = &
%%%%% begin : DD11Paper
%%%%%%%% x/y %%%%%%%%%%
\biggl[
-\biggl( {8 \over 9} + {29 \over 36} x \biggr) X^2
- \pi^2 \biggl( {11 \over 72} x + {1 \over 12} \biggr) X
\biggr] {x \over y}
-{1 \over 9} x X^3
%%%%%%%%% 1/y %%%%%%%%%
{} \PlusBreak{}
\biggl[
  {1 \over 18} X^3
  -{19 \over 36} X^2
  -{\pi^2 \over 24} X
\biggr] {1 \over y}
%%%%%%%% 1 %%%%%%%%%
-{1 \over 3} y X \li2(-x)
-{1 \over 27} (7 + 31 x) X
{} \MinusBreak{}
{17 \over 72} \pi^2 - {25 \over 72} \pi^2 x
%%%%%%%% y %%%%%%
+\biggl[
  {1 \over 3} \li3(-x)
  -{1 \over 6} Y X^2
  -\biggl( {455 \over 108} + {49 \over 36} \zeta_3 \biggr)
\biggr] y
%
%%%%%%%% Imaginary part %%%%%%%%%
%
{} \PlusBreak{}
i \pi {} \Biggl\{
%%%%%%%%% x/y %%%%%%%%%
-\biggl[ {29 \over 18} x + {16 \over 9} \biggr] {x \over y} X
-{1 \over 3} x X^2
%%%%%%%% 1/y %%%%%%%%
+ \biggl[ {1 \over 6} X^2 - {19 \over 18} X \biggr] {1 \over y}
%%%%%%%% 1 %%%%%%%%%%
{} \MinusBreak{ \null +  i \pi \BigglP }
{1 \over 3} y \li2(-x)
-{31 \over 27} x
-{7 \over 27}
%%%%%%%% y/x %%%%%%%
-\biggl[ {1 \over 3} X Y - {5 \over 72} \pi^2 \biggr] y
\Biggr\}
%%%%% end : DD11Paper
\,, \label{DD11}\\[1pt plus 4pt]
%___________________________________________________________________________
%
E^{[1]}_1 & = &
%%%%% begin : EE11Paper
%%%%%%% x/y %%%%%%%
\biggl[
  \biggl( {16 \over 9} + {29 \over 18} x \biggr) X^2
+\biggl( {11 \over 36} \pi^2 x + {\pi^2 \over 6} \biggr) X
\biggr] {x \over y}
+{2 \over 9} x X^3
%%%%%%% 1/y %%%%%%%
{} \MinusBreak{}
\biggl[
   {1 \over 9} X^3
  -{19 \over 18} X^2
  -{\pi^2 \over 12} X
\biggr] {1 \over y}
%%%%%%% 1 %%%%%%%%%
+{2 \over 3} y X \li2(-x)
+{1 \over 27} (14 + 62 x) X
%%%%%% y %%%%%%%%
{} \PlusBreak{}
\biggl[
   {2 \over 3} ( \li3(-y) - \li3(-x) + Y \li2(-x) )
  +{1 \over 3} Y X^2 + {1 \over 3} Y^2 X
  +{2 \over 9} Y^3 - {29 \over 18} Y^2
{} \PlusBreak{\null + \bigglP}
   \biggl( {62 \over 27} - {5 \over 12} \pi^2 \biggr) Y
  -\biggl( {685 \over 162} + {35 \over 36} \zeta_3 \biggr)
\biggr] y
-{\pi^2 \over 8} + {7 \over 72} \pi^2 x
%
%%%%%%%% Imaginary part %%%%%%%%%%
%
{} \PlusBreak{}
i \pi {} \Biggl\{
%%%%%%% x/y %%%%%%%%%
  {1 \over 9} (32 + 29 x) {x \over y} X
+ {2 \over 3} x X^2
%%%%%%% 1/y %%%%%%%%%
-\biggl[ {1 \over 3} X^2 - {19 \over 9} X \biggr] {1 \over y}
%%%%%% 1 %%%%%%
+{4 \over 3} y \li2(-x)
{} \MinusBreak{ \null + i \pi \BigglP }
{16 \over 9}
%%%%%% y/x %%%%%%%%%
+\biggl[
   {2 \over 3} X Y
  +{2 \over 3} Y^2
  -{29 \over 9} Y
  -{\pi^2 \over 9}
\biggr] y
\Biggr\}
%%%%% end : EE11Paper
\,, \label{EE11}\\[1pt plus 4pt]
%___________________________________________________________________________
%
F^{[1]}_1 & = &
%%%%% begin : FF11Paper
{25 \over 81} y
%%%%% end : FF11Paper
\,, \label{FF11} %\\[1pt plus 4pt]
%___________________________________________________________________________
%
\end{eqnarray}

For $h=1$ in \eqn{h1} and color factor $\trc^{[2]}$ in \eqn{basis12}:

\begin{eqnarray}
G^{[2]}_1 & = &
%%%%% begin : GG12Paper
%%%%%%% x/y %%%%%%%%
\biggl[
  (2 - 4 x) \li4 \biggl( - {x \over y} \biggr)
+(2 x+8) \li4(-x)
-(12+x) \li4(-y)
{} \MinusBreak{\null \bigglP}
  {5 \over 24} x Y^4
-\biggl( (2 - 3 x) Y
          + {3 \over 2} x \biggr) \li3(-x)
-\biggl( {1 \over 6} - {1 \over 12} x \biggr) Y X^3
{} \MinusBreak{\null \bigglP}
  \biggl(   (2+3 x) X - (6+3 x) Y
          - {38 \over 3} - {10 \over 3} x \biggr) \li3(-y)
  +\biggl( {5 \over 6}+{5 \over 4} x \biggr) Y^3 X
{} \PlusBreak{\null \bigglP}
  \biggl( {3 \over 2} x X
         + {1 \over 3} (38 + 10 x) Y
%{} \PlusBreak{\bigglP \null + \bigglP}
         + 3 \pi^2 + {5 \over 3} \pi^2 x \biggr) \li2(-x)
{} \MinusBreak{\null \bigglP}
  \biggl(  \biggl( {5 \over 4} + {13 \over 8} x \biggr) Y^2
          - \biggl( 1 + {3 \over 4} x \biggr) Y
          + \biggl( {1 \over 4} + {\pi^2 \over 8} \biggr) x
          - 1 - {\pi^2 \over 2} \biggr) X^2
{} \MinusBreak{\null \bigglP}
  \biggl( \biggl( {5 \over 8} - 2 \pi^2 \biggr) x
         -{7 \over 3} \pi^2 + 3 \biggr) Y X
-\biggl( \biggl( {\pi^2 \over 4} - 3 \zeta_3 \biggr) x
          +{23 \over 12} \pi^2 - 2 \zeta_3 \biggr) X
{} \MinusBreak{\null \bigglP}
  \pi^2 \biggl( {3 \over 8} x - {1 \over 12} \biggr) Y^2
+\biggl( \biggl( {79 \over 27} - {5 \over 2} \zeta_3 \biggr) x
         +3 \zeta_3 + {158 \over 27} \biggr) Y
- {49 \over 144} \pi^4 + {95 \over 18} \pi^2
{} \PlusBreak{\null \bigglP}
  \biggl( - {131 \over 480} \pi^4
    - {23213 \over 5184} + {89 \over 36} \pi^2
    - {197 \over 72} \zeta_3 \biggr) x
- {485 \over 36} \zeta_3 - {23213 \over 2592}
\biggr] {x \over y}
%%%%%%%%%% 1/y %%%%%%%%%%
{} \PlusBreak{}
\biggl[
  -2 \li4 \biggl( - {x \over y} \biggr)
  +\li4(-x)
  -3 \li4(-y)
  +(2 Y+1-2 X) \li3(-x)
{} \PlusBreak{\null + \bigglP}
   \biggl( 3 Y + {28 \over 3} - X \biggr) \li3(-y)
  +\biggl( {1 \over 2} X^2 - (1+Y) X + \pi^2
          + {28 \over 3} Y \biggr) \li2(-x)
{} \MinusBreak{\null + \bigglP}
   {1 \over 12} Y X^3
  +\biggl( - {5 \over 8} Y^2 + {\pi^2 \over 4} + {1 \over 8} Y \biggr) X^2
  -{1 \over 8} Y^4 - {5 \over 24} \pi^2 Y^2
{} \PlusBreak{\null + \bigglP}
   \biggl( {79 \over 27} - {3 \over 2} \zeta_3 \biggr) Y
  +\biggl( {11 \over 12} Y^3
          -\biggl( {19 \over 8} - {7 \over 6} \pi^2 \biggr) Y {}
          -\pi^2 + \zeta_3 \biggr) X
{} \MinusBreak{\null + \bigglP}
   {737 \over 72} \zeta_3
  +{101 \over 36} \pi^2 - {161 \over 1440} \pi^4
  -{23213 \over 5184}
\biggr] {1 \over y}
%%%%%%%% 1 %%%%%%%%%%
{} \MinusBreak{}
  \biggl( {65 \over 12} + {29 \over 12} x \biggr) Y^2 X {}
+ 4 x X \li3(-x) - x {} ( X^2 - 2 Y X ) \li2(-x)
{} \PlusBreak{}
  \pi^2 \biggl( {49 \over 16} + {139 \over 48} x \biggr) Y
+ {3 y \over 2 x} Y^2
%%%%%%% y/x %%%%%%%%%
+\biggl[
    {1 \over 2} Y^2 \li2(-x)
  + {11 \over 18} Y^3
  - {277 \over 72} Y^2
\biggr] y
%
%%%%%%%% Imaginary part %%%%%%%%%
%
{} \PlusBreak{}
i \pi {} \Biggl\{
%%%%%%%% x/y %%%%%%%%
\biggl[
-(6+x) \li3(-x)
+4 \li3(-y)
+\biggl( {38 \over 3} + {29 \over 6} x - x Y \biggr) \li2(-x)
{} \MinusBreak{\null + i \pi \BigglP  + \null }
  \biggl( {1 \over 6} - {1 \over 12} x \biggr) X^3 + X^2
- \biggl( \biggl( {599 \over 72} + {\pi^2 \over 4} \biggr) x
   + {277 \over 18}+ {5 \over 6} \pi^2 \biggr) Y
{} \PlusBreak{\null + i \pi \BigglP  + \null }
  \biggl( \biggl( 1 - {1 \over 2} x \biggr) Y^2
         + {1 \over 2} Y + \biggl( {7 \over 12} \pi^2 - {9 \over 8} \biggr) 
x
         + \pi^2-1 \biggr) X
{} \PlusBreak{\null + i \pi \BigglP  + \null }
  \biggl( {1 \over 2} \zeta_3 - {61 \over 144} \pi^2
   + {79 \over 27} \biggr) x
+ 5 \zeta_3 + {158 \over 27}
- {31 \over 72} \pi^2
\biggr] {x \over y}
%%%%%%% 1/y %%%%%%%%%%
{} \PlusBreak{ i \pi \BigglP \null + \null}
\biggl[
   2 \li3(-y)
  + {25 \over 3} \li2(-x)
  -{1 \over 12} X^3 + {5 \over 8} X^2
  -\biggl( {293 \over 72} + {5 \over 12} \pi^2 \biggr) Y
{} \PlusBreak{\null + i \pi \BigglP \,\,\,\, + \null }
   \biggl( {\pi^2 \over 2} + {1 \over 4} Y
          -{19 \over 8} + {1 \over 2} Y^2 \biggr) X
  -{1 \over 2} \zeta_3 + {79 \over 27}
  -{13 \over 144} \pi^2
\biggr] {1 \over y}
%%%%%%%%% 1 %%%%%%%%%%
{} \MinusBreak{ i \pi \BigglP \null + \null}
\biggl( 3 x + {3 \over 2} \biggr) Y X
%%%%%%% y/x %%%%%%%%%%
+\biggl[
  -{1 \over 2} Y X^2
  -{1 \over 4} Y^3
  +{31 \over 12} Y^2
\biggr] y
  +{3 \over x y} Y
\Biggr\}
%%%%% end : GG12Paper
\,, \label{GG12}\\[1pt plus 4pt]
%____________________________________________________________________________
%
H ^{[2]}_1 & = &
%%%%% begin : HH12Paper
%%%%%%%  x/y %%%%%%%%
\biggl[
4 x \li4 \biggl( - {x \over y} \biggr)
-(3 x+9) \li4(-x)
-4 x \li4(-y)
%{} \PlusBreak{\null \bigglP}
  +\biggl( {1 \over 24} x - {1 \over 4} \biggr) Y^4
{} \PlusBreak{\null \bigglP}
   \biggl( (6+5 x) X - (2+3 x) Y
        -{1 \over 3} x - {13 \over 6} \biggr) \li3(-x)
{} \PlusBreak{\null \bigglP}
  \biggl( - (2 - 3 x) X + (4+2 x) Y
          + {1 \over 3} x + {2 \over 3} \biggr) \li3(-y)
   +\biggl( {5 \over 12} x + {13 \over 6} \biggr) Y^3 X
{} \PlusBreak{\null \bigglP}
  \biggl(   \biggl( 2 x Y + {x \over 3} + {13 \over 6} \biggr) X
          + {1 \over 3} (2 + x) Y
          - \pi^2 (4 + 3 x) \biggr) \li2(-x)
{} \MinusBreak{\null \bigglP}
  \biggl(  \biggl( {1 \over 4} x - {1 \over 6} \biggr) Y
           + {11 \over 18} x + {25 \over 36} \biggr) X^3
+\biggl( \biggl( {241 \over 27} + {7 \over 2} \zeta_3 \biggr) x
         +3 \zeta_3 + {482 \over 27} \biggr) Y
{} \PlusBreak{\null \bigglP}
  \biggl( \biggl( -{1 \over 4} + {15 \over 8} x \biggr) Y^2
         -{1 \over 12} (7 x + 5) Y
         +\biggl( {277 \over 72} + {5 \over 24} \pi^2 \biggr) x
         +{37 \over 18}
         -{\pi^2 \over 4} \biggr) X^2
{} \PlusBreak{\null \bigglP}
    \biggl( \biggl( -{10 \over 3} \pi^2 + {23 \over 8} \biggr) x
           +{31 \over 4} - {10 \over 3} \pi^2 \biggr) Y X
+\biggl( {9 \over 8} \pi^2 x + {19 \over 12} \pi^2 \biggr) Y^2
{} \PlusBreak{\null \bigglP}
    \biggl( \biggl( -{79 \over 27} + {409 \over 144} \pi^2
                   -{7 \over 2} \zeta_3 \biggr) x
          +{103 \over 24} \pi^2 - 3 \zeta_3
          -{62 \over 27} \biggr) X
+ {551 \over 36} \zeta_3
+ {19 \over 24} \pi^2
{} \PlusBreak{\null \bigglP}
  \biggl( -{30659 \over 1296} + {443 \over 72} \zeta_3
          +{91 \over 144} \pi^2 + {179 \over 480} \pi^4 \biggr) x
+ {209 \over 720} \pi^4
- {30659 \over 648}
\biggr] {x \over y}
%%%%%%%% 1/y %%%%%%%%%
{} \PlusBreak{}
\biggl[
  - {9 \over 2} \li4(-x)
  - \biggl( Y + {7 \over 3} - 3 X \biggr) \li3(-x)
  + \biggl( 2 Y + {1 \over 3} - X \biggr) \li3(-y)
{} \PlusBreak{ \null + \bigglP}
   \biggl( {1 \over 3} Y - {3 \over 4} X^2
          + {7 \over 3} X - 2 \pi^2 \biggr) \li2(-x)
  +{1 \over 48} X^4
  -\biggl( {7 \over 18} - {1 \over 12} Y \biggr) X^3
{} \PlusBreak{ \null + \bigglP}
   \biggl( {125 \over 72} - {1 \over 8} Y^2
          - {\pi^2 \over 8} + {5 \over 12} Y \biggr) X^2
  - {1 \over 8} Y^4 + {19 \over 24} \pi^2 Y^2
  + \biggl( {241 \over 27} + {3 \over 2} \zeta_3 \biggr) Y
{} \PlusBreak{ \null + \bigglP}
   \biggl( {13 \over 12} Y^3
          - \biggl( {5 \over 3} \pi^2 - {39 \over 8} \biggr) Y
          + {17 \over 27} - {3 \over 2} \zeta_3
          + {95 \over 48} \pi^2 \biggr) X
  + {23 \over 144} \pi^2
{} \PlusBreak{ \null + \bigglP}
    {209 \over 1440} \pi^4 + {695 \over 72} \zeta_3
  - {30659 \over 1296}
\biggr] {1 \over y}
-{1 \over 24} x X^4 + {3 \over 2} x X^2 \li2(-x)
%%%%%%% 1 %%%%%%%%%
{} \PlusBreak{}
\biggl[
  2 Y^2 \li2(-x)
  -{4 \over 3} Y^2 X
  +{103 \over 36} Y^3
  - {157 \over 18} Y^2
  -{139 \over 48} \pi^2 Y
\biggr] y
+ 2 {y \over x} Y^2
%
%%%%%%% Imaginary part %%%%%%%%
%
{} \PlusBreak{}
i \pi {} \Biggl\{
%%%%%%%% x/y %%%%%%%%
\biggl[
  (4 + 2 x) \li3(-x)
+(5 x + 2) \li3(-y)
+\biggl( {1 \over 3} + {1 \over 12} x \biggr) X^3
{} \PlusBreak{\null + i \pi \BigglP \bigglP \null }
  \biggl( - (x+3) X + (8+6 x) Y
          + {17 \over 6} + {2 \over 3} x \biggr) \li2(-x)
- {1 \over 12} (43 + 31 x) X^2
{} \PlusBreak{\null + i \pi \BigglP  + \null }
  \biggl( {7 \over 2} x + 3 \biggr) Y^2 X
  -\biggl( {25 \over 6} x + {41 \over 6} \biggr) Y X
  +\biggl( {761 \over 72} - {3 \over 4} \pi^2 \biggr) x X
{} \PlusBreak{\null + i \pi \BigglP  + \null }
   \biggl( {427 \over 36} - {7 \over 6} \pi^2 \biggr) X
- \biggl( \biggl( {\pi^2 \over 4} + {1049 \over 72} \biggr) x
          -{\pi^2 \over 6} + {833 \over 36} \biggr) Y
{} \PlusBreak{\null + i \pi \BigglP  + \null }
  \biggl( 6 - {\pi^2 \over 18} \biggr) x
+ {140 \over 9} - {4 \over 9} \pi^2
\biggr] {x \over y}
%%%%%%%% 1/y %%%%%%%%%
{} \PlusBreak{\null + i \pi \BigglP }
\biggl[
  2 \li3(-x)
  +\li3(-y)
  +\biggl( {8 \over 3} - {3 \over 2} X + 4 Y \biggr) \li2(-x)
  +{1 \over 6} X^3 - {23 \over 12} X^2
{} \PlusBreak{\null + i \pi \BigglP  + \, \bigglP \null }
   \biggl( -{13 \over 6} Y - {7 \over 12} \pi^2
           + {601 \over 72} + {3 \over 2} Y^2 \biggr) X
  +\biggl( {\pi^2 \over 12} - {329 \over 72} \biggr) Y
{} \PlusBreak{\null + i \pi \BigglP  + \, \bigglP \null }
   {86 \over 9} - {17 \over 36} \pi^2
\biggr] {1 \over y}
%%%%%%%% 1 %%%%%%%%%%
{} \PlusBreak{\null + i \pi \BigglP }
\biggl[
  {1 \over 2} Y X^2
  -{5 \over 12} Y^3
  +{85 \over 12} Y^2
\biggr] y
  +{4 \over x y} Y
\Biggr\}
%%%%% end : HH12Paper
\,, \label{HH12}\\[1pt plus 4pt]
%__________________________________________________________________________
%
I^{[2]}_1 &=&
%%%%% begin : II12Paper
%%%%%%%%% x/y %%%%%%%%%%%%
\biggl[
-4 \li4 \biggl( -{x \over y} \biggr)
-(7 + x) \li4(-x)
-(3 x + 2) \li4(-y)
-(4 X - (6 + 3 x) Y) \li3(-y)
{} \PlusBreak{\null \bigglP}
    \biggl( (2 + x) X + {3 \over 2} + 2 Y \biggr) \li3(-x)
  -\biggl( {1 \over 24} + {1 \over 12} x \biggr) X^4
  + \biggl( {3 \over 4} x + {7 \over 12} - {1 \over 6} x Y \biggr) X^3
{} \MinusBreak{\null \bigglP}
    {1 \over 6} Y^4  + Y^3 X
   -\biggl(  \biggl( 2 Y + {3 \over 2} \biggr) X
            +{2 \over 3} \pi^2 + {2 \over 3} \pi^2 x
    \biggr) \li2(-x)
  + \biggl( {\pi^2 \over 12} x - {\pi^2 \over 6} \biggr) Y^2
{} \PlusBreak{\null \bigglP}
    \biggl( -\biggl( {3 \over 2} - {1 \over 4} x \biggr) Y^2
   -{1 \over 4} (3 x + 5) Y
    + \biggl( {2 \over 3} \pi^2 + {7 \over 8} \biggr) x
   -{1 \over 2} + {7 \over 12} \pi^2 \biggr) X^2
{} \MinusBreak{\null \bigglP}
   \biggl( \biggl( {4 \over 3} \pi^2 - {9 \over 4} \biggr) x
          - {19 \over 4} + \pi^2
   \biggr) Y X
   -\biggl( 9 + 6 x-{\pi^2 \over 4} \biggr) X
  + {17 \over 2} \zeta_3
  + {13 \over 60} \pi^4
{} \PlusBreak{\null \bigglP}
    (12 + 6 x-2 \zeta_3) Y +
\biggl( -{511 \over 64} + {7 \over 40} \pi^4
   + {15 \over 4} \zeta_3 - {29 \over 48} \pi^2 \biggr) x
-{511 \over 32}
-{41 \over 24} \pi^2
\biggr] {x \over y}
%%%%%%%%%%%%%% 1/y %%%%%%%%%%%%%%%%%
{} \PlusBreak{}
   \biggl[
  -2 \li4 \biggl( -{x \over y} \biggr)
  -{7 \over 2} \li4(-x)
  -\li4(-y)
  -{1 \over 48} X^4
  -{1 \over 12} Y^4
  - (2 X-3 Y) \li3(-y)
{} \PlusBreak{\null + \bigglP}
   \biggl( {1 \over 2} + Y + X \biggr) \li3(-x)
  -\biggl( {1 \over 4} X^2
          + \biggl( {1 \over 2} + Y \biggr) X
          + {\pi^2 \over 3}
   \biggr) \li2(-x)
   + {5 \over 24} X^3
{} \PlusBreak{\null + \bigglP}
     \biggl( {7 \over 24} \pi^2 -{3 \over 4} Y^2
            -{3 \over 8} Y + {5 \over 8}
     \biggr) X^2
   + \biggl( {2 \over 3} Y^3
            + \biggl( {5 \over 2} - {\pi^2 \over 2} \biggr) Y-3
            + {\pi^2 \over 12}
     \biggr) X
{} \MinusBreak{\null + \bigglP}
     {\pi^2 \over 12} Y^2
   + (6-\zeta_3) Y - {53 \over 48} \pi^2
   + {23 \over 4} \zeta_3 + {13 \over 120} \pi^4 - {511 \over 64}
\biggr] {1 \over y}
%%%%%%%%%%%%%%% other terms %%%%%%%%%%%%
+{1 \over 2} x X^2 \li2(-x)
{} \MinusBreak{}
   \biggl( {2 \over 3} + x \biggr) Y^3 X
+ {\pi^2 \over 6} Y
+\biggl[ {3 \over 2} Y^2 \li2(-x)
         -{3 \over 4} Y^2 X + {3 \over 4} Y^3
         +\biggl( {1 \over 2 x} - {25 \over 8} \biggr) Y^2
\biggr] y
%
%%%%%%%%%% Imaginary part %%%%%%%%%%%%%%
%
{} \PlusBreak{}
  i \pi {} \Biggl\{
\biggl[
     (4 + x) \li3(-x)
   + (3 x + 2) \li3(-y)
   + \biggl( {3 \over 2} x + {5 \over 4} \biggr) X^2
  -\biggl( {1 \over 3} x + {1 \over 6} \biggr) X^3
{} \MinusBreak{\null + i \pi \BigglP \bigglP \null }
     \biggl( (x + 3) X + {3 \over 2} - (3 x + 4) Y \biggr) \li2(-x)
   - \biggl( 3 x + {11 \over 2} \biggr) Y X
   - 2 \zeta_3
{} \PlusBreak{\null + i \pi \BigglP \bigglP \null }
     \biggl( {15 \over 4} - {\pi^2 \over 6} + 4 x \biggr) X
   - \biggl(  \biggl( 4 + {\pi^2 \over 2} \biggr) x
             +{27 \over 4} + {2 \over 3} \pi^2
     \biggr) Y
   + 3 + {\pi^2 \over 4}
\biggr] {x \over y}
{} \PlusBreak{\null + i \pi \BigglP }
  \biggl[
   2 \li3(-x)
  +\li3(-y)
  -\biggl( {3 \over 2} X + {1 \over 2} - 2 Y \biggr) \li2(-x)
  -{1 \over 12} X^3 + {1 \over 2} X^2
{} \PlusBreak{\null + i \pi \BigglP  + \, \bigglP \null }
   \biggl( {15 \over 4} - {9 \over 4} Y - {\pi^2 \over 12} + Y^2 \biggr) X
  -\biggl( {7 \over 4} + {\pi^2 \over 3} \biggr) Y
  -\zeta_3 + 3 + {\pi^2 \over 12}
\biggr]{1 \over y}
{} \MinusBreak{\null + i \pi \BigglP }
   2 x Y^2 X
+ \biggl( {3 \over 2} x Y^2 + {1 \over y^2} Y \biggr) {y \over x}
\Biggr\}
%%%%% end : II12Paper
\,, \label{II12} \\[1pt plus 4pt]
%_____________________________________________________________________________
%
J^{[2]}_1  & = &
%%%%% begin : JJ12Paper
%%%%%%%% y %%%%%%%
\biggl[
  -{1 \over 3} \li3(-y)
  -{1 \over 3} Y \li2(-x)
  -{1 \over 6} Y^2 X
  -{1 \over 9} Y^3  + {29 \over 36} Y^2
{} \PlusBreak{\null \bigglP}
   \biggl( {5 \over 24} \pi^2 - {31 \over 27} \biggr) Y
  + {49 \over 36} \zeta_3 - {25 \over 72} \pi^2
\biggr] y
%%%%%%% 1 %%%%%%%%
-{455 \over 108} (1+x)
%
%%%%%%% Imaginary Part %%%%%%%
%
{} \PlusBreak{}
i \pi {} \biggl\{
-{1 \over 3} \li2(-x)
-{1 \over 3} Y^2 + {29 \over 18} Y
-\biggl( {\pi^2 \over 72} + {31 \over 27} \biggr)
\biggr\} y
%%%%% end : JJ12Paper
\,, \label{JJ12}\\[1pt plus 4pt]
%____________________________________________________________________________
%
K^{[2]}_1  & = &
%%%%% begin : KK12Paper
%%%%%%% x/y %%%%%%%%
\biggl[
- \biggl( {8 \over 9} + {29 \over 36} x \biggr) X^2
- \pi^2 \biggl( {11 \over 72} x + {1 \over 12} \biggr) X
\biggr] {x \over y}
-{1 \over 9} x X^3
-{1 \over 3} y X \li2(-x)
%%%%%%% 1/y %%%%%%%%%
{} \PlusBreak{}
\biggl[
   {1 \over 18} X^3
  -{19 \over 36} X^2
  -{\pi^2 \over 24} X
\biggr] {1 \over y}
%%%%%%%% 1 %%%%%%%%%
-{1 \over 27} (7 + 31 x) X
+ {\pi^2 \over 72} (1 - 7 x)
{} \PlusBreak{}
\biggl[
   {1 \over 3} \li3(-x)
  -{1 \over 3}  \li3(-y)
  -{1 \over 3}  Y \li2(-x)
  -{1 \over 6} Y X^2
  -{1 \over 6}  Y^2 X
  -{1 \over 9}  Y^3
{} \PlusBreak{\null \bigglP}
    {29 \over 36}  Y^2
  +\biggl( {5 \over 24} \pi^2 - {31 \over 27} \biggr)  Y
  +\biggl( {685 \over 162} + {35 \over 36} \zeta_3 \biggr)
\biggr] y
%
%%%%%%% Imaginary part %%%%%%%
%
{} \PlusBreak{}
i \pi {} \Biggl\{
%%%%%% x/y %%%%%%%%%
-\biggl( {16 \over 9} + {29 \over 18} x \biggr) {x \over y} X
-{1 \over 3} x X^2
%%%%%%% 1 %%%%%%%
+{8 \over 9}
-\biggl[ {19 \over 18} X - {1 \over 6} X^2 \biggr] {1 \over y}
{} \MinusBreak{\null + i \pi \BigglP \null }
  {2 \over 3} y \li2(-x)
+\biggl[
  -{1 \over 3} Y^2
  +{29 \over 18} Y
  -{1 \over 3} X Y
  +{\pi^2 \over 18}
\biggr] y
\Biggr\}
%%%%% end : KK12Paper
\,, \label{KK12}\\[1pt plus 4pt]
%__________________________________________________________________________
%
L^{[2]}_1  & = &
%%%%% begin : LL12Paper
-{25 \over 81} y
%%%%% end : LL12Paper
\,, \label{LL12} %\\[1pt plus 4pt]
%________________________________________________________________________
%
\end{eqnarray}

For $h=2$ in \eqn{h2} and color factor $\trc^{[1]}$ in \eqn{basis12}:

\begin{eqnarray}
A^{[1]}_2 & = &
%%%%% begin : AA21Paper
%%%%%% x/y %%%%%%
-\biggl( {56 \over 9} + {85 \over 18} x \biggr) {x \over y} X^2
%%%%% 1 %%%%%%
-3 \li4 \biggl( - {x \over y} \biggr)
- (3 - x) \li4(-x)
+{1 \over 2} x X^2 \li2(-x)
{} \PlusBreak{}
  3 \li4(-y)
-\biggl( {11 \over 6} x - 3 + x X \biggr) \li3(-x)
-3 X \li3(-y)
+\biggl( -{49 \over 36} x + {1 \over 6} Y x \biggr) X^3
{} \PlusBreak{}
\biggl( \biggl( {11 \over 6} x - 3 \biggr) X - {\pi^2 \over 2}
         -{\pi^2 \over 3} x \biggr) \li2(-x)
+ {1 \over 12} x X^4
- {\pi^2 \over 3} x Y X
{} \PlusBreak{}
\biggl( \biggl( {11 \over 12} x - {3 \over 2} \biggr) Y
          -{5 \over 12} \pi^2 x \biggr) X^2
+\biggl( -\biggl( {1 \over 2} \zeta_3
                 -{373 \over 144} \pi^2 + {79 \over 27} \biggr) x
         + 3 \zeta_3 + {\pi^2 \over 2} \biggr) X
{} \PlusBreak{}
\biggl( {23213 \over 5184} + {113 \over 1440} \pi^4
        - {49 \over 18} \pi^2 + {197 \over 72} \zeta_3 \biggr) x
- {\pi^4 \over 30} - 3 \zeta_3 - {\pi^2 \over 2}
%%%%%% 1/x %%%%%%%%%%%
{} \PlusBreak{}
\biggl[
   -3 \li4(-x)
   +3 X \zeta_3
   +3 \li4(-y)
   -3 \li4 \biggl( - {x \over y} \biggr)
    -{\pi^4 \over 30}
{} \MinusBreak{\null + \bigglP}
     {\pi^2 \over 2} \li2(-x)
    -3 X \li3(-y)
\biggr] {1 \over x}
%%%%%%%% y/x %%%%%%%%%%
+\biggl[
    {3 \over 4} Y^2 X^2 + {1 \over 8} Y^4
  + {\pi^2 \over 4} Y^2 - {1 \over 2} X Y^3
\biggr] {y \over x}
%
%%%%%%%%% Imaginary part %%%%%%%%%%%%
%
{} \PlusBreak{}
i \pi {} \Biggl\{
%%%%%%%% x/y %%%%%%%%%%%%
-\biggl( {85 \over 9} x + {112 \over 9} \biggr) {x \over y} X
%%%%%%% 1 %%%%%%%%%%
- x \li3(-x)
- 3 \li3(-y)
+ x X \li2(-x)
{} \PlusBreak{\null + i \pi \BigglP }
\biggl( {11 \over 6} x - 3 \biggr) \li2(-x)
+ {1 \over 3} x X^3
-\biggl( {49 \over 12} - {1 \over 2} Y \biggr) x X^2
+\biggl( {11 \over 6} x - 3 \biggr) Y X
{} \MinusBreak{\null + i \pi \BigglP }
{\pi^2 \over 6} x X
-\biggl( {19 \over 144} \pi^2
        +{79 \over 27} + {1 \over 2} \zeta_3 \biggr) x
+{\pi^2 \over 2} + 3 \zeta_3
{} \PlusBreak{\null + i \pi \BigglP }
\biggl( 3 \zeta_3 - 3 \li3(-y) \biggr) {1 \over x}
\Biggr\}
%%%%% end : AA21Paper
\,, \label{AA21}\\[1pt plus 4pt]
%_________________________________________________________________________
%
B^{[1]}_2 & = &
%%%%% begin : BB21Paper
%%%%%%%%%%% x/y %%%%%%%%
\biggl( {869 \over 72} + {617 \over 72} x \biggr) {x \over y} X^2
%%%%%%%%% 1 %%%%%%%%%%%
-{3 \over 2} x X^2 \li2(-x)
+ 4 \zeta_3
-{8 \over 9} \pi^2
-{\pi^4 \over 45}
+2 \li4(-y)
{} \PlusBreak{}
2 \li4 \biggl( - {x \over y} \biggr)
-{1 \over 4} x X^4
%%%%%%%%%% 1 %%%%%%%%%%%%
-(3 x - 2) \li4(-x)
- \biggl( 2 Y + 2 + {11 \over 3} x - 4 X \biggr) \li3(-y)
{} \PlusBreak{}
\biggl( -6 + {11 \over 3} x + 4 Y + 3 x X \biggr) \li3(-x)
+\biggl( - {1 \over 2} Y x + {125 \over 36} x \biggr) X^3
{} \PlusBreak{}
\biggl( \biggl( 6 - {11 \over 3} x - 2 Y \biggr) X
         -\biggl( 2 + {11 \over 3} x \biggr) Y + \pi^2 x
         + {5 \over 3} \pi^2 \biggr) \li2(-x)
{} \PlusBreak{}
\biggl( \biggl( {1 \over 2} + {1 \over 4} x \biggr) Y^2
        +\biggl( 3 - {31 \over 12} x \biggr) Y
        +{3 \over 2} \pi^2 x \biggr) X^2
- {2 \over 3} Y^3 X - \biggl( {31 \over 12} x + 2 \biggr) Y^2 X
{} \PlusBreak{}
\biggl( \biggl( -{\pi^2 \over 3} + {599 \over 72} \biggr) x
        -{5 \over 12} \pi^2 + {271 \over 36} \biggr) Y^2
-\biggl( {1 \over 4}
          - \biggl( {9 \over 4} - {2 \over 3} \pi^2 \biggr) x
          \biggr) Y X
{} \PlusBreak{}
\biggl( \biggl( -{373 \over 72} \pi^2 - {4 \over 27} + \zeta_3 \biggr) x
         -{5 \over 6} \pi^2 - 4 \zeta_3
\biggr) X
+\biggl( {1 \over 4} + {1 \over 8} x \biggr) Y^4
{} \MinusBreak{}
\biggl( {17 \over 9} + {71 \over 36} x \biggr) Y^3
+\biggl( \biggl( {139 \over 24} \pi^2 - {158 \over 27} - \zeta_3 \biggr) x
         -{64 \over 9} + {211 \over 36} \pi^2+2 \zeta_3 \biggr) Y
{} \PlusBreak{}
\biggl( -{55 \over 144} \pi^2
         +{30659 \over 1296}
         +{79 \over 1440} \pi^4
         -{443 \over 72} \zeta_3 \biggr) x
%%%%%%%% 1/x %%%%%%%%%%%%%
{} \PlusBreak{}
\biggl[
  4 \li4 \biggl( - {x \over y} \biggr)
  +4 \li4(-x)
  -2 \li4(-y)
  +2 Y \li3(-x)
  -(Y+1-5 X) \li3(-y)
{} \MinusBreak{\null + \bigglP}
   \biggl( Y + Y X - {4 \over 3} \pi^2 \biggr) \li2(-x)
  +Y^2 X^2
  -\biggl( 5 \zeta_3 + {7 \over 8} Y^2
           + {5 \over 6} Y^3 \biggr) X
  +{1 \over 4} Y^4
{} \MinusBreak{\null + \bigglP}
   {59 \over 72} Y^3
  +\biggl( {73 \over 18} + {\pi^2 \over 24} \biggr) Y^2
  +\biggl( \zeta_3 + {43 \over 18} \pi^2 \biggr) Y
  +{\pi^4 \over 45} + \zeta_3
\biggr]{1 \over x}
%
%%%%%%% Imaginary part
%
{} \PlusBreak{}
i \pi {} \Biggl\{
%%%%%%%%%% x/y %%%%%%%%%%
\biggl[
   (3 x + 5) X \li2(-x)
  +\biggl( {797 \over 36} + {134 \over 9} x \biggr) X
\biggr] {x \over y}
%%%%%%%%%% 1/y %%%%%%%%%%%%%%
+\biggl[ {1 \over 4} X + 3 X \li2(-x) \biggr] {1 \over y}
%%%%%%%%% 1 %%%%%%%%%%%%%%
{} \MinusBreak{\null + i \pi \BigglP }
{64 \over 9}
+(3 x + 4) \li3(-x)
+2 \li3(-y)
-\biggl( {22 \over 3} x - 4 + 2 Y \biggr) \li2(-x)
{} \MinusBreak{\null + i \pi \BigglP }
x X^3 + \biggl( {29 \over 3} - Y \biggr) x X^2
+\biggl( {2 \over 3} + {1 \over 2} x \biggr) Y^3
+{20 \over 3} y Y^2
{} \PlusBreak{\null + i \pi \BigglP }
\biggl( \biggl( - 2 + {1 \over 2} x \biggr) Y^2
        + \biggl( -{20 \over 3} x + 4 \biggr) Y
        +{\pi^2 \over 3} x \biggr) X
{} \PlusBreak{\null + i \pi \BigglP }
\biggl( \biggl( {170 \over 9} - {\pi^2 \over 3} \biggr) x
        +{533 \over 36} + {\pi^2 \over 6} \biggr) Y
+\biggl( {11 \over 18} \pi^2 - 6 \biggr) x
-2 \zeta_3 - {3 \over 4} \pi^2
%%%%%%%%%% 1/x %%%%%%%%%%%%%%
{} \PlusBreak{\null + i \pi \BigglP }
\biggl[
     2 \li3(-x)
    +4 \li3(-y)
    -(Y + 1) \li2(-x)
    - \biggl( Y^2 + {3 \over 4} Y \biggr) X + {1 \over 3} Y^3
{} \MinusBreak{\null + i \pi \BigglP  + \, \bigglP \null }
     {17 \over 6} Y^2
    +\biggl( {\pi^2 \over 12} + {73 \over 9} \biggr) Y
    -4 \zeta_3
\biggr] {1 \over x}
%%%%%%%%%%% 1 %%%%%%%%%%%%%%
+{1 \over x y} X \li2(-x)
\Biggr\}
%%%%% end : BB21Paper
\,, \label{BB21}\\[1pt plus 4pt]
%__________________________________________________________________________
%
C^{[1]}_2 & = &
%%%%% begin : CC21Paper
%%%%%%%% x/y %%%%%%%%%
{x \over 8 y} (5 x - 7) X^2
%%%%%%%%% 1 %%%%%%%%%%
+12 \li4 \biggl( - {x \over y} \biggr)
+(4 x + 12) \li4(-x)
+2 x X^2 \li2(-x)
{} \PlusBreak{}
  8 x \li4(-y)
-(4 x X + 12 Y) \li3(-x)
+(6 X + 2 - (8 x+10) Y) \li3(-y)
{} \PlusBreak{}
  \biggl( 6 Y X -4 (1+x) Y^2
         +2 Y + {4 \over 3} \pi^2 x
         -2 \pi^2 \biggr) \li2(-x)
+{5 \over 24} x X^4
{} \PlusBreak{}
  \biggl( {2 \over 3} Y x - {9 \over 4} x \biggr) X^3
+\biggl( \biggl( -{3 \over 4} x + 3 \biggr) Y^2 + {9 \over 4} x Y
         -{11 \over 6} \pi^2 x \biggr) X^2
{} \MinusBreak{}
  \biggl( {8 \over 3} x + 5 \biggr) Y^3 X
+\biggl( 4 + {9 \over 4} x \biggr) Y^2 X
+\biggl( {3 \over 4}
         +\biggl( {11 \over 3} \pi^2 -{27 \over 4} \biggr) x \biggr) Y X
{} \PlusBreak{}
  \biggl( -{\pi^2 \over 2} + 12 x - 6 \zeta_3 \biggr) X
+\biggl( {1 \over 24} x + {2 \over 3} \biggr) Y^4
-\biggl( {9 \over 4} x + {7 \over 3} \biggr) Y^3
{} \PlusBreak{}
  \biggl( \biggl( {59 \over 8} - {\pi^2 \over 2} \biggr) x
         + {3 \over 4} + {13 \over 12} \pi^2 \biggr) Y^2
+\biggl( - {5 \over 4} \pi^2 - 6 + 10 \zeta_3 - 12 x \biggr) Y
{} \PlusBreak{}
  \biggl( {511 \over 64} + {29 \over 48} \pi^2
         - {49 \over 120} \pi^4 - {15 \over 4} \zeta_3 \biggr) x
+{\pi^2 \over 6} - 2 \zeta_3
%%%%%%%%%% 1/x %%%%%%%%%%%%
{} \PlusBreak{}
  \biggl[
   6 \li4 \biggl( - {x \over y} \biggr)
   +6 \li4(-x)
   -6 Y \li3(-x)
   +(3 - 5 Y + 3 X) \li3(-y)
   +{3 \over 2} Y^2 X^2
{} \PlusBreak{\null + \bigglP}
    (3 Y X - \pi^2 + 3 Y-2 Y^2) \li2(-x)
   +\biggl( {21 \over 8} Y^2 - {5 \over 2} Y^3 - 3 \zeta_3 \biggr) X
   +{1 \over 3} Y^4
{} \MinusBreak{\null + \bigglP}
    {7 \over 24} Y^3
   +\biggl( {13 \over 24} \pi^2 + {3 \over 2} \biggr) Y^2
   +\biggl( - {5 \over 3} \pi^2 + 5 \zeta_3 \biggr) Y
   -3 \zeta_3
\biggr] {1 \over x}
%
%%%%%%%% Imaginary part %%%%%%%%%
%
{} \PlusBreak{}
i \pi {} \Biggl\{
%%%%%%%% x/y %%%%%%%%
\biggl[
  -(4 x + 10) X \li2(-x)
  +\biggl( {17 \over 4} + 8 x \biggr) X
\biggr] {x \over y}
%%%%%%%%% 1/y %%%%%%%
-\biggl[
  {3 \over 4} X
  + 9 X \li2(-x)
\biggr] {1 \over y}
%%%%%%%% 1 %%%%%%%%%%
{} \MinusBreak{\null + i \pi \BigglP }
(12 + 4 x) \li3(-x)
-(8 x + 4) \li3(-y)
+(2 - (8 x+2) Y) \li2(-x)
{} \PlusBreak{\null + i \pi \BigglP }
{5 \over 6} x X^3
+{1 \over 2} (Y - 9) x X^2
-\biggl( \biggl( {11 \over 2} x + 2 \biggr) Y^2
        -(6 + 9 x) Y \biggr) X
{} \PlusBreak{\null + i \pi \BigglP }
\biggl( {2 \over 3} + {1 \over 6} x \biggr) Y^3
-\biggl( {9 \over 2} x + 4 \biggr) Y^2
+\biggl( \biggl( 8 + {4 \over 3} \pi^2 \biggr) x
        + {9 \over 4} + {3 \over 2} \pi^2 \biggr) Y
+4 \zeta_3
%%%%%% 1/x %%%%%%%%%%
{} \PlusBreak{\null + i \pi \BigglP }
\biggl[
   -6 \li3(-x)
   -2 \li3(-y)
   -(Y - 3) \li2(-x)
   -\biggl( Y^2 - {9 \over 4} Y \biggr) X
   +{1 \over 3} Y^3
{} \PlusBreak{\null + i \pi \BigglP  + \, \bigglP \null }
     {1 \over 4} Y^2
   +\biggl( {3 \over 4} \pi^2 + 3 \biggr) Y
   +2 \zeta_3
\biggr]{1 \over x}
%%%%%%% 1/xy %%%%%%%%%
-{3 \over x y} X \li2(-x)
- {5 \over 12} \pi^2 - 6
\Biggr\}
%%%%% end : CC21Paper
\,, \label{CC21}\\[1pt plus 4pt]
%___________________________________________________________________________
%
D^{[1]}_2 & = &
%%%%% begin : DD21Paper
\biggl[
{1 \over 3} (\li3(-x) - X \li2(-x))
+{1 \over 9} X^3
-\biggl( {29 \over 36} + {1 \over 6} Y \biggr) X^2
+ {25 \over 72} \pi^2 - {455 \over 108} - {49 \over 36} \zeta_3
{} \MinusBreak{\null \bigglP  }
\biggl( {11 \over 72} \pi^2 - {31 \over 27} \biggr) X
\biggr] x
%
%%%%%% Imaginary part %%%%%%%%
%
{} \PlusBreak{}
  i {\pi \over 3} x {} \biggl\{
  X^2 - \li2(-x)
-\biggl( {29 \over 6}  + Y \biggr)  X
+ {31 \over 9} + {5 \over 24} \pi^2
\biggr\}
%%%%% end : DD21Paper
\,, \label{DD21}\\[1pt plus 4pt]
%___________________________________________________________________________
%
E^{[1]}_2 & = &
%%%%% begin : EE21Paper
{2 \over 3} x {} (\li3(-y) - \li3(-x) + (X + Y) \li2(-x))
-{2 \over 9} x X^3
+\biggl( {1 \over 3} Y + {29 \over 18} \biggr) x X^2
{} \PlusBreak{}
\biggl( {1 \over 3} Y^2
        - {62 \over 27} + {11 \over 36} \pi^2 \biggr) x X
-{2 \over 9} y Y^3
-\biggl( {13 \over 9} + {29 \over 18} x \biggr) Y^2
{} \PlusBreak{}
\biggl( \biggl( - {5 \over 12} \pi^2 + {62 \over 27} \biggr) x
        + {16 \over 9} - {4 \over 9} \pi^2 \biggr) Y
+\biggl(
   - {685 \over 162} - {35 \over 36} \zeta_3
   - {7 \over 72} \pi^2 \biggr) x
{} \PlusBreak{}
  {2 \over 9} \pi^2
+ {1 \over 9} \biggl( Y^3 - 8 Y^2 - 2 \pi^2 Y \biggr) {1 \over x}
%
%%%%%%%% Imaginary part %%%%%%%%%
%
{} \PlusBreak{}
i \pi {} \biggl\{
{16 \over 9}
+{4 \over 3} x \li2(-x)
-{2 \over 3} x X^2
+\biggl( {29 \over 9} + {2 \over 3} Y \biggr) x X
-{2 \over 3} y Y^2
{} \MinusBreak{\null + i \pi \BigglP \null }
{1 \over 9} (26 + 29 x) Y
-{\pi^2 \over 9} x
-\biggl( {16 \over 9} Y - {1 \over 3} Y^2 \biggr) {1 \over x}
\biggr\}
%%%%% end : EE21Paper
\,, \label{EE21}\\[1pt plus 4pt]
%___________________________________________________________________________
%
F^{[1]}_2 & = &
%%%%% begin : FF21Paper
{25 \over 81} x
%%%%% end : FF21Paper
\,, \label{FF21} %\\[1pt plus 4pt]
%_________________________________________________________________________
%
\end{eqnarray}

For $h=2$ in \eqn{h2} and color factor $\trc^{[2]}$ in \eqn{basis12}:

\begin{eqnarray}
G^{[2]}_2 & = &
%%%%% begin : GG22Paper
%%%%%%%% x/y %%%%%%%%%%
-{x \over 4 y} (3 + x) X^2
+ x X^2 \li2(-x)
%%%%%%%%% 1 %%%%%%%%%%
-(3 + 4 x) \li4 \biggl( - {x \over y} \biggr)
+{1 \over 12} x X^3 Y
{} \PlusBreak{}
(2 x-3) \li4(-x)
-(4 + x) \li4(-y)
+\biggl( 3 - {3 \over 2} x - 4 x X + 3 x Y \biggr) \li3(-x)
{} \PlusBreak{}
\biggl( 3 y X + (3 x + 4) Y
         + {9 \over 2} + {10 \over 3} x \biggr) \li3(-y)
+\biggl( 1 + {5 \over 4} x \biggr) Y^3 X
{} \PlusBreak{}
\biggl( \biggl( {3 \over 2} x - 3 - 2 Y x \biggr) X
        -{1 \over 2} y Y^2
        +\biggl( {9 \over 2} + {10 \over 3} x \biggr) Y
{} \PlusBreak{\null + \,\, \bigglP}
         {\pi^2 \over 6} (10 x + 1) \biggr) \li2(-x)
+\biggl( \biggl( -{13 \over 8} + 2 \pi^2 \biggr) x
        + {2 \over 3} \pi^2 - {3 \over 2} \biggr) Y X
{} \MinusBreak{}
\biggl(  \biggl( {13 \over 8} x + {3 \over 4} \biggr) Y^2
         + \biggl( {3 \over 2} - {3 \over 4} x \biggr) Y
         + {\pi^2 \over 8} x \biggr) X^2
+\biggl( {17 \over 4} + {29 \over 12} x \biggr) Y^2 X
{} \PlusBreak{}
\biggl( \biggl( -{\pi^2 \over 4} + 3 \zeta_3 \biggr) x
          + {\pi^2 \over 6} + 3 \zeta_3 \biggr) X
-\biggl( {1 \over 6} + {5 \over 24} x \biggr) Y^4
+\biggl( {19 \over 36} + {11 \over 18} x \biggr) Y^3
{} \MinusBreak{}
\biggl( \biggl( {241 \over 72} + {3 \over 8} \pi^2 \biggr) x
         + {95 \over 36} - {\pi^2 \over 2} \biggr) Y^2
-\biggl( {131 \over 480} \pi^4 + {23213 \over 5184}
         - {107 \over 36} \pi^2 + {197 \over 72} \zeta_3 \biggr) x
{} \PlusBreak{}
\biggl( \biggl( {79 \over 27} - {139 \over 48} \pi^2
                - {5 \over 2} \zeta_3 \biggr) x
        -{179 \over 36} \pi^2 - 4 \zeta_3 + {32 \over 9} \biggr) Y
+ {61 \over 36} \pi^2
+ {2 \over 45} \pi^4 - 4 \zeta_3
%%%%%%%%%% 1/x %%%%%%%%%%
{} \PlusBreak{}
\biggl[
  -3 \li4 \biggl( - {x \over y} \biggr)
  -3 \li4(-x)
  -{1 \over 2} \li4(-y)
  -\biggl( 3 X - {3 \over 2} - 2 Y \biggr) \li3(-y)
{} \PlusBreak{\null + \bigglP}
   \biggl( {1 \over 4} Y^2 + {3 \over 2} Y - {\pi^2 \over 6} \biggr) 
\li2(-x)
  -{3 \over 4} Y^2 X^2
  -{7 \over 48} Y^4 + {11 \over 36} Y^3
  -{3 \over 2} \zeta_3 + {\pi^4 \over 180}
{} \PlusBreak{\null + \bigglP}
   \biggl( 3 \zeta_3 + {3 \over 4} Y^3
          +{\pi^2 \over 3} Y + {3 \over 2} Y^2 \biggr) X
  -\biggl( {37 \over 36} - {\pi^2 \over 8} \biggr) Y^2
  -\biggl( {19 \over 9} \pi^2 + 2 \zeta_3 \biggr) Y
\biggr] {1 \over x}
%
%%%%%% Imaginary part %%%%%%%%
%
{} \PlusBreak{}
i \pi {} \Biggl\{
%%%%%%%%%%%%%% 1 %%%%%%%%%%%%
{x \over 8 y} (13 + 9 x) X
+{3 \over 2 y} X
-x \li3(-x) + \li3(-y)
+{1 \over 12} x X^3 - {1 \over 2} x X^2 Y
{} \PlusBreak{\null + i \pi \BigglP \null }
\biggl( (1 - x) Y + {3 \over 2} + {29 \over 6} x \biggr) \li2(-x)
+ \biggl( 1 - {1 \over 2} x \biggr) Y^2 X
+ (3 x + 1) Y X
{} \PlusBreak{\null + i \pi \BigglP \null }
{7 \over 12} \pi^2 x X
+{1 \over 12} (31 x + 43) Y^2
-\biggl(  \biggl( {599 \over 72} + {\pi^2 \over 4} \biggr) x
         +{\pi^2 \over 6} + {61 \over 9} \biggr) Y
{} \PlusBreak{\null + i \pi \BigglP \null }
\biggl( {1 \over 2} \zeta_3 + {79 \over 27} - {61 \over 144} \pi^2 \biggr) x
- \zeta_3 + {\pi^2 \over 4} + {32 \over 9}
%%%%%%%%%%% 1/x %%%%%%%%%%%%%
{} \PlusBreak{\null + i \pi \BigglP \null }
\biggl[
  -\li3(-y)
  +{1 \over 2} (Y + 3) (\li2(-x) + Y X)
  +{5 \over 3} Y^2
{} \MinusBreak{\null + i \pi \BigglP  + \,\, \bigglP \null }
   \biggl( {\pi^2 \over 12} + {37 \over 18} \biggr) Y
  + \zeta_3
\biggr] {1 \over x}
%%%%%% y^2/x (?) %%%%%%%%%%%%
-{y^2 \over 4 x} Y^3
\Biggr\}
%%%%% end : GG22Paper
\,, \label{GG22}\\[1pt plus 4pt]
%_________________________________________________________________________
%
H^{[2]}_2 & = &
%%%%% begin : HH22Paper
%%%%%%%%% x/y %%%%%%%%%%%
-{1 \over 72} \biggl( 277 + 241  x \biggr) {x \over y} X^2
-{3 \over 2} x X^2 \li2(-x)
%%%%%%%%%%%% 1 %%%%%%%%%%%%%%
+4 (x-1) \li4 \biggl( - {x \over y} \biggr)
{} \MinusBreak{}
(3 x + 4) \li4(-x)
-4 x \li4(-y)
+\biggl( 5 x X + (4 - 3 x) Y - {1 \over 3} x \biggr) \li3(-x)
{} \PlusBreak{}
\biggl( (3 x - 2) X + (2 x + 4) Y + {1 \over 3} x - 4 \biggr) \li3(-y)
+{1 \over 24} x X^4
-\biggl( {1 \over 4} Y + {11 \over 18} \biggr) x X^3
{} \PlusBreak{}
\biggl( \biggl( 2 (x - 1) Y + {1 \over 3} x \biggr) X
        -2 y Y^2
        + \biggl( {1 \over 3} x - 4 \biggr) Y
        - 3 \pi^2 x \biggr) \li2(-x)
{} \PlusBreak{}
\biggl( \biggl( -1 + {15 \over 8} x \biggr) Y^2
        -{7 \over 12} x Y + {5 \over 24} \pi^2 x \biggr) X^2
+\biggl( {7 \over 3} + {5 \over 12}x \biggr) Y^3 X
{} \MinusBreak{}
\biggl( {4 \over 3} x + 5 \biggr) Y^2 X
+\biggl( \biggl( {31 \over 8} - {10 \over 3} \pi^2 \biggr) x
                - {2 \over 3} \pi^2 + {5 \over 4} \biggr) Y X
-\biggl( {5 \over 12} - {1 \over 24} x \biggr) Y^4
{} \PlusBreak{}
\biggl( \biggl( {409 \over 144} \pi^2 - {7 \over 2} \zeta_3
                - {79 \over 27} \biggr) x
        +{\pi^2 \over 2} + 2 \zeta_3 \biggr) X
+\biggl( {103 \over 36} x + {53 \over 18} \biggr) Y^3
{} \PlusBreak{}
\biggl( \biggl( {9 \over 8} \pi^2 - {83 \over 9} \biggr) x
        - {5 \over 12} \pi^2 - {44 \over 9} \biggr) Y^2
+\biggl( {19 \over 144} \pi^2 + {443 \over 72} \zeta_3
        -{30659 \over 1296} + {179 \over 480} \pi^4 \biggr) x
{} \PlusBreak{}
\biggl( \biggl( {7 \over 2} \zeta_3 + {241 \over 27}
                - {139 \over 48} \pi^2 \biggr) x
        +{59 \over 9} - {5 \over 36} \pi^2 - 4 \zeta_3 \biggr) Y
-{11 \over 36} \pi^2 + 4 \zeta_3
%%%%%%%%%%%%%%%% 1/x %%%%%%%%%%%%
{} \PlusBreak{}
\biggl[
  -2 \li4 \biggl( - {x \over y} \biggr)
  -2 \li4(-x)
  +2 Y \li3(-x)
  -\biggl( {5 \over 2} + X - 2 Y \biggr) \li3(-y)
{} \PlusBreak{\null + \bigglP}
   \biggl( Y^2 - {5 \over 2} Y - Y X \biggr) \li2(-x)
  -{1 \over 2} Y^2 X^2
  -{5 \over 24} Y^4
  +{55 \over 72} Y^3
{} \PlusBreak{\null + \bigglP}
   \biggl( -{\pi^2 \over 3} Y + {7 \over 6} Y^3 + \zeta_3
           -{19 \over 8} Y^2 \biggr) X
  -\biggl( {127 \over 36} + {5 \over 24} \pi^2 \biggr) Y^2
{} \PlusBreak{\null + \bigglP}
   \biggl( {13 \over 18} \pi^2 - 2 \zeta_3 \biggr) Y
  +{5 \over 2} \zeta_3
\biggr] {1 \over x}
%
%%%%%%%% Imaginary part %%%%%%%
%
{} \PlusBreak{}
i \pi {} \Biggl\{
%%%%%%%%%% x/y %%%%%%%%%%
\biggl[
   (x+3) X \li2(-x)
  -\biggl( {923 \over 72} + {761 \over 72} x \biggr) X
\biggr] {x \over y}
%%%%%%%%%% 1 %%%%%%%%%%%%%
- X {} \biggl[ {5 \over 4} - 3 \li2(-x) \biggr] {1 \over y}
{} \PlusBreak{\null + i \pi \BigglP \null }
(2 x + 4) \li3(-x)
+(5 x + 2) \li3(-y)
+\biggl( (2 + 6 x) Y + {2 \over 3} x -4 \biggr) \li2(-x)
{} \PlusBreak{\null + i \pi \BigglP \null }
{1 \over 12} x X^3
-\biggl( {31 \over 12} - {1 \over 2} Y \biggr) x X^2
+\biggl( 2 + {7 \over 2} x \biggr) Y^2 X
-\biggl( {25 \over 6} x + 6 \biggr) Y X
{} \MinusBreak{\null + i \pi \BigglP \null }
{3 \over 4} \pi^2 x X
-\biggl( {5 \over 12} x + {2 \over 3} \biggr) Y^3
+\biggl( {85 \over 12} x + {35 \over 6} \biggr) Y^2
-2 \zeta_3
{} \MinusBreak{\null + i \pi \BigglP \null }
\biggl( \biggl( {\pi^2 \over 4} + {1049 \over 72} \biggr) x
        + {5 \over 6} \pi^2 + {307 \over 36} \biggr) Y
+\biggl( -{\pi^2 \over 18} + 6 \biggr) x
+ {59 \over 9}
+{\pi^2 \over 4}
%%%%%%%% 1/x %%%%%%%%%%
{} \PlusBreak{\null + i \pi \BigglP \null }
\biggl[
  2 \li3(-x)
  +\li3(-y)
  +\biggl( Y - {5 \over 2} \biggr) \li2(-x)
  +\biggl( Y^2 - {9 \over 4} Y \biggr) X
  -{1 \over 3} Y^3
{} \PlusBreak{\null + i \pi \BigglP  + \,\, \bigglP \null }
   {7 \over 6} Y^2
  -\biggl( {5 \over 12} \pi^2 + {127 \over 18} \biggr) Y
  -\zeta_3
\biggr] {1 \over x}
%%%%%%%% 1/xy %%%%%%%%%%%%%
+{1 \over x y} X \li2(-x)
\Biggr\}
%%%%% end : HH22Paper
\,, \label{HH22}\\[1pt plus 4pt]
%__________________________________________________________________________
%
I^{[2]}_2 &=&
%%%%% begin : II22Paper
\biggl[
  -2 \li4 \biggl( -{x \over y} \biggr)
  -2 \li4(-x)
  -{1 \over 2} \li4(-y)
   + 2 Y \li3(-x)
   + (2 Y - X - 1) \li3(-y)
{} \PlusBreak{\null \bigglP}
   \biggl( {3 \over 4} Y^2 + {\pi^2 \over 3} - Y X - Y \biggr) \li2(-x)
  -{1 \over 2} Y^2 X^2
  -\biggl( {7 \over 8} Y^2 - \zeta_3 - {11 \over 12} Y^3 \biggr) X
{} \MinusBreak{\null \bigglP}
   {5 \over 48} Y^4 + {1 \over 24} Y^3
  -\biggl( {\pi^2 \over 4} + 1 \biggr) Y^2
   + \biggl( {2 \over 3} \pi^2 - 2 \zeta_3 \biggr) Y
   + {\pi^4 \over 180} + \zeta_3
\biggr] {1 \over x}
{} \MinusBreak{}
   4 \li4 \biggl( -{x \over y} \biggr)
-(4 + x) \li4(-x)
-(3 x + 1) \li4(-y)
+ (x X + 4 Y) \li3(-x)
{} \MinusBreak{}
   \biggl( 2 X + {1 \over 2} - (3 x + 4) Y \biggr) \li3(-y)
  -{1 \over 2} x {} X^2 \li2(-x)
  -{x \over 8 y} (7 x + 3) X^2
+ \zeta_3 - {\pi^2 \over 6}
{} \PlusBreak{}
   {\pi^4 \over 90}
+ \biggl( 3 - 4 \zeta_3 + {\pi^2 \over 2} + 6 x \biggr) Y
+ \biggl( \biggl( {\pi^2 \over 12} - {25 \over 8} \biggr) x
          - {\pi^2 \over 2} - {3 \over 4}
   \biggr) Y^2
{} \PlusBreak{}
   \biggl( {1 \over 4} x - 1 \biggr) Y^2 X^2
  +\biggl( - {3 \over 4} x Y + {2 \over 3} \pi^2 x \biggr) X^2
  +\biggl( {11 \over 6} + x \biggr) Y^3 X
  -{1 \over 4} (3 x + 5) Y^2 X
{} \MinusBreak{}
    \biggl( {1 \over 4} + \biggl( {4 \over 3} \pi^2 - {9 \over 4} \biggr) x
\biggr) Y X
   -\biggl( 6 x - 2 \zeta_3 - {\pi^2 \over 6} \biggr) X
   - {3 \over 4} y Y^3
   + x {} \biggl( {3 \over 4} -{1 \over 6} Y \biggr) X^3
{} \PlusBreak{}
     \biggl( -2 Y X + {3 \over 2} (1 + x) Y^2
             -{1 \over 2} Y-{2 \over 3} \pi^2 x
             + {2 \over 3} \pi^2
     \biggr) \li2(-x)
-{1 \over 12} x X^4
-{5 \over 24} Y^4
{} \PlusBreak{}
   \biggl( -{511 \over 64} + {7 \over 40} \pi^4
           + {15 \over 4} \zeta_3 - {29 \over 48} \pi^2
   \biggr) x
%
%%%%%%%%%%% Imaginary part %%%%%%%%%%%%%
%
{} \PlusBreak{}
i \pi {} \Biggl\{ 3
+\biggl[
   (3 + x) X \li2(-x)
  -\biggl( 4 x + {11 \over 4} \biggr) X
  \biggl] {x \over y}
+ \biggl[ {1 \over 4} X + 3 X \li2(-x) \biggr] {1 \over y}
{} \PlusBreak{\null + i \pi \BigglP }
   (4 + x) \li3(-x)
+ (3 x + 2) \li3(-y)
-\biggl( {1 \over 2} - (3 x + 1) Y \biggr) \li2(-x)
{} \MinusBreak{\null + i \pi \BigglP }
   {1 \over 3} x X^3 + {3 \over 2} x X^2
+ \biggl( (1 + 2 x) Y^2 -(3 x + 2) Y \biggr) X
-{1 \over 6} Y^3
{} \MinusBreak{\null + i \pi \BigglP }
   \biggl( \biggl( 4 + {\pi^2 \over 2} \biggr) x
           + {7 \over 4} + {2 \over 3} \pi^2
   \biggr) Y
+ {\pi^2 \over 6}-2 \zeta_3
+ \biggl( {y \over 4 x} - {3 \over 2} y \biggr) Y^2
{} \PlusBreak{\null + i \pi \BigglP }
  \biggl[
   2 \li3(-x)
  + \li3(-y)
  -\biggl( 1 - {1 \over 2} Y \biggr) \li2(-x)
   + \biggl( {1 \over 2} Y^2 - {3 \over 4} Y \biggr) X
  -{1 \over 12} Y^3
{} \MinusBreak{\null + i \pi \BigglP  + \, \bigglP \null }
   \biggl( {\pi^2 \over 3} + 2 \biggr) Y
  -\zeta_3
  \biggr] {1 \over x}
+ {1 \over x y} X \li2(-x)
\Biggr\}
%%%%% end : II22Paper
\,, \label{II22} \\[1pt plus 4pt]
%_____________________________________________________________________________
%
J^{[2]}_2 & = &
%%%%% begin : JJ22Paper
-{1 \over 3} x {} (\li3(-y) + Y \li2(-x))
-{1 \over 6} x Y^2 X
+{1 \over 9} y Y^3
+\biggl( {13 \over 18} + {29 \over 36} x \biggr) Y^2
{} \PlusBreak{}
\biggl( \biggl( -{31 \over 27} + {5 \over 24} \pi^2 \biggr) x
        + {2 \over 9} \pi^2 - {8 \over 9} \biggr) Y
+\biggl( {455 \over 108} + {49 \over 36} \zeta_3
        - {25 \over 72} \pi^2 \biggr) x - {\pi^2 \over 9}
{} \PlusBreak{}
\biggl( -{1 \over 18} Y^3 + {4 \over 9} Y^2 + {\pi^2 \over 9} Y
        \biggr) {1 \over x}
%
%%%%%%%%%% Imaginary part %%%%%%%%%%
%
{} \MinusBreak{}
i {\pi \over 3} {} \biggl\{
x \li2(-x)
-y Y^2
-\biggl( {13 \over 3} + {29 \over 6} x \biggr) Y
+\biggl( {31 \over 9} + {1 \over 24} \pi^2 \biggr) x
{} \PlusBreak{\null + i \pi \BigglP \null }
\biggl( {1 \over 2} Y^2 - {8 \over 3} Y \biggr) {1 \over x}
+{8 \over 3}
\biggr\}
%%%%% end : JJ22Paper
\,, \label{JJ22}\\[1pt plus 4pt]
%__________________________________________________________________________
%
K^{[2]}_2 & = &
%%%%% begin : KK22Paper
{1 \over 3} x {} (\li3(-x) - \li3(-y) - (Y + X) \li2(-x))
+{1 \over 9} x X^3
-\biggl( {29 \over 36} + {1 \over 6} Y \biggr) x X^2
{} \PlusBreak{}
\biggl( -{1 \over 6} Y^2
         -{11 \over 72} \pi^2 + {31 \over 27} \biggr) x X
+{1 \over 9} y Y^3
+\biggl( {13 \over 18} + {29 \over 36} x \biggr) Y^2
{} \PlusBreak{}
\biggl( \biggl( -{31 \over 27} + {5 \over 24} \pi^2 \biggr) x
        +{2 \over 9} \pi^2 - {8 \over 9} \biggr) Y
+\biggl( {685 \over 162} + {35 \over 36} \zeta_3
        + {7 \over 72} \pi^2 \biggr) x
{} \PlusBreak{}
\biggl( - {1 \over 18} Y^3 + {4 \over 9} Y^2
         + {\pi^2 \over 9} Y \biggr) {1 \over x}
-{\pi^2 \over 9}
%
%%%%%%%% Imaginary part %%%%%%%%%
%
{} \PlusBreak{}
i {\pi \over 3} {} \biggl\{
- 2 x \li2(-x)
+ x X^2
-\biggl( {29 \over 6} + Y \biggr) x X
+ y Y^2
{} \PlusBreak{\null + i \pi \BigglP \null }
\biggl( {13 \over 3} + {29 \over 6} x \biggr) Y
+{\pi^2 \over 6} x
-\biggl( {1 \over 2} Y^2 - {8 \over 3} Y \biggr) {1 \over x}
-{8 \over 3}
\biggl\}
%%%%% end : KK22Paper
\,, \label{KK22}\\[1pt plus 4pt]
%___________________________________________________________________________
%
L^{[2]}_2 & = &
%%%%% begin : LL22Paper
-{25 \over 81} x
%%%%% end : LL22Paper
\,, \label{LL22} %\\[1pt plus 4pt]
%___________________________________________________________________________
%
\end{eqnarray}

For $h=3$ in \eqn{h3} and color factor $\trc^{[1]}$ in \eqn{basis34}:

\begin{eqnarray}
A^{[1]}_3 & = &
%%%%% begin : AA31Paper
%%%%%%% x/y %%%%%%%%
\biggl[
-3 \li3(-y)
-\biggl( {\pi^2 \over 2} x + 3 Y + {\pi^2 \over 2} \biggr) \li2(-x)
- {3 \over 2} Y^2 X
{} \MinusBreak{\null \bigglP}
  {1 \over 2} Y^3 + \pi^2 Y
+{\pi^2 \over 2}
+ 3 \zeta_3
\biggr] {x \over y}
+ x Y^3 X - {1 \over 4} x Y^4 + {\pi^4 \over 8} x
{} \MinusBreak{}
3 x {} \biggl( \li4 \biggl( - {x \over y} \biggr) + \li4(-x) +
             \li4(-y) - Y {} (\li3(-x) + \li3(-y)) \biggr)
%%%%%%%%%%% 1/y %%%%%%%%%%
{} \PlusBreak{}
\biggl[
-\li4(-y)
-\biggl( {29 \over 6} - Y \biggr) \li3(-y)
+\biggl( {1 \over 2} Y^2 - {\pi^2 \over 3} - {29 \over 6} Y \biggr) \li2(-x)
+{1 \over 8} Y^4
{} \PlusBreak{\null \bigglP + \bigglP }
  \biggl( {1 \over 3} Y^3 - {29 \over 12} Y^2 \biggr) X
-{23 \over 18} Y^3
+\biggl( -{2 \over 3} \pi^2 + {49 \over 12} \biggr) Y^2
+ {413 \over 72} \zeta_3
{} \PlusBreak{\null \bigglP + \bigglP }
  \biggl( -{1513 \over 432} - 3 \zeta_3 + {47 \over 18} \pi^2 \biggr) Y
+{13 \over 288} \pi^4 - {\pi^2 \over 12}
+{23213 \over 5184}
\biggr] {1 \over y}
%%%%%%%%%% 1/xy %%%%%%%%%%%%%%%
+{3 \over 2 x y} Y^2
%
%%%%%%% Imaginary part %%%%%%%%
%
{} \PlusBreak{}
i \pi {} \Biggl\{
%%%%%%%%% x/y %%%%%%%%%%%
-\biggl[
  3 \li2(-x)
+{3 \over 2} Y^2
\biggr] {x \over y}
- 3 x {} \biggl( {\pi^2 \over 6} Y - \li3(-x) - \li3(-y)
                   -{1 \over 2} Y^2 X \biggr)
%%%%%%%%%%% 1/y %%%%%%%%%%%%%
{} \PlusBreak{\null + i \pi \BigglP }
\biggl[
  \li3(-y)
  -\biggl( {29 \over 6} - Y \biggr) \li2(-x)
  +{1 \over 2} Y^2 X + {1 \over 2} Y^3 - {23 \over 6} Y^2
{} \MinusBreak{\null + i \pi \BigglP  + \,\, \bigglP \null }
   \biggl( {\pi^2 \over 3} - {49 \over 6} \biggr) Y
  -{1513 \over 432} - 3 \zeta_3
  +{\pi^2 \over 18}
\biggr] {1 \over y}
%%%%%%%%%% 1 %%%%%%%%%%%%
-{1 \over 2} x Y^3
+{3 \over x y} Y
\Biggr\}
%%%%% end : AA31Paper
\,, \label{AA31}\\[1pt plus 4pt]
%__________________________________________________________________________
%
B^{[1]}_3 & = &
%%%%% begin : BB31Paper
%%%%%%%%%%% x/y %%%%%%%%%
\biggl[
-(6 + 2 x) \biggl( \li4 \biggl( - {x \over y} \biggr) - Y \li3(-y) \biggr)
+(3 x X + x + 6) \li3(-y)
{} \MinusBreak{\null \bigglP}
  (4 x + 6) (\li4(-x) + \li4(-y) - Y \li3(-x))
+(X + 1) x \li3(-x)
{} \PlusBreak{\null \bigglP}
  \biggl( - x {} (1 - Y) X - x Y^2 + (x + 6) Y
          +{\pi^2 \over 3} x + \pi^2 \biggr) \li2(-x)
+{1 \over 12} x X^4
{} \MinusBreak{\null \bigglP}
  \biggl( {1 \over 6} x Y - {1 \over 4} + {59 \over 72} x \biggr) X^3
+ \biggl( 2 + {1 \over 6} x \biggr) Y^3 X
+ \biggl( {17 \over 4} + {35 \over 24} x \biggr) Y^2 X
{} \PlusBreak{\null \bigglP}
  \biggl( {3 \over 2} x Y^2
         + \biggl( {1 \over 2} x - 1 \biggr) Y + {7 \over 12}
         + \biggl( {5 \over 24} \pi^2 + {73 \over 18} \biggr) x \biggr) X^2
{} \MinusBreak{\null \bigglP}
  \biggl(  {61 \over 12}
          + \biggl(   {73 \over 9}
                    + {11 \over 12} \pi^2 \biggr) x \biggr) Y X
+\biggl( {64 \over 9} - {71 \over 72} \pi^2 x
         + {\pi^2 \over 6} \biggr) X
{} \PlusBreak{\null \bigglP}
  \biggl( {9 \over 2}
         + \biggl( {73 \over 18} + {3 \over 8} \pi^2 \biggr) x \biggr) Y^2
-\biggl( {64 \over 9} + {9 \over 4} \pi^2 + {41 \over 36} \pi^2 x \biggr) Y
{} \PlusBreak{\null \bigglP}
  \biggl( {77 \over 360} \pi^4 + {73 \over 18} \pi^2 \biggr) x
+{31 \over 18} \pi^2 + {\pi^4 \over 4}
-2 \zeta_3
\biggr] {x \over y}
%%%%%%%%%%% 1/y %%%%%%%%%%%
{} \PlusBreak{}
\biggl[
  -4 \li4 \biggl( - {x \over y} \biggr)
  -2 \li4(-x)
  +\li4(-y)
  +\biggl( 2 Y + {8 \over 3} - X \biggr) \li3(-x)
{} \PlusBreak{\null + \bigglP}
   \biggl( Y + {37 \over 3} - 3 X \biggr) \li3(-y)
  +{1 \over 24} X^4
  -\biggl( {1 \over 3} Y + {65 \over 72} \biggr) X^3
  + {179 \over 24} Y^2 X
{} \MinusBreak{\null + \bigglP}
   \biggl( \biggl( Y + {8 \over 3} \biggr) X - {37 \over 3} Y
           +{1 \over 2} Y^2 - {5 \over 3} \pi^2 \biggr) \li2(-x)
-\biggl( {175 \over 36} - {3 \over 4} \pi^2 \biggr) Y X
{} \PlusBreak{\null + \bigglP}
   \biggl( {5 \over 24} \pi^2 - {1 \over 2} Y^2
          - {1 \over 3} Y + {349 \over 72} \biggr) X^2
-\biggl( \zeta_3 - {34 \over 27} + {13 \over 12} \pi^2 \biggr) X
-{1 \over 6} Y^3 X
{} \PlusBreak{\null + \bigglP}
   \biggl( {15 \over 8} \pi^2 - {19 \over 18} \biggr) Y^2
  -\biggl(  {811 \over 54} - {1 \over 2} \zeta_3
           + {263 \over 48} \pi^2 \biggr) Y
  -{851 \over 72} \zeta_3 - {53 \over 288} \pi^4
{} \MinusBreak{\null + \bigglP}
   {151 \over 48} \pi^2 + {30659 \over 1296}
\biggr] {1 \over y}
%%%%%%%%%% 1 %%%%%%%%%%%
-{1 \over 36} (47 - 29 x) Y^3
-{1 \over 4} y Y^4
-{7 \over 2 x y} Y^2
%
%%%%%%%%% Imaginary part %%%%%%%%
%
{} \PlusBreak{}
i \pi {} \Biggl\{
%%%%%%%%%% x/y %%%%%%%%%
\biggl[
  (5 x+6) (\li3(-x) + \li3(-y))
+(x X + 6 - x Y) \li2(-x)
{} \MinusBreak{\null + i \pi \BigglP \bigglP \null }
  \biggl( {1 \over 4} + {35 \over 24} x - {1 \over 2} x Y \biggr) X^2
+ (2 x + 3) Y^2 X + \biggl( {35 \over 12} x + {1 \over 2} \biggr) Y X
{} \PlusBreak{\null + i \pi \BigglP \bigglP \null }
  \biggl( {\pi^2 \over 6} x - {47 \over 12} \biggr) X
-\biggl( 1 + {5 \over 6} x \biggr) Y^3
+\biggl( {11 \over 4} - {35 \over 24} x \biggr) Y^2
{} \MinusBreak{\null + i \pi \BigglP \bigglP \null }
  \biggl( \pi^2 - {47 \over 12} + {\pi^2 \over 2} x \biggr) Y
- {35 \over 24} \pi^2 x -{\pi^2 \over 12}
\biggr] {x \over y}
%%%%%%%%%% 1/y %%%%%%%%%%%
{} \PlusBreak{\null + i \pi \BigglP }
\biggl[
  \li3(-x)
  -2 \li3(-y)
  -\biggl( X + 2 Y - {29 \over 3} \biggr) \li2(-x)
  -{41 \over 24} X^2
  -{7 \over 6} Y^3
{} \PlusBreak{\null + i \pi \BigglP  + \, \bigglP \null }
   \biggl( -{3 \over 2} Y^2 + {29 \over 6}
           + {23 \over 12} Y + {\pi^2 \over 2} \biggr) X
  + {125 \over 24} Y^2
  +\biggl( {5 \over 6} \pi^2 - {251 \over 36} \biggr) Y
{} \MinusBreak{\null + i \pi \BigglP  + \, \bigglP \null }
   {169 \over 144} \pi^2
  -{1 \over 2} \zeta_3 - {743 \over 54}
\biggr] {1 \over y}
%%%%%%%%% 1 %%%%%%%%%%
-{7 \over x y} Y
\Biggr\}
%%%%% end : BB31Paper
\,, \label{BB31}\\[1pt plus 4pt]
%__________________________________________________________________________
%
C^{[1]}_3 & = &
%%%%% begin : CC31Paper
%%%%%%%%% x/y %%%%%%%%%%%
\biggl[
-6 x {} ( \li4(-x) + \li4(-y) )
+(5 x X - 4 - 2 x Y - 3 x) \li3(-x)
{} \MinusBreak{\null \bigglP}
  ( 3 x + 4 - 4 x Y + x X) \li3(-y)
+{1 \over 12} x X^4
+\biggl( {7 \over 4} - {7 \over 24} x - {5 \over 6} x Y \biggr) X^3
{} \MinusBreak{\null \bigglP}
  ( 2 x X^2 - (4 + x Y + 3 x) X
   - Y^2 x + (4 + 3 x) Y) \li2(-x)
{} \PlusBreak{\null \bigglP}
  \biggl( {1 \over 2} x Y^2
         + \biggl( {5 \over 4} x - {5 \over 2} \biggr) Y
         + {9 \over 4}
         + \biggl( {5 \over 24} \pi^2 + {3 \over 2} \biggr) x \biggr) X^2
-{1 \over 12} x Y^4
{} \PlusBreak{\null \bigglP}
   \biggl( {5 \over 6} x Y^3
         - \biggl( {15 \over 4}
                  + \biggl( 3 + {\pi^2 \over 4} \biggr) x \biggr) Y
         +{7 \over 3} \pi^2 + {5 \over 24} \pi^2 x + 6 \biggr) X
+ {4 \over 3} \pi^2
{} \PlusBreak{\null \bigglP}
  {\pi^2 \over 24} x Y^2
+\biggl( {7 \over 12} \pi^2 - 6 + {2 \over 3} \pi^2 x \biggr) Y
+\biggl( {3 \over 2} \pi^2 + {19 \over 40} \pi^4 \biggr) x
\biggr] {x \over y}
%%%%%%%%%% 1/y %%%%%%%%%%
{} \PlusBreak{}
\biggl[
  8 \li4 \biggl( - {x \over y} \biggr)
+6 \li4(-x)
+2 \li4(-y)
-(1 + 6 Y - 3 X) \li3(-x)
-{1 \over 24} X^4
{} \PlusBreak{\null + \bigglP}
{5 \over 12} Y^4
-(8 Y + 1 - 9 X) \li3(-y)
+\biggl( {3 \over 4} Y + 5 Y^2 - {9 \over 8} \pi^2 + {65 \over 8} \biggr) 
X^2
{} \MinusBreak{\null + \bigglP}
  \biggl( 2 X^2 - (7 Y + 1) X + Y
          +{8 \over 3} \pi^2 + 3 Y^2 \biggr) \li2(-x)
-\biggl( {5 \over 24} + {2 \over 3} Y \biggr) X^3
{} \MinusBreak{\null + \bigglP}
  \biggl( {7 \over 2} Y^3
          +\biggl( {31 \over 12} \pi^2 + {35 \over 4} \biggr) Y
          +6 + {\pi^2 \over 8} + 8 \zeta_3 \biggr) X
-\biggl( {13 \over 24} \pi^2 - {3 \over 2} \biggr) Y^2
{} \MinusBreak{\null + \bigglP}
  \biggl( {93 \over 16} + {29 \over 6} \pi^2 - 11 \zeta_3 \biggr) Y
+{511 \over 64} - {15 \over 4} \zeta_3
+{169 \over 360} \pi^4 + {17 \over 16} \pi^2
\biggr] {1 \over y}
%%%%%%%%% 1 %%%%%%%%%%
{} \MinusBreak{}
{1 \over 8} (11 - 13 x) Y^2 X
+{1 \over 3} x Y^3
-{3 \over 2} x Y^2
+{3 \over 2 x y} Y^2
%
%%%%%%%% Imaginary part %%%%%%%%%
%
{} \PlusBreak{}
i \pi {} \Biggl\{
%%%%%%%%% x/y %%%%%%%%%
\biggl[
  3 x {} (\li3(-x) + \li3(-y) + (Y - X) \li2(-x))
-\biggl( {9 \over 8} x + {3 \over 2} x Y - {3 \over 4} \biggr) X^2
{} \PlusBreak{\null + i \pi \BigglP \bigglP \null }
  \biggl( 3 x Y^2 - \biggl( {3 \over 2} - {9 \over 4} x \biggr) Y
         + {3 \over 4} - {\pi^2 \over 2} x \biggr) X
-\biggl( {3 \over 4} + {3 \over 2} \pi^2 x \biggr) Y
{} \MinusBreak{\null + i \pi \BigglP \bigglP \null }
  {9 \over 8} \pi^2 x
+{\pi^2 \over 4}
\biggr] {x \over y}
%%%%%%%%%%% 1/y %%%%%%%%%%%
{} \PlusBreak{\null + i \pi \BigglP }
\biggl[
  -3 \li3(-x)
  +\li3(-y)
  +(Y + 3 X) \li2(-x)
  -{3 \over 8} X^2
{} \PlusBreak{\null + i \pi \BigglP  + \, \bigglP \null }
   \biggl( 2 Y^2 + {15 \over 2} + {21 \over 4} Y - {3 \over 2} \pi^2 \biggr) 
X
  -\biggl( {23 \over 4} + {2 \over 3} \pi^2 \biggr) Y
  -{9 \over 8} \pi^2
{} \MinusBreak{\null + i \pi \BigglP  + \, \bigglP \null }
  {189 \over 16} + 3 \zeta_3
\biggr] {1 \over y}
%%%%%%%%%%% 1 %%%%%%%%%%%%%%%%
+{1 \over 2}(x-1) Y^3
-{1 \over 8} (15 - 9 x) Y^2
+{3 \over x y} Y
\Biggr\}
%%%%% end : CC31Paper
\,, \label{CC31}\\[1pt plus 4pt]
%__________________________________________________________________________
%
D^{[1]}_3 & = &
%%%%% begin : DD31Paper
\biggl[
  {1 \over 3} \li3(-y)
+{1 \over 3} Y \li2(-x)
+{1 \over 6} Y^2 X
+{5 \over 18} Y^3
-{37 \over 36} Y^2
-\biggl( {11 \over 18} \pi^2 - {145 \over 54} \biggr) Y
{} \PlusBreak{\null \bigglP}
  {41 \over 72} \pi^2
-{49 \over 36} \zeta_3
-{455 \over 108}
\biggr] {1 \over y}
%
%%%%%%%% Imaginary part %%%%%%%%%
%
% {} \PlusBreak{}
- i {\pi \over 3 y} {} \biggl\{
{\pi^2 \over 6}
-{5 \over 2} Y^2
+{37 \over 6} Y
-\li2(-x)
-{145 \over 18}
\biggr\}
%%%%% end : DD31Paper
\,, \label{DD31}\\[1pt plus 4pt]
%_________________________________________________________________________
%
E^{[1]}_3 & = &
%%%%% begin : EE31Paper
%%%%%%%%% x/y %%%%%%%%%%
\biggl[
{1 \over 9} x X^3
-\biggl( {8 \over 9} x + {1 \over 3} \biggr) X^2
+\biggl( -{1 \over 3} x Y^2
          + \biggl( {4 \over 3} + {16 \over 9} x \biggr) Y
          + {\pi^2 \over 9} x - {16 \over 9} \biggr) X
{} \PlusBreak{\null \bigglP}
  {1 \over 9} (2 \pi^2 x + 16) Y
-{8 \over 9} \pi^2 x - {5 \over 9} \pi^2
\biggr] {x \over y}
%%%%%%%%%%% 1/y %%%%%%%%
{} \PlusBreak{}
\biggl[
  -{2 \over 3} \li3(-x)
  -{4 \over 3} \li3(-y)
  -\biggl( {4 \over 3} Y - {2 \over 3} X \biggr) \li2(-x)
  +{1 \over 9} X^3
  +\biggl( {1 \over 3} Y - {19 \over 18} \biggr) X^2
{} \PlusBreak{\null + \bigglP}
   \biggl( {14 \over 27} - {4 \over 3} Y^2
          + {7 \over 9} Y + {\pi^2 \over 4} \biggr) X
  +\biggl( {41 \over 24} \pi^2 + {107 \over 27} \biggr) Y
  -{685 \over 162}
{} \MinusBreak{\null + \bigglP}
   {11 \over 36} \zeta_3
  +{5 \over 8} \pi^2
\biggr] {1 \over y}
%%%%%%%%%% 1 %%%%%%%%%%%
+{2 \over 9} (1 - x) Y^3
+{1 \over 9} (1 + 8 x) Y^2
%
%%%%%%%% I,aginary part %%%%%%%%
%
{} \PlusBreak{}
i {\pi \over 3} {} \Biggl\{
%%%%%%%%% x/y %%%%%%%%%%%
\biggl[
  x X^2
+2 (1 - x Y) X
-2 Y + \pi^2 x
+ x Y^2
\biggr] {x \over y}
%%%%%%%%% 1/y %%%%%%%%%%%
{} \PlusBreak{\null + i \pi \BigglP }
\biggl[
  -2 \li2(-x)
  + X^2
  - (2 Y + 4) X
  +{121 \over 9} + {29 \over 24} \pi^2
  +{5 \over 3} Y - 4 Y^2
\biggr] {1 \over y}
\Biggr\}
%%%%% end : EE31Paper
\,, \label{EE31}\\[1pt plus 4pt]
%_____________________________________________________________________________
%
F^{[1]}_3 & = &
%%%%% begin : FF31Paper
\biggl[
  {1 \over 9} Y^2
-{10 \over 27} Y
-{\pi^2 \over 9}
+{25 \over 81}
\biggr] {1 \over y}
%%%%%%% Imaginary part %%%%%
+i \pi {} \biggl[
{2 \over 9} Y - {10 \over 27}
\biggr] {1 \over y}
%%%%% end : FF31Paper
\,, \label{FF31} %\\[1pt plus 4pt]
%_____________________________________________________________________________
%
\end{eqnarray}

For $h=3$ in \eqn{h3} and color factor $\trc^{[2]}$ in \eqn{basis34}:

\begin{eqnarray}
G^{[2]}_3 & = &
%%%%% begin : GG32Paper
%%%%%%%%%% x/y %%%%%%%%%
\biggl[
  \biggl( 3 - {1 \over 2} x \biggr) \li4 \biggl( - {x \over y} \biggr)
% +3 (1 + x) (\li4(-x) + \li4(-y))
- \biggl( 2 x X + (3 + x) Y
%         + {3 \over 2} (1 + x)
  \biggr) \li3(-y)
-{1 \over 48} x X^4
{} \MinusBreak{\null \bigglP}
  \biggl(  2 x X + (3 + x) Y
          - {3 \over 2} (1 - x)
  \biggr) \li3(-x)
+\biggl( {1 \over 12} (1 + x Y) + {11 \over 36} x \biggr) X^3
%{} \MinusBreak{\null \bigglP}
{} \PlusBreak{\null \bigglP}
  \biggl( {1 \over 4} x X^2
         -\biggl( {1 \over 2} x Y + {3 \over 2} (1 - x) \biggr) X
         +{1 \over 4} x Y^2
% - {3 \over 2} (1 + x) Y
%{} \MinusBreak{\null \bigglP + \,\,\, \bigglP }
        - {\pi^2 \over 2} + {\pi^2 \over 12} x \biggr) \li2(-x)
{} \MinusBreak{\null \bigglP}
  \biggl(   x Y^2 + {1 \over 2} Y - {7 \over 12}
          + \biggl( {37 \over 36} + {5 \over 24} \pi^2 \biggr) x \biggr) X^2
+\biggl( {1 \over 4} + {1 \over 48} x \biggr) Y^4
-{1 \over 2} \zeta_3
{} \MinusBreak{\null \bigglP}
  \biggl( 1 - {1 \over 6} x \biggr) Y^3 X
-\biggl( {11 \over 12} x + {3 \over 4} \biggr) Y^2 X
+\biggl( -{5 \over 6}
          + \biggl( {3 \over 4} \pi^2
                   + {37 \over 18} \biggr) x \biggr) Y X
{} \PlusBreak{\null \bigglP}
  \biggl( -{32 \over 9} - {\pi^2 \over 4}
          +{5 \over 9} \pi^2 x \biggr) X
+ \biggl( {1 \over 9} x - {1 \over 3} \biggr) Y^3
+\biggl( {1 \over 4} - \biggl( {37 \over 36}
         + {\pi^2 \over 4} \biggr) x \biggr) Y^2
{} \PlusBreak{\null \bigglP}
  \biggl( {11 \over 18} \pi^2 x + {5 \over 12} \pi^2 + {32 \over 9} \biggr) 
Y
-\biggl( {157 \over 720} \pi^4 + {37 \over 36} \pi^2 \biggr) x
+ {7 \over 18} \pi^2
-{\pi^4 \over 8}
\biggr] {x \over y}
%%%%%%%%%% 1/y %%%%%%%%%%
{} \PlusBreak{}
\biggl[
   {5 \over 2} \li4 \biggl( - {x \over y} \biggr)
  +4 \li4(-x)
  -2 \li4(-y)
  -\biggl( {1 \over 3} + 2 Y + X \biggr) \li3(-x)
  - {1 \over 48} X^4
{} \MinusBreak{\null + \bigglP}
   \biggl( {29 \over 6} - 2 X + Y \biggr) \li3(-y)
  +\biggl(  {5 \over 12} Y + {5 \over 8} Y^2 + {\pi^2 \over 4}
           -{125 \over 72} \biggr) X^2
{} \PlusBreak{\null + \bigglP}
   \biggl(  {1 \over 4} X^2
          + \biggl( {1 \over 3} + {3 \over 2} Y \biggr) X
          -{29 \over 6} Y - {3 \over 4} Y^2
          -{5 \over 12} \pi^2 \biggr) \li2(-x)
  + {7 \over 18} X^3
{} \MinusBreak{\null + \bigglP}
   \biggl(  {5 \over 12} Y^3
           +{17 \over 4} Y^2
           -\biggl( {\pi^2 \over 12} - {5 \over 72} \biggr) Y
           -{3 \over 16} \pi^2
           +{17 \over 27}
           -{1 \over 2} \zeta_3 \biggr) X
  +{1 \over 16} Y^4
{} \PlusBreak{\null + \bigglP}
   {1 \over 3} Y^3
  -\biggl( {29 \over 36} + {13 \over 24} \pi^2 \biggr) Y^2
  +\biggl( {43 \over 18} \pi^2 + 3 \zeta_3 + {3049 \over 432} \biggr) Y
{} \MinusBreak{\null + \bigglP}
   {43 \over 1440} \pi^4 + {7 \over 72} \zeta_3
  +{25 \over 9} \pi^2 - {23213 \over 5184}
\biggr] {1 \over y}
%%%%%%%%% 1 %%%%%%%%%%
{} \PlusBreak{}
{1 \over 2 x y} Y^2
- 3 x {} (\li4(-x) + \li4(-y))
+ {3 \over 2} x {} ( \li3(-y) + Y \li2(-x) )
%+ {3 \over 2} x Y \li2(-x)
%
%%%%%%%%%%% Imaginary part %%%%%%%%%%%%
%
{} \PlusBreak{}
i \pi \Biggl\{
%%%%%%%%%% x/y %%%%%%%%%%%
\biggl[
% -3 (1 + x) (\li3(-x) + \li3(-y))
-3 \li2(-x)
+{1 \over 6} x X^3
+\biggl( {1 \over 2} + {1 \over 3} x \biggr) Y^3
%{} \MinusBreak{\null + i \pi \BigglP \bigglP \null }
-\biggl( {1 \over 2} x Y - {1 \over 2} - {1 \over 6} x \biggr) X^2
{} \MinusBreak{\null + i \pi \BigglP \bigglP \null }
  \biggl( Y^2 + \biggl( {1 \over 3} x + 1 \biggr) Y
         - {\pi^2 \over 6} x - {1 \over 3} \biggr) X
+\biggl( {\pi^2 \over 2} + {\pi^2 \over 3} x - {1 \over 3} \biggr) Y
{} \MinusBreak{\null + i \pi \BigglP \bigglP \null }
  \biggl( 1 - {1 \over 6} x \biggr) Y^2
% +{\pi^2 \over 6} (1 + x)
\biggr] {x \over y}
+3 x {} (\li3(-x) + \li3(-y))
-{\pi^2 \over 6} x
%%%%%%%%%% 1/y %%%%%%%%%%%
{} \PlusBreak{\null + i \pi \BigglP }
\biggl[
  -3 \li3(-x)
  +\li3(-y)
  +\biggl( 2 X - {9 \over 2} \biggr) \li2(-x)
  -{1 \over 12} X^3
  +{1 \over 4} Y^3
{} \PlusBreak{\null + i \pi \BigglP  + \, \bigglP \null }
   \biggl( {17 \over 12} - {1 \over 2} Y \biggr) X^2
  +\biggl( {1 \over 2} Y^2 - {85 \over 24}
          -{17 \over 6} Y + {\pi^2 \over 4} \biggr) X
  -{5 \over 6} Y^2
  +{41 \over 48} \pi^2
{} \MinusBreak{\null + i \pi \BigglP  + \, \bigglP \null }
   \biggl( {\pi^2 \over 4} + {121 \over 72} \biggr) Y
  +{7 \over 2} \zeta_3 + {2777 \over 432}
\biggr] {1 \over y}
%%%%%%%%% 1 %%%%%%%%%%%%%
+\biggl( x - {1 \over 2} \biggr) Y^2 X
+{1 \over x y} Y
\Biggr\}
%%%%% end : GG32Paper
\,, \label{GG32}\\[1pt plus 4pt]
%____________________________________________________________________________
%
H^{[2]}_3 & = &
%%%%% begin : HH32Paper
%%%%%%%%%%% x/y %%%%%%%%%%%%
\biggl[
  2 x {} (\li4(-x) + \li4(-y))
+\biggl( 1 + x Y - 2 x X + {5 \over 2} x \biggr) \li3(-x)
-{1 \over 8} x X^4
{} \PlusBreak{\null \bigglP}
  \biggl( {5 \over 2} x - x Y + 1 \biggr) \li3(-y)
-\biggl( {17 \over 12} - {55 \over 72} x - {2 \over 3} x Y \biggr) X^3
  -\biggl( {7 \over 4} - {17 \over 24} x \biggr) Y^2 X
{} \PlusBreak{\null \bigglP}
  {1 \over 24} x Y^4
+\biggl( x X^2
         -\biggl( {5 \over 2} x + x Y + 1 \biggr) X
         +\biggl( 1 + {5 \over 2} x \biggr) Y \biggr) \li2(-x)
{} \PlusBreak{\null \bigglP}
  \biggl(  \biggl( 3 - {3 \over 2} x \biggr) Y
          -{13 \over 6}
          -\biggl(  {\pi^2 \over 8} + {127 \over 36}
                   +{3 \over 4} Y^2 \biggr) x \biggr) X^2
-\biggl( {31 \over 180} \pi^4 + {127 \over 36} \pi^2 \biggr) x
{} \PlusBreak{\null \bigglP}
  \biggl( \biggl( {89 \over 12}
                 +\biggl( {\pi^2 \over 4} + {127 \over 18} \biggr) x \biggr) 
Y
          -{59 \over 9} + {25 \over 72} \pi^2 x
          - {4 \over 3} \pi^2 \biggr) X
{} \MinusBreak{\null \bigglP}
  \biggl( {21 \over 4}
         + \biggl( {\pi^2 \over 8} + {127 \over 36} \biggr) x \biggr) Y^2
+\biggl( {5 \over 12} \pi^2 + {59 \over 9} + {\pi^2 \over 36} x \biggr) Y
-{28 \over 9} \pi^2
\biggr] {x \over y}
%%%%%%%%%%% 1/y %%%%%%%%%%%%%
{} \PlusBreak{}
\biggl[
  -4 \li4 \biggl( - {x \over y} \biggr)
  -6 \li4(-x)
  +\li4(-y)
  +\biggl( 4 Y - {11 \over 6} \biggr) \li3(-x)
  -{5 \over 24} Y^4
{} \MinusBreak{\null + \bigglP}
   \biggl( 5 X - 4 Y + {13 \over 6} \biggr) \li3(-y)
  -\biggl( {7 \over 12} Y + {283 \over 36}
          + {27 \over 8} Y^2 - {\pi^2 \over 4} \biggr) X^2
{} \PlusBreak{\null + \bigglP}
   \biggl( X^2 - \biggl( 5 Y - {11 \over 6} \biggr) X
     + \pi^2 + {5 \over 2} Y^2
     - {13 \over 6} Y \biggr) \li2(-x)
  +\biggl( {3 \over 4} Y + {49 \over 72} \biggr) X^3
{} \PlusBreak{\null + \bigglP}
  \biggl( {29 \over 12} Y^3 - {43 \over 24} Y^2
         +\biggl( {11 \over 12} \pi^2 + {679 \over 72} \biggr) Y
         +{11 \over 2} \zeta_3 + {17 \over 16} \pi^2
         +{64 \over 27} \biggr) X
{} \MinusBreak{\null + \bigglP}
   \biggl( {\pi^2 \over 4} + {29 \over 9} \biggr) Y^2
  +\biggl( {57 \over 16} \pi^2 + {781 \over 54}
          - {11 \over 2} \zeta_3 \biggr) Y
  -{30659 \over 1296} + {467 \over 72} \zeta_3
{} \PlusBreak{\null + \bigglP}
   {101 \over 144} \pi^2 - {263 \over 1440} \pi^4
\biggr] {1 \over y}
%%%%%%%%%% 1 %%%%%%%%%%
+{1 \over 36} (13 - 31 x) Y^3
+{1 \over 2 x y} Y^2
%
%%%%%%%%%% Imaginary part %%%%%%%%%%
%
{} \PlusBreak{}
i \pi \Biggl\{
%%%%%%%%%% x/y %%%%%%%%%%
\biggl[
-x {} (\li3(-x) + \li3(-y) - (X - Y) \li2(-x))
-{1 \over 6} x X^3
+{1 \over 3} x Y^3
{} \PlusBreak{\null + i \pi \BigglP \bigglP \null }
  \biggl( {49 \over 24} x + x Y - {3 \over 4} \biggr) X^2
+\biggl( {37 \over 12} - {3 \over 2} x Y^2
         + \biggl( {3 \over 2} - {49 \over 12} x \biggr) Y \biggr) X
{} \PlusBreak{\null + i \pi \BigglP \bigglP \null }
  \biggl( {49 \over 24} x - {3 \over 4} \biggr) Y^2
+\biggl( {2 \over 3} \pi^2 x - {37 \over 12} \biggr) Y
+{49 \over 24} \pi^2 x - {\pi^2 \over 4}
\biggr] {x \over y}
%%%%%%%%%% 1/y %%%%%%%%%
{} \PlusBreak{\null + i \pi \BigglP }
\biggl[
  4 \li3(-x)
  -\li3(-y)
  -\biggl( 3 X + {1 \over 3} \biggr) \li2(-x)
  +{1 \over 12} X^3
  +{1 \over 12} Y^3
{} \PlusBreak{\null + i \pi \BigglP  + \, \bigglP \null }
   \biggl( {13 \over 24} + {1 \over 2} Y \biggr) X^2
  -\biggl( {3 \over 2} Y^2 + {151 \over 24}
          +{31 \over 12} Y - {\pi^2 \over 4} \biggr) X
  -{43 \over 24} Y^2
{} \PlusBreak{\null + i \pi \BigglP  + \, \bigglP \null }
   \biggl( {\pi^2 \over 4} + {215 \over 72} \biggr) Y
  +{101 \over 6} + {61 \over 72} \pi^2
\biggr] {1 \over y}
%%%%%%%%%% 1 %%%%%%%%%%%
+{1 \over x y} Y
\Biggr\}
%%%%% end : HH32Paper
\,, \label{HH32}\\[1pt plus 4pt]
%_____________________________________________________________________________
%
I^{[2]}_3 &=&
%%%%% begin : II32Paper
%%%%%%%%%%%%%%% x/y %%%%%%%%%%%%%%%
\biggl[
  - {1 \over 2} x \li4 \biggl( -{x \over y} \biggr)
  + 2 x \li4(-x)
  + 2 x \li4(-y)
  + \biggl( {3 \over 2} + x Y + x - 2 x X \biggr) \li3(-x)
{} \PlusBreak{\null \bigglP}
    \biggl( x - x Y + {3 \over 2} \biggr) \li3(-y)
   -\biggl( {2 \over 3} - {1 \over 4} x Y - {1 \over 24} x \biggr) X^3
   +\biggl( 1 - {1 \over 4} x (1 + Y) \biggr) Y X^2
{} \PlusBreak{\null \bigglP}
    \biggl( {3 \over 4} x X^2
           - \biggl( {3 \over 2} + {1 \over 2} x Y + x \biggr) X
           - {1 \over 4} x Y^2 + \biggl( {3 \over 2} + x \biggr) Y
           + {\pi^2 \over 12} x
    \biggr) \li2(-x)
{} \MinusBreak{\null \bigglP}
    \biggl( {5 \over 4} + \biggl( {\pi^2 \over 12} + 1 \biggr) x \biggr) X^2
   + {1 \over 48} x Y^4
   - {1 \over 6} x Y^3 X
   + \biggl( {9 \over 4} + \biggl( {\pi^2 \over 6} + 2 \biggr) x \biggr) Y X
{} \MinusBreak{\null \bigglP}
   \biggl( {\pi^2 \over 8} x + {11 \over 12} \pi^2 + 3 \biggr) X
  - \biggl( 1 + \biggl( {\pi^2 \over 24} + 1 \biggr) x \biggr) Y^2
-\biggl( {\pi^2 \over 6} x - 3 + {\pi^2 \over 6} \biggr) Y
{} \MinusBreak{\null \bigglP}
  \biggl( \pi^2 + {25 \over 144} \pi^4 \biggr) x
-{5 \over 6} \pi^2 - {1 \over 2} \zeta_3
\biggr] {x \over y}
%%%%%%%%%%%%%%%%%% 1/y %%%%%%%%%%%%%%%%%%%
{} \PlusBreak{}
   \biggl[
  -{5 \over 2} \li4 \biggl( -{x \over y} \biggr)
  -2 \li4(-x)
  -\li4(-y)
  -{7 \over 48} Y^4
   + \biggl( 2 Y + {1 \over 2} - X \biggr) \li3(-x)
{} \PlusBreak{\null + \bigglP}
     \biggl( {1 \over 2} - 3 X + 3 Y \biggr) \li3(-y)
  -\biggl( {1 \over 4} Y + {7 \over 4} Y^2
          + {27 \over 8}-{5 \over 12} \pi^2
   \biggr) X^2
   + {4 \over 3} Y^3 X
{} \PlusBreak{\null + \bigglP}
     \biggl( {3 \over 4} X^2
      - {1 \over 2} (1 + 5 Y) X
      + {5 \over 4} Y^2 + {3 \over 4} \pi^2
      + {1 \over 2} Y \biggr) \li2(-x)
   + \biggl( {1 \over 24} + {1 \over 4} Y \biggr) X^3
{} \PlusBreak{\null + \bigglP}
     \biggl( \biggl( {17 \over 4} + {5 \over 6} \pi^2 \biggr) Y
            + 3 \zeta_3 + 3
            - {\pi^2 \over 24}
     \biggr) X
   + \biggl( {\pi^2 \over 8} - {1 \over 2} \biggr) Y^2
{} \PlusBreak{\null + \bigglP}
     \biggl( - 6 \zeta_3 + {7 \over 4} \pi^2
             + {45 \over 16} \biggr) Y
  -{17 \over 144} \pi^4-{511 \over 64}
  -{21 \over 16} \pi^2 + {13 \over 4} \zeta_3
\biggr] {1 \over y}
%%%%%%%%%%%%% others %%%%%%%%%%%%%%%
{} \MinusBreak{}
{1 \over 48} (1-x) X^4
+ {3 \over 8} (1 - x) Y^2 X
-{1 \over 6} x Y^3
-{1 \over 2 x y} Y^2
%
%%%%%%%%%%%%% Imaginary part %%%%%%%%%%%%%
%
{} \PlusBreak{}
  i \pi {} \Biggl\{
%%%%%%%%%%% x/y %%%%%%%%%%%%%
\biggl[
  -x \li3(-x)
  -x \li3(-y)
  + x {} (X-Y) \li2(-x)
  + \biggl( {3 \over 8} x + {1 \over 2} x Y - {1 \over 4} \biggr) X^2
{} \MinusBreak{\null + i \pi \BigglP \bigglP \null }
    \biggl(   x Y^2
            - \biggl( {1 \over 2} - {3 \over 4} x \biggr) Y
            + {1 \over 4} - {\pi^2 \over 6} x
    \biggr) X
  + \biggl( {1 \over 4} + {\pi^2 \over 2} x \biggr) Y
-{\pi^2 \over 12}
  + {3 \over 8} \pi^2 x
\biggr] {x \over y}
%%%%%%%%%%%% 1/y %%%%%%%%%%%%%%%
{} \PlusBreak{\null + i \pi \BigglP }
\biggl[
   \li3(-x)
  -X \li2(-x)
   + {1 \over 8} X^2
  -\biggl( {5 \over 2} + {7 \over 4} Y
     -{\pi^2 \over 2}
      + {1 \over 2} Y^2 \biggr) X
{} \PlusBreak{\null + i \pi \BigglP  + \, \bigglP \null }
     \biggl( {13 \over 4} + {\pi^2 \over 6} \biggr) Y
   + {93 \over 16} + {13 \over 24} \pi^2-3 \zeta_3
\biggr] {1 \over y}
%%%%%%%%%%%%%% others %%%%%%%%%%%%
{} \PlusBreak{\null + i \pi \BigglP }
   {1 \over 6} (1 - x) Y^3
+ {1 \over 8} (5 - 3 x) Y^2
-{1 \over y x} Y
\Biggr\}
%%%%% end : II32Paper
\,, \label{II32} \\[1pt plus 4pt]
%_____________________________________________________________________________
%
J^{[2]}_3 & = &
%%%%% begin : JJ32Paper
%%%%%%%%%% x/y %%%%%%%%%
\biggl[
-{1 \over 18} x X^3
+\biggl( {4 \over 9} x + {1 \over 6} \biggr) X^2
+\biggl( {1 \over 6} x Y^2
         -\biggl( {8 \over 9} x + {2 \over 3} \biggr) Y + {8 \over 9}
         -{\pi^2 \over 18} x \biggr) X
{} \MinusBreak{\null \bigglP}
  {1 \over 9} x Y^3
+\biggl( {1 \over 2} + {4 \over 9} x \biggr) Y^2
-{1 \over 9} (\pi^2 x + 8) Y
+{4 \over 9} \pi^2 x + {5 \over 18} \pi^2
\biggr] {x \over y}
%%%%%%%%%% 1/y %%%%%%%%%%%
{} \PlusBreak{}
\biggl[
   {1 \over 3} (\li3(-x) + \li3(-y) + (Y -  X) \li2(-x))
  -{1 \over 18} X^3
  -\biggl( {1 \over 6} Y - {19 \over 36} \biggr) X^2
{} \PlusBreak{\null + \bigglP}
   \biggl(  {1 \over 2} Y^2 - {\pi^2 \over 8}
          - {7 \over 27} - {7 \over 18} Y \biggr) X
  -{1 \over 6} Y^3 + {13 \over 12} Y^2
  -\biggl( {2 \over 9} \pi^2 + {193 \over 54} \biggr) Y
{} \MinusBreak{\null + \bigglP}
   {67 \over 72} \pi^2 + {37 \over 36} \zeta_3
  +{455 \over 108}
\biggr] {1 \over y}
%
%%%%%%%%%%% Imaginary part %%%%%%%%%%%
%
{} \PlusBreak{}
i \pi \Biggl\{
%%%%%%%%%% x/y %%%%%%%%%
\biggl[
-{1 \over 6} x X^2
-{1 \over 3} (1 - x Y) X
-{\pi^2 \over 6} x - {1 \over 6} x Y^2
+{1 \over 3} Y
\biggr] {x \over y}
%%%%%%%%%% 1/y %%%%%%%%%%%%
{} \PlusBreak{\null + i \pi \BigglP }
\biggl[
  -{1 \over 6} X^2
  +{1 \over 3} (2 + Y) X
  -{\pi^2 \over 8} - {23 \over 6}
  -{1 \over 6} Y^2 + {16 \over 9} Y
\biggr] {1 \over y}
\Biggr\}
%%%%% end : JJ32Paper
\,, \label{JJ32}\\[1pt plus 4pt]
%_____________________________________________________________________________
%
K^{[2]}_3 & = &
%%%%% begin : KK32Paper
%%%%%%%%%% x/y %%%%%%%%%%
\biggl[
-{1 \over 18} x X^3
+\biggl( {4 \over 9} x + {1 \over 6} \biggr) X^2
+\biggl( {1 \over 6} x Y^2
         -\biggl( {8 \over 9} x + {2 \over 3} \biggr) Y
         +{8 \over 9} - {\pi^2 \over 18} x \biggr) X
{} \MinusBreak{\null \bigglP}
  {1 \over 9} (\pi^2 x + 8) Y
+{4 \over 9} \pi^2 x
+{5 \over 18} \pi^2
\biggr] {x \over y}
%%%%%%%%%% 1/y %%%%%%%%%%
{} \PlusBreak{}
\biggl[
   {1 \over 3} (\li3(-x) + 2 \li3(-y) + (2 Y - X) \li2(-x))
  -{1 \over 18} X^3
  -\biggl( {1 \over 6} Y - {19 \over 36} \biggr) X^2
{} \MinusBreak{\null + \bigglP}
   \biggl( {7 \over 27} - {2 \over 3} Y^2
          +{7 \over 18} Y + {\pi^2 \over 8} \biggr) X
  - \biggl( {7 \over 8} \pi^2 + {83 \over 27} \biggr) Y
  + {685 \over 162}
{} \PlusBreak{\null + \bigglP}
   {23 \over 36} \zeta_3
  -{19 \over 72} \pi^2
\biggr] {1 \over y}
%%%%%%%%% 1 %%%%%%%%%%
+{1 \over 9} (x - 1) Y^3
-\biggl( {1 \over 18} + {4 \over 9} x \biggr) Y^2
%
%%%%%%%% Imaginary part %%%%%%%%
%
{} \PlusBreak{}
i \pi \Biggl\{
%%%%%%% x/y %%%%%%%%%
\biggl[
-{1 \over 6} x X^2
-{1 \over 3}(1 - x Y) X
-{\pi^2 \over 6} x
-{1 \over 6} x Y^2 + {1 \over 3} Y
\biggr] {x \over y}
%%%%%%%%% 1/y %%%%%%%%%
{} \PlusBreak{\null + i \pi \BigglP }
\biggl[
   \li2(-x)
  -{1 \over 2} X^2
  +(2 + Y) X
  -{5 \over 6} Y - {2 \over 3} \pi^2
  -10 + 2 Y^2
\biggr] {1 \over 3 y}
\Biggr\}
%%%%% end : KK32Paper
\,, \label{KK32}\\[1pt plus 4pt]
%_________________________________________________________________________
%
L^{[2]}_3 & = &
%%%%% begin : LL32Paper
\biggl[
{\pi^2 \over 9}
-{1 \over 9} Y^2
+{10 \over 27} Y
-{25 \over 81}
\biggr] {1 \over y}
%%%%%%%% Imaginary part %%%%%%%%
-i \pi {} \biggl\{
{2 \over 9} Y - {10 \over 27}
\biggr\} {1 \over y}
%%%%% end : LL32Paper
\,, \label{LL32} %\\[1pt plus 4pt]
%__________________________________________________________________________
%
\end{eqnarray}

For $h=4$ in \eqn{h4} and color factor $\trc^{[1]}$ in \eqn{basis34}:
\begin{eqnarray}
A^{[1]}_4 & = &
%%%%% begin : AA41Paper
%%%%%%%% x/y %%%%%%%%%%%%
\biggl[
  {1 \over 2} \li4(-y)
-{4 \over 3} \li3(-y)
-\biggl( {\pi^2 \over 3} + {4 \over 3} Y
         -{1 \over 4} Y^2 \biggr) \li2(-x)
-\biggl( {2 \over 3} Y^2 - {1 \over 12} Y^3 \biggr) X
{} \PlusBreak{\null \bigglP}
  {1 \over 12} Y^4 - {5 \over 8} Y^3
+\biggl( {101 \over 36} - {3 \over 8} \pi^2 \biggr) Y^2
+\biggl( -2 \zeta_3 + {25 \over 18} \pi^2 - {3049 \over 432} \biggr) Y
{} \PlusBreak{\null \bigglP}
  {23213 \over 5184} - {25 \over 18} \pi^2
+{269 \over 72} \zeta_3 + {41 \over 1440} \pi^4
\biggr] {x \over y}
%%%%%%%%%%%%% 1/y %%%%%%%%%%%
{} \PlusBreak{}
\biggl[
   {1 \over 2} \li3(-y)
  +{1 \over 2} Y \li2(-x)
  +{1 \over 4} Y^2 X - {1 \over 4} Y^3
  +2 Y^2 + \biggl( {7 \over 12} \pi^2 - {32 \over 9} \biggr) Y
{} \MinusBreak{\null + \bigglP}
  {29 \over 36} \pi^2 + \zeta_3
\biggr] {1 \over y}
%%%%%%%%%%%% 1/x/y %%%%%%%%%%%
{} \PlusBreak{}
\biggl[
  -{3 \over 2} \li4(-y)
  +Y \li3(-y)
  +{1 \over 4} Y^2 \li2(-x)
  +{1 \over 4} Y^3 X
  +{1 \over 24} Y^4
  -{29 \over 72} Y^3
{} \PlusBreak{\null + \bigglP}
   \biggl( {16 \over 9} - {7 \over 24} \pi^2 \biggr) Y^2
  +\biggl( {29 \over 36} \pi^2 - \zeta_3 \biggr) Y
  +{\pi^4 \over 60}
\biggr] {1 \over x y}
%
%%%%%%%%%%% Imaginary part %%%%%%%%%%%%
%
{} \PlusBreak{}
i \pi {} \Biggl\{
%%%%%%%%%%% x/y %%%%%%%%%%%%%%
\biggl[
  \biggl( {1 \over 2} Y - {4 \over 3} \biggr) \li2(-x)
+{1 \over 3} Y^3
-{15 \over 8} Y^2
+\biggl( {101 \over 18} - {\pi^2 \over 12} \biggr) Y
+{5 \over 36} \pi^2
{} \MinusBreak{\null + i \pi \BigglP \bigglP \null }
  2 \zeta_3
-{3049 \over 432}
\biggr] {x \over y}
%%%%%%%%%%%% 1/y %%%%%%%%%%%%%
+\biggl[
   {\pi^2 \over 12}
  -{32 \over 9}
  -{3 \over 4} Y^2 + 4 Y
  +{1 \over 2} \li2(-x)
\biggr] {1 \over y}
%%%%%%%%%%%%%% 1/x/y %%%%%%%%%%%
{} \PlusBreak{\null + i \pi \BigglP }
\biggr[
   \li3(-y)
  +{1 \over 2} Y \li2(-x)
  +{1 \over 2} Y^2 X + {1 \over 6} Y^3
  -{29 \over 24} Y^2
{} \MinusBreak{\null + i \pi \BigglP  + \, \bigglP \null }
   \biggl( {\pi^2 \over 4} - {32 \over 9} \biggr) Y
  -\zeta_3
\biggr] {1 \over x y}
\Biggr\}
%%%%% end : AA41Paper
\,, \label{AA41}\\[1pt plus 4pt]
%__________________________________________________________________________
%
B^{[1]}_4 & = &
%%%%% begin : BB41Paper
%%%%%%%%%%% x/y %%%%%%%%%%%
\biggl[
  2 \li4 \biggl( - {x \over y} \biggr)
+2 \li4(-x)
+{5 \over 2} \li4(-y)
-\biggl( 2 Y - {29 \over 3} \biggr) \li3(-x)
+{1 \over 8} X^4
{} \MinusBreak{\null \bigglP}
  \biggl( 4 Y - X - {77 \over 6} \biggr) \li3(-y)
+\biggl( \biggl( Y - {29 \over 3} \biggr) X - {7 \over 4} Y^2
         +{\pi^2 \over 3} + {77 \over 6} Y \biggr) \li2(-x)
{} \MinusBreak{\null \bigglP}
  \biggl( {71 \over 36} + {1 \over 2} Y \biggr) X^3
+\biggl( {779 \over 72} + {3 \over 2} Y^2
         -{11 \over 6} Y + {5 \over 2} x + {5 \over 12} \pi^2 \biggr) X^2
- {29 \over 12} Y^3 X
{} \PlusBreak{\null \bigglP}
   {265 \over 24} Y^2 X
- \biggl( {473 \over 36} + 5 x + {\pi^2 \over 2} \biggr) Y X
- \biggl(  2 \zeta_3
          + {158 \over 27}
          + {245 \over 72} \pi^2 \biggr) X
{} \PlusBreak{\null \bigglP}
  \biggl( {5 \over 2} x + {17 \over 12} \pi^2 + {73 \over 18} \biggr) Y^2
+\biggl( {3 \over 2} \zeta_3 - {205 \over 54}
         -{1177 \over 144} \pi^2 \biggr) Y
-{707 \over 72} \zeta_3
{} \PlusBreak{\null \bigglP}
  {30659 \over 1296}
+{583 \over 144} \pi^2 + {5 \over 2} \pi^2 x
-{103 \over 480} \pi^4
\biggr] {x \over y}
-{5 \over 24} Y^4 + {47 \over 36} Y^3
%%%%%%%%%%%% 1/y %%%%%%%%%%%%%
{} \PlusBreak{}
\biggl[
  -6 \biggl( \li4 \biggl( - {x \over y} \biggr)
            +\li4(-x) + \li4(-y) - \li3(-x) + (X - Y) \li3(-y) \biggr)
{} \PlusBreak{\null + \bigglP}
   {9 \over 2} \li3(-y)
  +\biggl( \pi^2 - 6 X + {9 \over 2} Y \biggr) \li2(-x)
  -\biggl( {3 \over 2} Y^2 + 3 Y \biggr) X^2
{} \PlusBreak{\null + \bigglP}
   Y^3 X
  +{11 \over 2} Y^2 X
  +\biggl( 2 \pi^2 - {1 \over 4} \biggr) Y X
  +\biggl( 6 \zeta_3 - {7 \over 6} \pi^2 \biggr) X
  -\biggl( 2 - {\pi^2 \over 2} \biggr) Y^2
{} \PlusBreak{\null + \bigglP}
   \biggl( {37 \over 9} - 6 \zeta_3 - {13 \over 3} \pi^2 \biggr) Y
  +{1 \over 15} \pi^4 + {85 \over 36} \pi^2
  +\zeta_3
\biggr] {1 \over y}
%%%%%%%%%%%% 1/x/y %%%%%%%%%%%%%%%
-\biggl( {23 \over 36} Y^3 - {5 \over 24} Y^4 \biggr) {1 \over x}
{} \PlusBreak{}
\biggl[
-8 \li4 \biggl( - {x \over y} \biggr)
-8 \li4(-x)
-{11 \over 2} \li4(-y)
+2 Y \li3(-x)
-2 X^2 Y^2 + {17 \over 12} Y^3 X
{} \PlusBreak{\null + \bigglP}
  (7 Y - 7 X - 1) \li3(-y)
-\biggl( Y - {1 \over 4} Y^2
         -{5 \over 3} \pi^2 + Y X \biggr) \li2(-x)
+ {7 \over 3} \pi^2 Y X
{} \MinusBreak{\null + \bigglP}
  {7 \over 8} Y^2 X
-\biggl( {14 \over 9} - {4 \over 3} \pi^2 \biggr) Y^2
+ 7 \zeta_3 X
-\biggl( {19 \over 36} \pi^2 + 7 \zeta_3 \biggr) Y
+ \zeta_3 + {11 \over 180} \pi^4
\biggr] {1 \over x y}
%
%%%%%%%%%%%%%% Imaginary part %%%%%%%%%%%
%
{} \PlusBreak{}
i \pi {} \Biggl\{
%%%%%%%%%%%% x/y %%%%%%%%%%%%%%%
\biggl[
-2 \li3(-x)
-3 \li3(-y)
-\biggl( {5 \over 2} Y - {19 \over 6} - X \biggr) \li2(-x)
-\biggl( {35 \over 12} - {1 \over 2} Y \biggr) X^2
{} \PlusBreak{\null + i \pi \BigglP \bigglP \null }
  \biggl( {2 \over 3} \pi^2 + {17 \over 2} - 2 Y^2
         +{67 \over 12} Y \biggr) X
-{1 \over 3} Y^3 + {17 \over 24} Y^2
+\biggl( \pi^2 - {181 \over 36} \biggr) Y
{} \MinusBreak{\null + i \pi \BigglP \bigglP \null }
  {521 \over 54} - {415 \over 144} \pi^2
-{1 \over 2} \zeta_3
\biggr] {x \over y}
%%%%%%%%%%%%% 1/y %%%%%%%%%%%%%%
{} \PlusBreak{\null + i \pi \BigglP }
\biggl[
  -{3 \over 2} \li2(-x)
  +\biggl( {1 \over 2} Y - {1 \over 4} \biggr) X
  -{\pi^2 \over 3} + {5 \over 4} Y^2
  -{17 \over 4} Y + {37 \over 9}
\biggr] {1 \over y}
%%%%%%%%%%%%% 1/x/y %%%%%%%%%%%%%
{} \PlusBreak{\null + i \pi \BigglP }
\biggl[
  2 \li3(-x)
  -\biggl( X + {1 \over 2} Y + 1 \biggr) \li2(-x)
  -\biggl( {3 \over 4} Y + {1 \over 2} Y^2 \biggr) X
  -{2 \over 3} Y^3
{} \PlusBreak{\null + i \pi \BigglP  + \, \bigglP \null }
   {37 \over 24} Y^2
  -\biggl( {28 \over 9} - {\pi^2 \over 3} \biggr) Y
\biggr] {1 \over x y}
\Biggr\}
%%%%% end : BB41Paper
\,, \label{BB41}\\[1pt plus 4pt]
%__________________________________________________________________________
%
C^{[1]}_4 & = &
%%%%% begin : CC41Paper
%%%%%%%%%%%% x/y %%%%%%%%%%%%%%
\biggl[
2 \li4 \biggl( - {x \over y} \biggr)
-6 \li4(-x)
-10 \li4(-y)
-(2 Y - 8 X) \li3(-x)
-\biggl( {3 \over 2} Y + {9 \over 4} \biggr) X^3
{} \PlusBreak{\null \bigglP}
  {1 \over 24} X^4
+(5 X + 4 Y - 1) \li3(-y)
-\biggl(  Y + {2 \over 3} \pi^2
          -5 Y X + 4 X^2 \biggr) \li2(-x)
{} \PlusBreak{\null \bigglP}
  \biggl( {3 \over 2} x - {11 \over 12} \pi^2
         +4 Y^2 + {9 \over 2} Y + {71 \over 8} \biggr) X^2
-{1 \over 6} Y^3 X
-\biggl( {11 \over 6} \pi^2 + {41 \over 4} + 3 x \biggr) Y X
{} \MinusBreak{\null \bigglP}
  \biggl( {7 \over 4} \pi^2 + 12 + 5 \zeta_3 \biggr) X
+\biggl( {3 \over 2} x + {\pi^2 \over 8} + {3 \over 2} \biggr) Y^2
+\biggl( 3 \zeta_3 + {99 \over 16} - {13 \over 12} \pi^2 \biggr) Y
{} \PlusBreak{\null \bigglP}
  {91 \over 90} \pi^4 - {35 \over 4} \zeta_3
+{511 \over 64} + {151 \over 48} \pi^2
+{3 \over 2} \pi^2 x
\biggr] {x \over y}
%%%%%%%%%%% 1/y %%%%%%%%%%%%%%%%
{} \PlusBreak{}
\biggl[
   2 \li3(-y)
  +2 Y \li2(-x)
  +\biggl( {\pi^2 \over 2} + {3 \over 4} Y \biggr) X
  -{9 \over 2} Y^2
  +\biggl( {3 \over 4} \pi^2 + 6 \biggr) Y
{} \PlusBreak{\null + \bigglP}
   {5 \over 12} \pi^2 - 8 \zeta_3
\biggr] {1 \over y}
%%%%%%%%% others %%%%%%%%%%%%%%%%%
+{19 \over 8} Y^2 X
-\biggl( {1 \over 3} Y^3 + {21 \over 8} Y^2 X \biggr) {1 \over x}
%%%%%%%%%%%%%% 1/x/y %%%%%%%%%%%%
{} \PlusBreak{}
\biggl[
    6 \biggl( \li4 \biggl( - {x \over y} \biggr)
             + \li4(-x) + \li4(-y) - Y \li3(-x) \biggr)
   -(8 Y - 3 X - 3) \li3(-y)
{} \MinusBreak{\null + \bigglP}
    (2 Y^2 - 3 Y - 3 Y X + 2 \pi^2) \li2(-x)
   +{3 \over 2} X^2 Y^2
   -\biggl( 3 \zeta_3 + {5 \over 2} Y^3 + \pi^2 Y \biggr) X
{} \PlusBreak{\null + \bigglP}
    {1 \over 3} Y^4
   -\biggl( {5 \over 8} \pi^2 + {9 \over 2} \biggr) Y^2
   -\biggl( {35 \over 12} \pi^2 - 8 \zeta_3 \biggr) Y
  -3 \zeta_3 - {\pi^4 \over 15}
\biggr] {1 \over x y}
%
%%%%%%%%%% Imaginary part %%%%%%%%%%%%%
%
{} \PlusBreak{}
i \pi {} \Biggl\{
%%%%%%%%%%% x/y %%%%%%%%%%%%%%%
\biggl[
  6 \li3(-x)
+9 \li3(-y)
+(5 Y - 1 - 3 X) \li2(-x)
-\biggl( {3 \over 2} Y + {9 \over 4} \biggr) X^2
-2 \zeta_3
{} \PlusBreak{\null + i \pi \BigglP \bigglP \null }
  \biggl( 6 Y^2 + {15 \over 2} - 2 \pi^2 + {21 \over 4} Y \biggr) X
-\biggl( {31 \over 12} \pi^2 + {29 \over 4} \biggr) Y
-{93 \over 16} - {25 \over 12} \pi^2
\biggr] {x \over y}
%%%%%%%%%%%% 1/y %%%%%%%%%%%%%%
{} \PlusBreak{\null + i \pi \BigglP }
\biggl[
  2 \li2(-x)
  -\biggl( {3 \over 2} Y - {3 \over 4} \biggr) X
  +{5 \over 12} \pi^2
  -{33 \over 4} Y + 6
\biggr] {1 \over y}
%%%%%%%%%%%%% others %%%%%%%%%%%%%
+ {5 \over 6} Y^3 + {15 \over 8} Y^2
%%%%%%%%%%%%% 1/x/y %%%%%%%%%%%%%
{} \PlusBreak{\null + i \pi \BigglP }
\biggl[
   -6 \li3(-x)
   -5 \li3(-y)
   -(Y - 3 X - 3) \li2(-x)
   +\biggl( {9 \over 4} Y - Y^2 \biggr) X
{} \MinusBreak{\null + i \pi \BigglP  + \, \bigglP \null }
    \biggl( 9 - {5 \over 12} \pi^2 \biggr) Y
   +5 \zeta_3
\biggr] {1 \over  x y}
-\biggl( {17 \over 8} Y^2 + {5 \over 6} Y^3 \biggr) {1 \over x}
\Biggr\}
%%%%% end : CC41Paper
\,, \label{CC41}\\[1pt plus 4pt]
%_____________________________________________________________________________
%
D^{[1]}_4 & = &
%%%%% begin : DD41Paper
%%%%%%%%%% x/y %%%%%%%%%%%%
\biggl[
  {1 \over 3} \li3(-y)
+{1 \over 3} Y \li2(-x)
+{1 \over 6} Y^2 X
+{1 \over 6} Y^3 - {13 \over 12} Y^2
-\biggl( {7 \over 18} \pi^2 - {193 \over 54} \biggr) Y
-{455 \over 108}
{} \PlusBreak{\null \bigglP}
  {19 \over 24} \pi^2
-{49 \over 36} \zeta_3
\biggr] {x \over y}
%%%%%%%%%%%% 1/y %%%%%%%%%%%
+\biggl[
  {2 \over 9} \pi^2 - {1 \over 2} Y^2
  +{8 \over 9} Y
\biggr] {1 \over y}
%%%%%%%%%%%% 1/x/y %%%%%%%%%%
+\biggl[
  {1 \over 9} Y^3 - {4 \over 9} Y^2
- {2 \over 9} \pi^2 Y
\biggr] {1 \over x y}
%
%%%%%%%%%%%% Imaginary part %%%%%%%%%%
%
{} \PlusBreak{}
i \pi {} \Biggl\{
%%%%%%%%%%%% x/y %%%%%%%%%%%
\biggl[
  {1 \over 3} \li2(-x)
+{1 \over 2} Y^2 - {13 \over 6} Y
+{193 \over 54} - {\pi^2 \over 18}
\biggr] {x \over y}
%%%%%%%%%%%% 1/x/y %%%%%%%%%%%
+\biggl[ {8 \over 9} - Y \biggr] {1\over y}
{} \MinusBreak{\null + i \pi \BigglP }
\biggl[ {8 \over 9} Y - {1 \over 3} Y^2 \biggr] {1 \over x y}
\Biggr\}
%%%%% end : DD41Paper
\,, \label{DD41}\\[1pt plus 4pt]
%___________________________________________________________________________
%
E^{[1]}_4 & = &
%%%%% begin : EE41Paper
%%%%%%%%%%%%% x/y %%%%%%%%%%%%%
\biggl[
-{2 \over 3} \li3(-x)
-{4 \over 3} \li3(-y)
-{1 \over 3} (4 Y - 2 X) \li2(-x)
+{2 \over 9} X^3
+\biggl( {1 \over 3} Y - {29 \over 18} \biggr) X^2
{} \PlusBreak{\null \bigglP}
  \biggl( {13 \over 36} \pi^2 - {5 \over 3} Y^2
         +{62 \over 27} + {11 \over 9} Y \biggr) X
+\biggl( {107 \over 72} \pi^2 + {11 \over 27} \biggr) Y
-{685 \over 162} - {11 \over 72} \pi^2
{} \MinusBreak{\null \bigglP}
  {11 \over 36} \zeta_3
\biggr] {x \over y}
%%%%%%%%%%%%% others %%%%%%%%%%%%
-{1 \over 9} (4 \pi^2 + 16 Y) {1 \over y}
-{2 \over 9} Y^3 - {1 \over 9} Y^2
-{1 \over 9} (8 Y^2 - 2 Y^3) {1 \over x}
+{4 \over 9} {\pi^2 \over x y} Y
%
%%%%%%%%%%%% Imaginary part %%%%%%%%%
%
{} \PlusBreak{}
i \pi {} \Biggl\{
%%%%%%%%%%%%% x/y %%%%%%%%%%%%
\biggl[
-{2 \over 3} \li2(-x)
+{2 \over 3} X^2 - \biggl( 2 + {4 \over 3} Y \biggr) X
+{53 \over 72} \pi^2 + {73 \over 27}
-{1 \over 3} Y^2 + {13 \over 9} Y
\biggr] {x \over y}
%%%%%%%%%%% others %%%%%%%%%%%
{} \PlusBreak{\null + i \pi \BigglP }
\biggl[ 2 Y - {16 \over 9} \biggr] {1 \over y}
+\biggl[ {16 \over 9} Y - {2 \over 3} Y^2 \biggr] {1 \over x y}
\Biggr\}
%%%%% end : EE41Paper
\,, \label{EE41}\\[1pt plus 4pt]
%___________________________________________________________________________
%
F^{[1]}_4 & = &
%%%%% begin : FF41Paper
\biggl[
  {1 \over 9} Y^2
-{10 \over 27} Y
-{\pi^2 \over 9}
+{25 \over 81}
\biggr] {x \over y}
%%%%%%%% Imaginary part %%%%%%%%
+i \pi {} \biggl[
{2 \over 9} Y - {10 \over 27}
\biggr] {x \over y}
%%%%% end : FF41Paper
\,, \label{FF41} %\\[1pt plus 4pt]
%__________________________________________________________________________
%
\end{eqnarray}

For $h=4$ in \eqn{h4} and color factor $\trc^{[2]}$ in \eqn{basis34}:
\begin{eqnarray}
G^{[2]}_4 & = &
%%%%% begin : GG42Paper
%%%%%%%%%% x/y %%%%%%%%%%%
\biggl[
-3 \li4 \biggl( - {x \over y} \biggr)
+2 \li4(-x)
-\li4(-y)
-\biggl( {28 \over 3} - 2 Y + 3 X \biggr) \li3(-x)
-{1 \over 24} X^4
{} \MinusBreak{\null \bigglP}
  {7 \over 24} Y^4
+\biggl( 2 Y - {25 \over 3} - X \biggr) \li3(-y)
-\biggl( {25 \over 3} Y - {\pi^2 \over 2}
         -{28 \over 3} X - {1 \over 2} X^2 \biggr) \li2(-x)
{} \PlusBreak{\null \bigglP}
  \biggl( {11 \over 18} + {1 \over 12} Y \biggr) X^3
-\biggl(  {7 \over 8} Y^2 - {47 \over 12} Y
          +{3 \over 2} x + {385 \over 72} - {\pi^2 \over 24} \biggr) X^2
+{17 \over 12} Y^3 X
{} \MinusBreak{\null \bigglP}
  {205 \over 24} Y^2 X
+\biggl( {545 \over 72} + {5 \over 6} \pi^2 + 3 x \biggr) Y X
+\biggl( {79 \over 27} + {3 \over 2} \zeta_3
          +{263 \over 144} \pi^2 \biggr) X
+ {113 \over 72} Y^3
{} \MinusBreak{\null \bigglP}
  \biggl( {67 \over 12} + {3 \over 2} x + {5 \over 6} \pi^2 \biggr) Y^2
+\biggl( 2 \zeta_3 + {1513 \over 432} + {29 \over 6} \pi^2 \biggr) Y
-{23213 \over 5184} - {193 \over 1440} \pi^4
{} \MinusBreak{\null \bigglP}
  {65 \over 72} \zeta_3 - {14 \over 9} \pi^2
-{3 \over 2} \pi^2 x
\biggr] {x \over y}
%%%%%%%%%%%%% 1/y %%%%%%%%%%%%%
{} \PlusBreak{}
\biggl[
   6 \biggl( \li4 \biggl( - {x \over y} \biggr)
            + \li4(-x) + \li4(-y) - \li3(-x) \biggr)
  -(4 + 6 Y - 6 X) \li3(-y)
{} \MinusBreak{\null + \bigglP}
   (4 Y + \pi^2 - 6 X) \li2(-x)
  +\biggl( 3 Y + {3 \over 2} Y^2 \biggr) X^2
  -Y^3 X - {19 \over 4} Y^2 X
{} \PlusBreak{\null + \bigglP}
   \biggl( {7 \over 4} - 2 \pi^2 \biggr) Y X
  +\biggl( {5 \over 6} \pi^2 - 6 \zeta_3 \biggr) X
  +{13 \over 12} Y^3
  -\biggl( {11 \over 4} + {\pi^2 \over 2} \biggr) Y^2
{} \PlusBreak{\null + \bigglP}
   \biggl( {29 \over 12} \pi^2 + 6 \zeta_3 \biggr) Y
  -{17 \over 12} \pi^2 - \zeta_3
  -{\pi^4 \over 15}
\biggr] {1 \over y}
%%%%%%%%%%% 1/x/y %%%%%%%%%%%%%
{} \PlusBreak{}
\biggl[
   8 \li4 \biggl( - {x \over y} \biggr)
  +8 \li4(-x)
  +5 \li4(-y)
  -2 Y \li3(-x)
  +2 X^2 Y^2 + {1 \over 8} Y^4
  + {7 \over 24} Y^3
{} \PlusBreak{\null + \bigglP}
   \biggl( 7 X - {1 \over 2} - 7 Y \biggr) \li3(-y)
  +\biggl( Y X - {4 \over 3} \pi^2
          -{1 \over 2} Y^2 - {1 \over 2} Y \biggr) \li2(-x)
{} \MinusBreak{\null + \bigglP}
   \biggl( {5 \over 8} Y^2 + {5 \over 3} Y^3
           +7 \zeta_3 + 2 \pi^2 Y \biggr) X
  -\biggl( {11 \over 24} \pi^2 + {5 \over 4} \biggr) Y^2
  +\biggl( 7 \zeta_3 + {\pi^2 \over 6} \biggr) Y
{} \PlusBreak{\null + \bigglP}
   {1 \over 2} \zeta_3 - {\pi^4 \over 18}
\biggr] {1 \over x y}
%
%%%%%%%%% Imaginary part %%%%%%%%%%%
%
{} \PlusBreak{}
i \pi {} \Biggl\{
%%%%%%%%% x/y %%%%%%%%%%%
\biggl[
-\li3(-x)
+\li3(-y)
+(1+X) \li2(-x)
+{1 \over 12} X^3
+\biggl( {13 \over 12} - Y \biggr) X^2
{} \PlusBreak{\null + i \pi \BigglP \bigglP \null }
  \biggl( {5 \over 12} \pi^2 + {3 \over 2} Y^2
         -{25 \over 8} - {11 \over 12} Y \biggr) X
-{1 \over 12} Y^3 + {1 \over 3} Y^2
-\biggl( {\pi^2 \over 2} + {259 \over 72} \biggr) Y
{} \PlusBreak{\null + i \pi \BigglP \bigglP \null }
  {2777 \over 432} + {7 \over 2} \zeta_3
+{37 \over 48} \pi^2
\biggr] {x \over y}
%%%%%%%%%%%% 1/y %%%%%%%%%
{} \PlusBreak{\null + i \pi \BigglP }
\biggl[
   2 \li2(-x)
  +\biggl( {1 \over 2} Y + {7 \over 4} \biggr) X
  -{\pi^2 \over 12} + {1 \over 2} Y^2
  -{15 \over 4} Y
\biggr] {1 \over y}
%%%%%%%%%%% 1/x/y %%%%%%%
{} \PlusBreak{\null + i \pi \BigglP }
\biggl[
  -2 \li3(-x)
  +\biggl( X - {1 \over 2} \biggr) \li2(-x)
  -{3 \over 4} Y X + {1 \over 6} Y^3
  +{1 \over 2} Y^2
{} \MinusBreak{\null + i \pi \BigglP  + \, \bigglP \null }
   \biggl( {5 \over 2} - {\pi^2 \over 12} \biggr) Y
\biggr] {1 \over x y}
\Biggl\}
%%%%% end : GG42Paper
\,, \label{GG42}\\[1pt plus 4pt]
%____________________________________________________________________________
%
H^{[2]}_4 & = &
%%%%% begin : HH42Paper
%%%%%%%%%%% x/y %%%%%%%%%
\biggl[
  {9 \over 2} \li4(-y)
-\biggl( 2 X + {1 \over 3} - Y \biggr) \li3(-x)
-\biggl( 3 X + {8 \over 3} + Y \biggr) \li3(-y)
+{5 \over 6} Y^3 X
{} \MinusBreak{\null \bigglP}
  {1 \over 8} X^4
+\biggl( 2 X^2 - \biggl( 4 Y - {1 \over 3} \biggr) X
         -{8 \over 3} Y + {5 \over 4} Y^2 \biggr) \li2(-x)
+\biggl( {103 \over 36} + {17 \over 12} Y \biggr) X^3
{} \PlusBreak{\null \bigglP}
  \biggl( {\pi^2 \over 8} - {193 \over 18}
         -{61 \over 12} Y - 2 x - {25 \over 8} Y^2 \biggr) X^2
- {1 \over 3} Y^2 X
+\biggl( {5 \over 6} \pi^2 + 4 x + {797 \over 72} \biggr) Y X
{} \PlusBreak{\null \bigglP}
  \biggl( {241 \over 27} + {395 \over 144} \pi^2
         +{7 \over 2} \zeta_3 \biggr) X
-\biggl( {\pi^2 \over 6} + {41 \over 18} + 2 x \biggr) Y^2
{} \PlusBreak{\null \bigglP}
  \biggl( {235 \over 54} - {3 \over 2} \zeta_3
         +{365 \over 144} \pi^2 \biggr) Y
-{97 \over 48} \pi^2 - {107 \over 288} \pi^4
-{30659 \over 1296} + {719 \over 72} \zeta_3
-2 \pi^2 x
\biggr] {x \over y}
%%%%%%%%%%%% 1/y %%%%%%%%%%
{} \PlusBreak{}
\biggl[
  -{5 \over 2} ( \li3(-y) + Y \li2(-x) )
  -\biggl( 2 Y + {5 \over 4} Y^2 \biggr) X
  +{21 \over 4} Y^2
  -\biggl( {32 \over 9} - {3 \over 4} \pi^2 \biggr) Y
{} \MinusBreak{\null + \bigglP}
   {11 \over 36} \pi^2
  +4 \zeta_3
\biggr] {1 \over y}
%%%%%%%%%%% others %%%%%%%%%%%%%
-{13 \over 36} Y^3
+{25 \over 36 x} Y^3
%%%%%%%%%%% 1/x/y %%%%%%%%%%%%%
{} \PlusBreak{}
\biggl[
  -4 \biggl( \li4 \biggl( - {x \over y} \biggr)
            +\li4(-x) - Y {} (\li3(-x) + \li3(-y)) \biggr)
  -{3 \over 2} \li4(-y)
{} \MinusBreak{\null + \bigglP}
   \biggl( 2 X + {1 \over 2} \biggr) \li3(-y)
  -\biggl( {1 \over 2} Y + 2 Y X
          -\pi^2 - {5 \over 4} Y^2 \biggr) \li2(-x)
  -X^2 Y^2
{} \MinusBreak{\null + \bigglP}
   {1 \over 6} Y^4
  +\biggl(  {19 \over 12} Y^3 - {1 \over 4} Y^2
           +{\pi^2 \over 3} Y + 2 \zeta_3 \biggr) X
  +\biggl( {127 \over 36} - {5 \over 24} \pi^2 \biggr) Y^2
{} \PlusBreak{\null + \bigglP}
   \biggl( {25 \over 18} \pi^2 - 4 \zeta_3 \biggr) Y
  +{1 \over 2} \zeta_3 + {\pi^4 \over 60}
\biggr] {1 \over x y}
%
%%%%%%%%%%% Imaginary part %%%%%%%%%%
%
{} \PlusBreak{}
i \pi {} \Biggl\{
%%%%%%%%%% x/y %%%%%%%%%%%%
\biggl[
-\li3(-x)
-4 \li3(-y)
-\biggl( {7 \over 3} + {3 \over 2} Y \biggr) \li2(-x)
-{1 \over 12} X^3
+{7 \over 12} Y^3
{} \PlusBreak{\null + i \pi \BigglP \bigglP \null }
  \biggl( {10 \over 3} + {3 \over 2} Y \biggr) X^2
-\biggl( {49 \over 6} Y + {7 \over 2} Y^2
         +{83 \over 8} - {\pi^2 \over 4} \biggr) X
+{25 \over 12} Y^2
{} \PlusBreak{\null + i \pi \BigglP \bigglP \null }
  \biggl( {469 \over 72} + {4 \over 3} \pi^2 \biggr) Y
+2 \zeta_3 + {29 \over 9} \pi^2
+{239 \over 18}
\biggr] {x \over y}
%%%%%%%%%%%%% 1/y %%%%%%%%%%%
{} \PlusBreak{\null + i \pi \BigglP }
\biggl[
   {17 \over 2} Y - Y^2
  +{\pi^2 \over 12} - {32 \over 9}
  -2 X - {5 \over 2} \li2(-x)
\biggr] {1 \over y}
%%%%%%%%%%% 1/x/y %%%%%%%%%%
{} \PlusBreak{\null + i \pi \BigglP }
\biggl[
   4 \li3(-x)
  +2 \li3(-y)
  -\biggl( {1 \over 2} - {1 \over 2} Y + 2 X \biggr) \li2(-x)
  +{1 \over 2} Y^2 X - {1 \over 6} Y^3
{} \MinusBreak{\null + i \pi \BigglP  + \, \bigglP \null }
   {25 \over 12} Y^2
  -\biggl( {5 \over 12} \pi^2 - {127 \over 18} \biggr) Y
  -2 \zeta_3
\biggr] {1 \over x y}
\Biggr\}
%%%%% end : HH42Paper
\,, \label{HH42}\\[1pt plus 4pt]
%____________________________________________________________________________
%
I^{[2]}_4 &=&
%%%%% begin : II42Paper
%%%%%%%%%%%%%%% x/y %%%%%%%%%%%%%%%
\biggl[
  -\li4 \biggl( -{x \over y} \biggr)
  + 2 \li4(-x)
  + {7 \over 2} \li4(-y)
  + \biggl( {3 \over 4} + {1 \over 2} Y \biggr) X^3
  + (Y-3 X) \li3(-x)
{} \MinusBreak{\null \bigglP}
    \biggl( Y + 2 X-{1 \over 2} \biggr) \li3(-y)
  + \biggl( {3 \over 2} X^2 + {\pi^2 \over 6}
       + {1 \over 2} Y + {1 \over 4} Y^2
       -2 Y X \biggr) \li2(-x)
{} \PlusBreak{\null \bigglP}
    \biggl( -{3 \over 2} Y^2 + {\pi^2 \over 3}
      -{29 \over 8} - {3 \over 2} Y - {1 \over 2} x \biggr) X^2
  + {1 \over 4} Y^3 X
  + \biggl( x + {19 \over 4} + {2 \over 3} \pi^2 \biggr) Y X
{} \PlusBreak{\null \bigglP}
  \biggl( 6 + {7 \over 12} \pi^2 + 2 \zeta_3 \biggr) X
-\biggl( {1 \over 2} x + {1 \over 2} + {\pi^2 \over 6} \biggr) Y^2
-\biggl( 3 \zeta_3 + {51 \over 16} - {\pi^2 \over 2} \biggr) Y
{} \MinusBreak{\null \bigglP}
    {23 \over 72} \pi^4 - {\pi^2 \over 2} x
  + {23 \over 4} \zeta_3
-{511 \over 64}-{107 \over 48} \pi^2
\biggr] {x \over y}
%%%%%%%%%%%%%%% 1/y %%%%%%%%%%%%%%%%%
{} \PlusBreak{}
\biggl[
  -{1 \over 2} \li3(-y)
  -{1 \over 2} Y \li2(-x)
  -\biggl( {1 \over 4} Y + {\pi^2 \over 6} \biggr) X
   + 2 Y^2 - \biggl( {\pi^2 \over 6} + 3 \biggr) Y
{} \PlusBreak{\null + \bigglP}
     3 \zeta_3 - {\pi^2 \over 4}
\biggr] {1 \over y}
%%%%%%%%%%%%%% others %%%%%%%%%%%%%
-{7 \over 8} Y^2 X
+\biggl( {1 \over 6} Y^3 + {7 \over 8} Y^2 X \biggr) {1 \over x}
{} \PlusBreak{}
\biggl[
    -2 \li4 \biggl( -{x \over y} \biggr)
    -2 \li4(-x)
    -{5 \over 2} \li4(-y)
     + 2 Y \li3(-x)
     + (3 Y - 1 - X) \li3(-y)
{} \MinusBreak{\null + \bigglP}
       \biggl( Y X - {2 \over 3} \pi^2
               - {3 \over 4} Y^2 + Y \biggr) \li2(-x)
    -{1 \over 2}  Y^2 X^2
     + \biggl( \zeta_3 + {\pi^2 \over 3} Y
           + {11 \over 12} Y^3 \biggr) X
{} \MinusBreak{\null + \bigglP}
      {1 \over 8}Y^4
     + \biggl( {\pi^2 \over 4} + 2 \biggr) Y^2
     + \biggl( {13 \over 12} \pi^2 - 3 \zeta_3 \biggr) Y
     + {\pi^4 \over 36} + \zeta_3
\biggr] {1 \over x y}
%
%%%%%%%%%%% Imaginary part %%%%%%%%%%%
%
{} \PlusBreak{}
  i \pi {} \Biggl\{
%%%%%%%%%%%%% x/y %%%%%%%%%%%%%%%%
\biggl[
-2 \li3(-x)
-3 \li3(-y)
-\biggl( {3 \over 2} Y - {1 \over 2} - X \biggr) \li2(-x)
  + \biggl( {3 \over 4} + {1 \over 2} Y \biggr) X^2
{} \PlusBreak{\null + i \pi \BigglP \bigglP \null }
    \biggl( {2 \over 3} \pi^2 - {5 \over 2}
   -2 Y^2 - {7 \over 4} Y \biggr) X
  + \biggl( {5 \over 6} \pi^2 + {15 \over 4} \biggr) Y
  + {45 \over 16} - \zeta_3
  + {5 \over 6} \pi^2
\biggr] {x \over y}
%%%%%%%%%%%%%%% 1/y %%%%%%%%%%%%%%%
{} \PlusBreak{\null + i \pi \BigglP }
\biggl[
   -{1 \over 2} \li2(-x)
   + \biggl( {1 \over 2} Y - {1 \over 4} \biggr) X
   + {15 \over 4} Y - {\pi^2 \over 6} - 3
\biggr] {1 \over y}
%%%%%%%%%%%%%%%% others %%%%%%%%%%%%%%%
-{1 \over 3} Y^3 - {5 \over 8} Y^2
{} \PlusBreak{\null + i \pi \BigglP }
\biggl[
     2 \li3(-x)
   + 2 \li3(-y)
   + \biggl( {1 \over 2} Y - 1 - X \biggr) \li2(-x)
   + \biggl( {1 \over 2} Y^2 - {3 \over 4} Y \biggr) X
{} \MinusBreak{\null + i \pi \BigglP  + \, \bigglP \null }
   \biggl( {\pi^2 \over 6} - 4 \biggr) Y
  -2 \zeta_3
\biggr] {1 \over x y}
+\biggl( {1 \over 3} Y^3 + {7 \over 8} Y^2 \biggr) {1 \over x}
\Biggr\}
%%%%% end : II42Paper
\,, \label{II42} \\[1pt plus 4pt]
%_____________________________________________________________________________
%
J^{[2]}_4 & = &
%%%%% begin : JJ42Paper
%%%%%%%%%%%%% x/y %%%%%%%%%%%%%
\biggl[
  {1 \over 3} \li3(-x)
+{1 \over 3} \li3(-y)
-{1 \over 3} (X - Y) \li2(-x)
-{1 \over 9} X^3
-\biggl( {1 \over 6} Y - {29 \over 36} \biggr) X^2
{} \MinusBreak{\null \bigglP}
  \biggl(  {31 \over 27} - {2 \over 3} Y^2
          +{13 \over 72} \pi^2 + {11 \over 18} Y \biggr) X
-{5 \over 18} Y^3 + {37 \over 36} Y^2
-\biggl( {145 \over 54} + {\pi^2 \over 3} \biggr) Y
{} \PlusBreak{\null \bigglP}
  {455 \over 108} + {37 \over 36} \zeta_3
-{55 \over 72} \pi^2
\biggr] {x \over y}
%
%%%%%%%%%% Imaginary part %%%%%%%%%%
%
{} \PlusBreak{}
i \pi {} \biggl\{
-{1 \over 3} X^2
+\biggl( {2 \over 3} Y + 1 \biggr) X+
{13 \over 9} Y - {23 \over 6}
-{1 \over 3} Y^2 - {7 \over 24} \pi^2
\biggr\} {x \over y}
%%%%% end : JJ42Paper
\,, \label{JJ42}\\[1pt plus 4pt]
%____________________________________________________________________________
%
K^{[2]}_4 & = &
%%%%% begin : KK42Paper
%%%%%%%%%%% x/y %%%%%%%%%%%%
\biggl[
  {1 \over 3} \li3(-x)
+{2 \over 3} \li3(-y)
+{1 \over 3} (2 Y- X) \li2(-x)
-{1 \over 9} X^3
-\biggl( {1 \over 6} Y - {29 \over 36} \biggr) X^2
{} \MinusBreak{\null \bigglP}
  \biggl( {13 \over 72} \pi^2 - {5 \over 6} Y^2
         +{31 \over 27} + {11 \over 18} Y \biggr) X
-\biggl( {55 \over 72} \pi^2 + {35 \over 27} \biggr) Y
+{685 \over 162} + {23 \over 36} \zeta_3
+{\pi^2 \over 8}
\biggr] {x \over y}
%%%%%%%%%%%% 1/y %%%%%%%%%%%%
{} \PlusBreak{}
{1 \over 9} (8 Y + 2 \pi^2) {1 \over y}
+{1 \over 18} Y^2 + {1 \over 9} Y^3
+{1 \over 9}(4 Y^2 - Y^3) {1 \over x}
-{2 \over 9} {\pi^2 \over x y} Y
%
%%%%%%%%%%% Imaginary part %%%%%%%%%%
%
{} \PlusBreak{}
i \pi {} \Biggl\{
%%%%%%%%%%%% x/y %%%%%%%%%%
\biggl[
  {1 \over 3} \li2(-x)
-{1 \over 3} X^2
+\biggl( {2 \over 3} Y + 1 \biggr) X
-{7 \over 18} \pi^2 + {1 \over 6} Y^2
-{13 \over 18} Y - {22 \over 9}
\biggr] {x \over y}
%%%%%%%%%%% others %%%%%%%%%%
{} \MinusBreak{\null + i \pi \BigglP }
\biggl[ Y - {8 \over 9} \biggr] {1 \over y}
-\biggl[ {8 \over 9} Y - {1 \over 3} Y^2 \biggr] {1 \over x y}
\Biggr\}
%%%%% end : KK42Paper
\,, \label{KK42}\\[1pt plus 4pt]
%____________________________________________________________________________
%
L^{[2]}_4 & = &
%%%%% begin : LL42Paper
-\biggl[
  {1 \over 9} Y^2
-{10 \over 27} Y
-{\pi^2 \over 9}
+{25 \over 81}
\biggr] {x \over y}
%%%%%%%% Imaginary part %%%%%%%%%%%
-i \pi {} \biggl[
{2 \over 9} Y - {10 \over 27}
\biggr] {x \over y}
%%%%% end : LL42Paper
\,, \label{LL42} %\\[1pt plus 4pt]
%_____________________________________________________________________________
%
\end{eqnarray}

For $h=5$ in \eqn{h5} and color factor $\trc^{[1]}$ in \eqn{basis56}:

\begin{eqnarray}
A^{[1]}_5 & = &
%%%%% begin : AA51Paper
%%%%%%%%%%%% x/y %%%%%%%%%%
\biggl[
  {3 \over 2} x \li4 \biggl( - {x \over y} \biggr)
-\biggl( x {} (Y -  X) + {1 \over 2} \biggr) (\li3(-x) + \li3(-y))
+{1 \over 16} x X^4 + {5 \over 48} x Y^4
{} \MinusBreak{\null \bigglP}
  \biggl(  {1 \over 4} x X^2
          -{1 \over 2}(x Y + 1) X
          +{\pi^2 \over 4} x
          +{1 \over 4} x Y^2
          +{1 \over 2} Y \biggr) \li2(-x)
-{2 \over 3} x Y^3 X
{} \MinusBreak{\null \bigglP}
  \biggl( {1 \over 4} x Y + {37 \over 72} x - {1 \over 3} \biggr) X^3
+\biggl( {3 \over 4} x Y^2
         +\biggl( {5 \over 8} x - {3 \over 4} \biggr) Y
         +{1 \over 6}
         +\biggl( {\pi^2 \over 8} + {16 \over 9} \biggr) x \biggr) X^2
{} \PlusBreak{\null \bigglP}
  \biggl( {1 \over 2} + {7 \over 24} x \biggr) Y^2 X
-\biggl( {13 \over 6}
         + \biggl( {32 \over 9} + {5 \over 12} \pi^2 \biggr) x \biggr) Y X
  +\biggl( {32 \over 9} + {\pi^2 \over 3} - {37 \over 72} \pi^2 x \biggr) X
{} \MinusBreak{\null \bigglP}
  \biggl( {1 \over 4} + {29 \over 72} x \biggr) Y^3
+\biggl( 2 + \biggl( {\pi^2 \over 6} + {16 \over 9} \biggr) x \biggr) Y^2
-\biggl( {29 \over 72} \pi^2 x + {32 \over 9} \biggr) Y
{} \PlusBreak{\null \bigglP}
  \biggl( {11 \over 240} \pi^4 + {16 \over 9} \pi^2 \biggr) x
+{3 \over 2} \zeta_3 + {43 \over 36} \pi^2
\biggr] {x \over y}
%%%%%%%%%%% 1/y %%%%%%%%%%%%
{} \PlusBreak{}
\biggl[
  -{1 \over 2} \li4 \biggl( - {x \over y} \biggr)
  +{4 \over 3} (\li3(-x) + \li3(-y))
  +{1 \over 48} X^4
  -\biggl( {37 \over 72} + {1 \over 4} Y \biggr) X^3
{} \PlusBreak{\null + \bigglP}
   {1 \over 16} Y^4
  -{1 \over 3} Y^3 X
  -\biggl(  {1 \over 4} X^2
           +\biggl( {4 \over 3} - {1 \over 2} Y \biggr) X
           -{4 \over 3} Y
           -{\pi^2 \over 12}
           +{1 \over 4} Y^2 \biggr) \li2(-x)
{} \PlusBreak{\null + \bigglP}
   \biggl(  {1 \over 2} Y^2 - {\pi^2 \over 24}
           +{28 \over 9} - {1 \over 24} Y \biggr) X^2
  +{13 \over 8} Y^2 X
  -\biggl( {\pi^2 \over 12} + {23 \over 9} \biggr) Y X
  -{5 \over 8} Y^3
{} \MinusBreak{\null + \bigglP}
   \biggl(  {3 \over 2} \zeta_3
           +{31 \over 48} \pi^2
           -{17 \over 27} \biggr) X
  +\biggl( {101 \over 36} + {\pi^2 \over 6} \biggr) Y^2
  -\biggl(  {3049 \over 432} + 2 \zeta_3
           +{67 \over 72} \pi^2 \biggr) Y
{} \PlusBreak{\null + \bigglP}
   {173 \over 72} \zeta_3 + {23213 \over 5184}
  -{35 \over 18} \pi^2 + {47 \over 1440} \pi^4
\biggr] {1 \over y}
%
%%%%%%%%%% Imaginary part %%%%%%%%%%%%
%
{} \PlusBreak{}
i \pi {} \Biggl\{
%%%%%%%%%% x/y %%%%%%%%%%
\biggl[
-{11 \over 12} x X^2
+{11 \over 6} (x Y - 1) X
+{11 \over 6} Y - {11 \over 12} \pi^2 x
-{11 \over 12} x Y^2
\biggr] {x \over y}
%%%%%%%%%%%% 1/y %%%%%%%%%
{} \PlusBreak{\null + i \pi \BigglP }
\biggl[
  -{11 \over 12} X^2
  +\biggl( {11 \over 3} + {11 \over 6} Y \biggr) X
  -{7 \over 2} \zeta_3 + {55 \over 18} Y
  -{11 \over 16} \pi^2 - {2777 \over 432}
{} \MinusBreak{\null + i \pi \BigglP  + \, \bigglP \null }
   {11 \over 12} Y^2
\biggr] {1 \over y}
\Biggl\}
%%%%% end : AA51Paper
\,, \label{AA51}\\[1pt plus 4pt]
%____________________________________________________________________________
%
B^{[1]}_5 & = &
%%%%% begin : BB51Paper
%%%%%%%%%%%% x/y %%%%%%%%%%%
\biggl[
  \biggl( 6 + {11 \over 2} x \biggr) \li4 \biggl( - {x \over y} \biggr)
+(8 x + 6) (\li4(-x) + \li4(-y))
-{3 \over 16} x X^4
{} \MinusBreak{\null \bigglP}
  \biggl( 2 x X + (5 x + 6) Y - {3 \over 2} - x \biggr) \li3(-x)
-\biggl( (7 x + 6) Y + {9 \over 2} - x \biggr) \li3(-y)
{} \PlusBreak{\null \bigglP}
  \biggl(  {3 \over 4} x X^2
          -\biggl( {1 \over 2} x Y + {3 \over 2} + x \biggr) X
          -{1 \over 4} x Y^2
          -\biggl( {9 \over 2} - x \biggr) Y
          -{11 \over 12} \pi^2 x - \pi^2 \biggr) \li2(-x)
{} \MinusBreak{\null \bigglP}
  \biggl( {5 \over 6} - {11 \over 12} x Y - {101 \over 72} x \biggr) X^3
+\biggl( -{5 \over 4} x Y^2 - {1 \over 2} (5 x - 3) Y
          +{17 \over 12} - {14 \over 9} x \biggr) X^2
{} \MinusBreak{\null \bigglP}
  \biggl(  {3 \over 2} Y^3
          +\biggl( {7 \over 2} - {19 \over 24} x \biggr) Y^2
          -\biggl( {7 \over 12}
                  + \biggl( {28 \over 9} + {\pi^2 \over 2} \biggr) x \biggr) 
Y
          +{37 \over 9}
          +{3 \over 4} \pi^2
          -{89 \over 72} \pi^2 x \biggr) X
{} \PlusBreak{\null \bigglP}
  \biggl( {1 \over 2} + {17 \over 48} x \biggr) Y^4
-\biggl( 2 + \biggl( {14 \over 9} + {5 \over 24} \pi^2 \biggr) x \biggr) Y^2
+\biggl( {5 \over 3} \pi^2 + {11 \over 36} \pi^2 x + {37 \over 9} \biggr) Y
{} \MinusBreak{\null \bigglP}
  \biggl( {61 \over 144} \pi^4 + {14 \over 9} \pi^2 \biggr) x
+{11 \over 2} \zeta_3 + {\pi^2 \over 9}
-{\pi^4 \over 4}
\biggr] {x \over y}
%%%%%%%%%%%%% 1/y %%%%%%%%%%%%
{} \PlusBreak{}
\biggl[
  -{5 \over 2} \li4 \biggl( - {x \over y} \biggr)
  -2 \li4(-x)
  -2 \li4(-y)
  +\biggl( 2 Y - {19 \over 6} - X \biggr) \li3(-x)
  +{1 \over 48} Y^4
{} \MinusBreak{\null + \bigglP}
   \biggl( {77 \over 6} - 4 Y + 3 X \biggr) \li3(-y)
  -{1 \over 16} X^4
  -\biggl( {1 \over 6} Y + {403 \over 72} + {9 \over 4} Y^2 \biggr) X^2
  + Y^3 X
{} \PlusBreak{\null + \bigglP}
   \biggl(  {3 \over 4} X^2
           +\biggl( {19 \over 6} - {5 \over 2} Y \biggr) X
           -{77 \over 6} Y + {7 \over 4} Y^2
           +{5 \over 12} \pi^2 \biggr) \li2(-x)
  - {215 \over 24} Y^2 X
{} \PlusBreak{\null + \bigglP}
   \biggl( {89 \over 72} + {7 \over 12} Y \biggr) X^3
  +\biggl( {7 \over 6} \pi^2 + {49 \over 36} \biggr) Y X
  -\biggl( {115 \over 27} - {7 \over 4} \pi^2 - 4 \zeta_3 \biggr) X
{} \MinusBreak{\null + \bigglP}
   \biggl( {31 \over 24} \pi^2 - {73 \over 18} \biggr) Y^2
  +\biggl(  {421 \over 48} \pi^2
           -{5 \over 2} \zeta_3 - {205 \over 54} \biggr) Y
  +{217 \over 72} \zeta_3 + {859 \over 144} \pi^2
  +{30659 \over 1296}
{} \MinusBreak{\null + \bigglP}
   {271 \over 1440} \pi^4
\biggr] {1 \over y}
%%%%%%%%%%%%%% others %%%%%%%%%%
-{\pi^2 \over 4} (1 - x) X^2
-\biggl( 1 - {3 \over 2} x \biggr) Y^3 X
-{1 \over 36} (23 x - 47) Y^3
+{5 \over 2 x y} Y^2
%
%%%%%%%%%% Imaginary part %%%%%%%%
%
{} \PlusBreak{}
i \pi {} \Biggl\{
%%%%%%%%%%%% x/y %%%%%%%%%%
\biggl[
-(7 x+6) (\li3(-x) + \li3(-y))
+(x {} (X - Y) - 6) \li2(-x)
+{53 \over 24} \pi^2 x
{} \PlusBreak{\null + i \pi \BigglP \bigglP \null }
  \biggl( {1 \over 2} x Y + {53 \over 24} x - {1 \over 4} \biggr) X^2
+\biggl( Y^2 - \biggl( {53 \over 12} x - {1 \over 2} \biggr) Y
         +{\pi^2 \over 6} x + {41 \over 12} \biggr) X
{} \PlusBreak{\null + i \pi \BigglP \bigglP \null }
  \biggl( {7 \over 6} x + 1 \biggr) Y^3
+\biggl( {53 \over 24} x - {13 \over 4} \biggr) Y^2
+\biggl( {3 \over 2} \pi^2 x + {\pi^2} - {41 \over 12} \biggr) Y
- {\pi^2 \over 12}
\biggr] {x \over y}
%%%%%%%%%%% 1/y %%%%%%%%%%
{} \PlusBreak{\null + i \pi \BigglP }
\biggl[
   \li3(-x)
  +\li3(-y)
  -\biggl( X - Y + {29 \over 3} \biggr) \li2(-x)
  +{47 \over 24} X^2
  +{5 \over 6} Y^3
{} \MinusBreak{\null + i \pi \BigglP  + \, \bigglP \null }
   {155 \over 24} Y^2
  -\biggl( {59 \over 6} + {65 \over 12} Y - {\pi^2 \over 2} \biggr) X
  +\biggl( {341 \over 36} - {\pi^2 \over 3} \biggr) Y
  +{3 \over 2} \zeta_3
{} \MinusBreak{\null + i \pi \BigglP  + \, \bigglP \null }
   {145 \over 18} + {259 \over 144} \pi^2
\biggr] {1 \over y}
%%%%%%%%%%%%% others %%%%%%%%%%%
+4 x Y^2 X
+{5 \over x y} Y
\Biggr\}
%%%%% end : BB51Paper
\,, \label{BB51}\\[1pt plus 4pt]
%____________________________________________________________________________
%
C^{[1]}_5 & = &
%%%%% begin : CC51Paper
%%%%%%%%%%%%% x/y %%%%%%%%%%%
\biggl[
-6 x {} \biggl( \li4 \biggl( - {x \over y} \biggr) + \li4(-x) + \li4(-y)
\biggr)
+(2 x Y + x X - 3 x - 2) \li3(-x)
{} \PlusBreak{\null \bigglP}
  (8 x Y - 3 x - 5 x X - 2) \li3(-y)
-{1 \over 6} x Y^4
-\biggl( {9 \over 2}
         + \biggl( {7 \over 24} \pi^2 + {9 \over 2} \biggr) x \biggr) Y^2
{} \PlusBreak{\null \bigglP}
  {1 \over 6} x X^4
-\biggl(  x X^2
          +(x Y - 2 - 3 x) X
          -2 x Y^2
          +(2 + 3 x) Y
          -\pi^2 x \biggr) \li2(-x)
{} \MinusBreak{\null \bigglP}
  \biggl( {23 \over 24} x - {3 \over 4} + {7 \over 6} x Y \biggr) X^3
-\biggl(  {1 \over 2} x Y^2
          -\biggl( {13 \over 4} x - {1 \over 2} \biggr) Y
          +{15 \over 4}
          + \biggl( {9 \over 2} - {\pi^2 \over 24} \biggr) x \biggr) X^2
{} \PlusBreak{\null \bigglP}
  \biggl(  {13 \over 6} x Y^3
          +\biggl( {33 \over 4}
                  + \biggl( 9 + {3 \over 4} \pi^2 \biggr) x \biggr) Y
          -{11 \over 24} \pi^2 x + {2 \over 3} \pi^2 - 6 \biggr) X
{} \PlusBreak{\null \bigglP}
  \biggl( 6 + {4 \over 3} \pi^2 x + {11 \over 12} \pi^2 \biggr) Y
+\biggl( -{9 \over 2} \pi^2 + {7 \over 24} \pi^4 \biggr) x
-6 \zeta_3 - {10 \over 3} \pi^2
\biggr] {x \over y}
%%%%%%%%%%%%% 1/y %%%%%%%%%%%%
{} \PlusBreak{}
\biggl[
   10 \li4 \biggl( - {x \over y} \biggr)
  +6 \li4(-x)
  -2 \li4(-y)
  +(1 + 3 X - 6 Y) \li3(-x)
  +{1 \over 24} X^4
{} \PlusBreak{\null + \bigglP}
   (1 - 4 Y + 9 X) \li3(-y)
  +\biggl( -X^2 - (1 - 5 Y) X
           +Y - {13 \over 3} \pi^2 \biggr) \li2(-x)
{} \MinusBreak{\null + \bigglP}
   {3 \over 2} Y^3 X
  +{1 \over 3} Y^4
  -\biggl( {1 \over 3} Y + {13 \over 24} \biggr) X^3
  -\biggl( {5 \over 8} \pi^2 - {3 \over 4} Y - {1 \over 8} - 4 Y^2 \biggr) 
X^2
{} \PlusBreak{\null + \bigglP}
    \biggl( {29 \over 4} - {35 \over 12} \pi^2 \biggr) Y X
  - \biggl( 4 \zeta_3 - 6 + {9 \over 8} \pi^2 \biggr) X
  -\biggl( {29 \over 24} \pi^2 - {3 \over 2} \biggr) Y^2
{} \PlusBreak{\null + \bigglP}
   \biggl( {99 \over 16} + 7 \zeta_3 - {31 \over 6} \pi^2 \biggr) Y
  -{39 \over 4} \zeta_3 + {511 \over 64}
  -{269 \over 48} \pi^2
  +{247 \over 360} \pi^4
\biggr] {1 \over y}
%%%%%%%%%%%%% others %%%%%%%%%%%%
{} \MinusBreak{}
{1 \over 8} (19 - 29 x) Y^2 X
-{1 \over 3} x Y^3
+{3 \over 2 x y} Y^2
%
%%%%%%%%%%% Imaginary part %%%%%%%%%
%
{} \PlusBreak{}
i \pi {} \Biggl\{
%%%%%%%%%%% x/y %%%%%%%%%%%
\biggl[
  3 x {} (\li3(-x) + \li3(-y) - (X - Y)\li2(-x))
-\biggl( {3 \over 2} x Y + {9 \over 8} x - {3 \over 4} \biggr) X^2
+{\pi^2 \over 4}
{} \PlusBreak{\null + i \pi \BigglP \bigglP \null }
  \biggl(  3 x Y^2 - \biggl( {3 \over 2} - {9 \over 4} x \biggr) Y
          -{\pi^2 \over 2} x + {3 \over 4} \biggr) X
-\biggl( {3 \over 2} \pi^2 x + {3 \over 4} \biggr) Y
- {9 \over 8} \pi^2 x
\biggr] {x \over y}
%%%%%%%%%%%% 1/y %%%%%%%%%%%
{} \PlusBreak{\null + i \pi \BigglP }
\biggl[
  -3 \li3(-x)
  +5 \li3(-y)
  +(3 X + 5 Y) \li2(-x)
  -{3 \over 8} X^2
  +4 Y^2 X
{} \PlusBreak{\null + i \pi \BigglP  + \, \bigglP \null }
   {21 \over 4} Y X
  +\biggl( {15 \over 2} - {3 \over 2} \pi^2 \biggr) X
  -\biggl( {4 \over 3} \pi^2 - {41 \over 4} \biggr) Y
  +3 \zeta_3 + {195 \over 16}
  -{9 \over 8} \pi^2
\biggr] {1 \over y}
%%%%%%%%%%%%%% others %%%%%%%%%%
{} \MinusBreak{\null + i \pi \BigglP }
{1 \over 2} (1 - x) Y^3
+{1 \over 8} (9 x - 15) Y^2
+{3 \over x y} Y
\Biggr\}
%%%%% end : CC51Paper
\,, \label{CC51}\\[1pt plus 4pt]
%___________________________________________________________________________
%
D^{[1]}_5 & = &
%%%%% begin : DD51Paper
%%%%%%%%%%%% x/y %%%%%%%%%%%%%
\biggl[
  {1 \over 18} x X^3
-\biggl( {4 \over 9} x + {1 \over 6} \biggr) X^2
-\biggl(  {1 \over 6} x Y^2
          -\biggl( {8 \over 9} x + {2 \over 3} \biggr) Y
          +{8 \over 9}
          -{\pi^2 \over 18} x \biggr) X
+{1 \over 9} x Y^3
{} \MinusBreak{\null \bigglP}
  \biggl( {4 \over 9} x + {1 \over 2} \biggr) Y^2
+{1 \over 9} (\pi^2 x + 8) Y
-{5 \over 18} \pi^2 - {4 \over 9} \pi^2 x
\biggr] {x \over y}
%%%%%%%%%%%% 1/y %%%%%%%%%%%%%
{} \PlusBreak{}
\biggl[
  -{1 \over 3} \li3(-x)
  -{1 \over 3} \li3(-y)
  +{1 \over 3} (X - Y) \li2(-x)
  +{1 \over 18} X^3
  +\biggl( {1 \over 6} Y - {19 \over 36} \biggr) X^2
{} \PlusBreak{\null + \bigglP}
   \biggl(  {7 \over 27} + {\pi^2 \over 8}
           +{7 \over 18} Y - {1 \over 2} Y^2 \biggr) X
  +{1 \over 6} Y^3 - {13 \over 12} Y^2
  +\biggl( {193 \over 54} + {2 \over 9} \pi^2 \biggr) Y
{} \PlusBreak{\null + \bigglP}
   {67 \over 72} \pi^2 - {455 \over 108}
  -{37 \over 36} \zeta_3
\biggr] {1 \over y}
%
%%%%%%%%%%% Imaginary part %%%%%%%%%
%
{} \PlusBreak{}
i \pi {} \Biggl\{
%%%%%%%%%%% x/y %%%%%%%%%%
\biggl[
  {1 \over 6} x X^2
+{1 \over 3} (1 - x Y) X
-{1 \over 3} Y + {1 \over 6} x Y^2
+{\pi^2 \over 6} x
\biggr] {x \over y}
%%%%%%%%%%%% 1/y %%%%%%%%%%%%
{} \PlusBreak{\null + i \pi \BigglP }
\biggl[
   {1 \over 6} X^2
  -{1 \over 3} (2 + Y) X
  -{16 \over 9} Y + {\pi^2 \over 8}
  +{1 \over 6} Y^2 + {23 \over 6}
\biggr] {1 \over y}
\Biggr\}
%%%%% end : DD51Paper
\,, \label{DD51}\\[1pt plus 4pt]
%_____________________________________________________________________________
%
E^{[1]}_5 & = &
%%%%% begin : EE51Paper
%%%%%%%%%%%%% x/y %%%%%%%%%%%%
\biggl[
-{1 \over 9} x X^3
+\biggl( {8 \over 9} x + {1 \over 3} \biggr) X^2
+\biggl(  {1 \over 3} x Y^2
          -\biggl( {4 \over 3} + {16 \over 9} x \biggr) Y
          +{16 \over 9} - {\pi^2 \over 9} x \biggr) X
{} \MinusBreak{\null \bigglP}
  {1 \over 9} (2 \pi^2 x + 16) Y
+{5 \over 9} \pi^2 + {8 \over 9} \pi^2 x
\biggr] {x \over y}
%%%%%%%%%%%% 1/y %%%%%%%%%%%
{} \PlusBreak{}
\biggl[
   {2 \over 3} \li3(-x)
  +{4 \over 3} \li3(-y)
  -{2 \over 3} (X - 2 Y) \li2(-x)
  -{1 \over 9} X^3
  +\biggl( {19 \over 18} - {1 \over 3} Y \biggr) X^2
{} \MinusBreak{\null + \bigglP}
   \biggl(  {\pi^2 \over 4} + {7 \over 9} Y
           -{4 \over 3} Y^2 + {14 \over 27} \biggr) X
  -\biggl( {13 \over 8} \pi^2 - {11 \over 27} \biggr) Y
  -{685 \over 162} - {59 \over 72} \pi^2
{} \MinusBreak{\null + \bigglP}
   {59 \over 36} \zeta_3
\biggr] {1 \over y}
%%%%%%%%%%%% others %%%%%%%%%%
-{2 \over 9} (1 - x) Y^3
-{1 \over 9}(1 + 8 x) Y^2
%
%%%%%%%%%%% Imaginary part %%%%%%%%%%
%
{} \PlusBreak{}
i \pi {} \Biggl\{
%%%%%%%%%% x/y %%%%%%%%%%%
\biggl[
-{1 \over 3} x X^2
-{2 \over 3} (1 - x Y) X
+{2 \over 3} Y - {\pi^2 \over 3} x
-{1 \over 3} x Y^2
\biggr] {x \over y}
%%%%%%%%%% 1/y %%%%%%%%%%
{} \PlusBreak{\null + i \pi \BigglP }
\biggl[
   {2 \over 3} \li2(-x)
  -{1 \over 3} X^2
  +{2 \over 3} (2 + Y) X
  -{1 \over 9} - {23 \over 72} \pi^2
  -{5 \over 9} Y + {4 \over 3} Y^2
\biggr] {1 \over y}
\Biggr\}
%%%%% end : EE51Paper
\,, \label{EE51}\\[1pt plus 4pt]
%____________________________________________________________________________
%
F^{[1]}_5 & = &
%%%%% begin : FF51Paper
\biggl[
  {1 \over 9} Y^2
-{10 \over 27} Y
-{\pi^2 \over 9}
+{25 \over 81}
\biggr] {1 \over y}
%%%%%%%%% Imaginary part %%%%%%%%%
+i \pi {} \biggl[
{2 \over 9} Y - {10 \over 27}
\biggr] {1 \over y}
%%%%% end : FF51Paper
\,, \label{FF51} %\\[1pt plus 4pt]
%_____________________________________________________________________________
%
\end{eqnarray}

For $h=5$ in \eqn{h5} and color factor $\trc^{[2]}$ in \eqn{basis56}:
\begin{eqnarray}
G^{[2]}_5 & = &
%%%%%%%%%%%%% x/y %%%%%%%%%%%
%%%%% begin : GG52Paper
\biggl[
-(5 x + 6) \li4 \biggl( - {x \over y} \biggr)
-(8 x + 6) (\li4(-x) + \li4(-y))
-\biggl( {5 \over 12} x + {1 \over 2} \biggr) Y^4
{} \PlusBreak{\null \bigglP}
  \biggl(  2 x X + (6 + 5 x) Y
          + {1 \over 2} x - 2 \biggr) \li3(-x)
+\biggl( (7 x + 6) Y + 4 + {1 \over 2} x \biggr) \li3(-y)
{} \MinusBreak{\null \bigglP}
  \biggl(  {1 \over 2} x X^2
          -\biggl( -{1 \over 2} x + 2 \biggr) X
          -{1 \over 2} x Y^2
          -\biggl( 4 + {1 \over 2} x \biggr) Y
          -\pi^2 - {5 \over 6} \pi^2 x \biggr) \li2(-x)
{} \MinusBreak{\null \bigglP}
  {1 \over 6} x X^3 Y
% +2 (1 + x) Y^3 X
+\biggl( {1 \over 4} x Y^2 + \biggl( {1 \over 8} x + {3 \over 4} \biggr) Y
         -1 + {\pi^2 \over 8} - {5 \over 4} x \biggr) X^2
{} \PlusBreak{\null \bigglP}
  \biggl( {3 \over 2} - {1 \over 4} x \biggr) Y^2 X
+\biggl( {15 \over 4}
         -\biggl( {\pi^2 \over 4} - {5 \over 2} \biggr) x \biggr) Y X
- {\pi^2 \over 12} (x - 5) X
{} \PlusBreak{\null \bigglP}
  \biggl( {13 \over 12} + {7 \over 24} x \biggr) Y^3
-\biggl( {11 \over 4}
         +\biggl( {5 \over 4} - {\pi^2 \over 24} \biggr) x \biggr) Y^2
-\biggl( {7 \over 6} \pi^2 - {\pi^2 \over 8} x \biggr) Y
{} \PlusBreak{\null \bigglP}
  \biggl( {17 \over 40} \pi^4 - {5 \over 4} \pi^2 \biggr) x
-{29 \over 12} \pi^2 + {\pi^4 \over 4}
-5 \zeta_3
\biggr] {x \over y}
-2 x Y^3 X
%%%%%%%%%%%%% 1/y %%%%%%%%%%%%
{} \PlusBreak{}
\biggl[
   \li4 \biggl( - {x \over y} \biggr)
  -2 \li4(-x)
  +3 \li4(-y)
  +(2 X - 1) \li3(-x)
  -{1 \over 8} Y^4
  +{1 \over 12} Y X^3
{} \PlusBreak{\null + \bigglP}
   \biggl( {25 \over 3} - 2 Y + X \biggr) \li3(-y)
  +\biggl( {25 \over 3} Y + X
          -{\pi^2 \over 2} - {1 \over 2} X^2 \biggr) \li2(-x)
{} \PlusBreak{\null + \bigglP}
   {113 \over 72} Y^3
  +\biggl( {3 \over 8} Y^2 - {\pi^2 \over 4} - {1 \over 8} Y \biggr) X^2
  -{5 \over 12} Y^3 X + {17 \over 3} Y^2 X
  +\biggl( {\pi^2 \over 4} - \zeta_3 \biggr) X
{} \PlusBreak{\null + \bigglP}
   \biggl( {19 \over 8} - {11 \over 12} \pi^2 \biggr) Y X
  -\biggl( {67 \over 12} - {3 \over 4} \pi^2 \biggr) Y^2
  +\biggl(  {1513 \over 432} - {407 \over 72} \pi^2
           + 4 \zeta_3 \biggr) Y
{} \MinusBreak{\null + \bigglP}
   {41 \over 24} \pi^2 + {223 \over 1440} \pi^4
  -{23213 \over 5184} - {665 \over 72} \zeta_3
\biggr] {1 \over y}
%%%%%%%%%%%%%% others %%%%%%%%%%%
-\biggl( {\pi^2 \over 8} x - {\pi^2 \over 4} \biggr) X^2
-{3 \over 2 x y} Y^2
%
%%%%%%%% Imaginary part %%%%%%%%%%%%
%
{} \PlusBreak{}
i \pi {} \Biggl\{
%%%%%%%%%%% x/y %%%%%%%%%%%%%
\biggl[
   (7 x+6) (\li3(-x) + \li3(-y))
  -(x X - 6 - x Y) \li2(-x)
  -{1 \over 6} x X^3
{} \PlusBreak{\null + i \pi \BigglP \bigglP \null }
   \biggl( \biggl( {1 \over 2} - {3 \over 4} x \biggr) Y
          -{\pi^2 \over 3} x + {7 \over 4} \biggr) X
%  -(1 + x) Y^3
  +\biggl( {3 \over 8} x + {11 \over 4} \biggr) Y^2
{} \MinusBreak{\null + i \pi \BigglP \bigglP \null }
   \biggl( \pi^2 + {7 \over 4} + {4 \over 3} \pi^2 x \biggr) Y
  +{3 \over 8} \pi^2 x - {\pi^2 \over 12}
\biggr] {x \over y}
-{3 \over x y} Y + x Y^3
%%%%%%%%%%%% 1/y %%%%%%%%%%%%
{} \PlusBreak{\null + i \pi \BigglP }
\biggl[
   2 \li3(-x)
  -\li3(-y)
  +\biggl( {28 \over 3} - X \biggr) \li2(-x)
  +{1 \over 12} X^3 + {1 \over 2} X^2 Y
{} \MinusBreak{\null + i \pi \BigglP  + \, \bigglP \null }
   {7 \over 12} Y^3
  +\biggl( {19 \over 8} - {3 \over 4} \pi^2 + {11 \over 4} Y \biggr) X
  + {149 \over 24} Y^2
  -\biggl( {211 \over 24} - {\pi^2 \over 4} \biggr) Y
{} \MinusBreak{\null + i \pi \BigglP  + \, \bigglP \null }
   {37 \over 72} \pi^2 + {1513 \over 432}
  +3 \zeta_3
\biggr] {1 \over y}
%%%%%%%%%%%%%%%% others %%%%%%%%%
+ {1 \over 8} (5 - 3 x) X^2
+{1 \over 2} (1 - 7 x) Y^2 X
\Biggr\}
%%%%% end : GG52Paper
\,, \label{GG52}\\[1pt plus 4pt]
%___________________________________________________________________________
%
H^{[2]}_5 & = &
%%%%% begin : HH52Paper
%%%%%%%%%% x/y %%%%%%%%%%
\biggl[
  {3 \over 2} \li4 \biggl( - {x \over y} \biggr) x
+4 x {} (\li4(-x) + \li4(-y))
+\biggl( {5 \over 2} + {1 \over 2} x - 2 x X \biggr) \li3(-x)
{} \PlusBreak{\null \bigglP}
  \biggl( {1 \over 2} x + 2 x X + {5 \over 2} - 4 x Y \biggr) \li3(-y)
-{7 \over 6} x Y^3 X
-\biggl( {1 \over 12} - {1 \over 4} x Y + {11 \over 36} x \biggr) X^3
{} \PlusBreak{\null \bigglP}
  \biggl(  {3 \over 4} x X^2
          -{1 \over 2} (x + 5 - x Y) X
          -{5 \over 4} x Y^2
          +{1 \over 2} (5 + x) Y
          -{\pi^2 \over 4} x \biggr) \li2(-x)
{} \PlusBreak{\null \bigglP}
  {1 \over 48} x X^4
+{1 \over 16} x Y^4
+\biggl(  {1 \over 2} x Y^2
          -\biggl( 1 + {1 \over 4} x \biggr) Y + {17 \over 12}
          -\biggl( {\pi^2 \over 24} - {127 \over 36} \biggr) x \biggr) X^2
{} \PlusBreak{\null \bigglP}
  \biggl( {9 \over 4} + {17 \over 12} x \biggr) Y^2 X
-\biggl( {20 \over 3}
         +\biggl( {127 \over 18}
                 +{\pi^2 \over 4} \biggr) x \biggr) Y X
+\biggl( {127 \over 36} \pi^2 - {37 \over 144} \pi^4 \biggr) x
{} \MinusBreak{\null \bigglP}
   \biggl(  {7 \over 18} \pi^2 x - {32 \over 9}
           +{7 \over 12} \pi^2 \biggr) X
+\biggl( {21 \over 4}
         +\biggl( {\pi^2 \over 6} + {127 \over 36} \biggr) x \biggr) Y^2
{} \MinusBreak{\null \bigglP}
  \biggl( {13 \over 12} \pi^2 + {32 \over 9} + {31 \over 36} \pi^2 x \biggr) 
Y
+{53 \over 18} \pi^2 + {3 \over 2} \zeta_3
\biggr] {x \over y}
+{1 \over 36} (25 x - 13) Y^3
-{2 \over x y} Y^2
%%%%%%%%%%%% 1/y %%%%%%%%%%
{} \PlusBreak{}
\biggl[
  -{9 \over 2} \li4 \biggl( - {x \over y} \biggr)
  +\biggl( {7 \over 3} - 3 X + 2 Y \biggr) \li3(-x)
  +\biggl( {8 \over 3} + Y - 4 X \biggr) \li3(-y)
{} \PlusBreak{\null + \bigglP}
   \biggl(  {3 \over 4} X^2
           -\biggl( {3 \over 2} Y + {7 \over 3} \biggr) X
           +{8 \over 3} Y - {5 \over 4} Y^2
           +{11 \over 4} \pi^2 \biggr) \li2(-x)
  +{1 \over 48} X^4
{} \MinusBreak{\null + \bigglP}
   \biggl( {7 \over 18} + {1 \over 6} Y \biggr) X^3
  +\biggl( {3 \over 4} \pi^2 - {1 \over 6} Y
          +{125 \over 72} - {11 \over 8} Y^2 \biggr) X^2
  -{1 \over 12} Y^3 X + {1 \over 6} Y^2 X
{} \MinusBreak{\null + \bigglP}
   \biggl( {337 \over 72} - {7 \over 4} \pi^2 \biggr) Y X
  +\biggl(  {17 \over 27} + {3 \over 2} \zeta_3
           -{11 \over 16} \pi^2 \biggr) X
  -{3 \over 16} Y^4
  +\biggl( {9 \over 8} \pi^2 - {41 \over 18} \biggr) Y^2
{} \PlusBreak{\null + \bigglP}
   \biggl( {17 \over 16} \pi^2 - {5 \over 2} \zeta_3
          + {235 \over 54} \biggr) Y
  -{661 \over 1440} \pi^4 + {61 \over 48} \pi^2
  -{30659 \over 1296} + {527 \over 72} \zeta_3
\biggr] {1 \over y}
%
%%%%%%%%%%% Imaginary part %%%%%%%%%%%%%%%
%
{} \PlusBreak{}
i \pi {} \Biggl\{
%%%%%%%%%%%% x/y %%%%%%%%%%%
\biggl[
-2 x {} (\li3(-x) + \li3(-y) - (X - Y) \li2(-x))
+{1 \over 6} x X^3
-{11 \over 12} x X^2
{} \PlusBreak{\null + i \pi \BigglP \bigglP \null }
  \biggl(  {\pi^2 \over 2} x - {3 \over 2} x Y^2
          +{11 \over 6} x Y - {23 \over 6} \biggr) X
+{1 \over 6} x Y^3 - {11 \over 12} x Y^2
{} \PlusBreak{\null + i \pi \BigglP \bigglP \null }
  {1 \over 6} (5 \pi^2 x + 23) Y
-{11 \over 12} \pi^2 x
\biggr] {x \over y}
%%%%%%%%%%%% 1/y %%%%%%%%%%%%
{} \PlusBreak{\null + i \pi \BigglP }
\biggl[
  -\li3(-x)
  -3 \li3(-y)
  +\biggl( {1 \over 3} - 4 Y \biggr) \li2(-x)
  -{1 \over 12} X^3  -{7 \over 12} Y^3 - {1 \over 6} X^2
{} \MinusBreak{\null + i \pi \BigglP  + \, \bigglP \null }
   \biggl( 2 Y^2 - {5 \over 4} \pi^2 + {29 \over 24} + {8 \over 3} Y \biggr) 
X
  - {1 \over 12} Y^2
  -\biggl( {665 \over 72} - {11 \over 12} \pi^2 \biggr) Y
  -{\pi^2 \over 72}
{} \MinusBreak{\null + i \pi \BigglP  + \, \bigglP \null }
   \zeta_3
  +{269 \over 54}
\biggr] {1 \over y}
%%%%%%%%%%%%% others %%%%%%%%%
-{1 \over 2} (x - 1) Y X^2
-{4\over x y} Y
\Biggr\}
%%%%% end : HH52Paper
\,, \label{HH52}\\[1pt plus 4pt]
%____________________________________________________________________________
%
I^{[2]}_5 &=&
%%%%% begin : II52Paper
%%%%%%%%%%%% x/y %%%%%%%%%%%%%
\biggl[
    {5 \over 2} x \li4 \biggl( -{x \over y} \biggr)
  + 2 x \li4(-x)
  + 2 x \li4(-y)
  -{1 \over 16} x X^4
  + \biggl( x - x Y + {1 \over 2} \biggr) \li3(-x)
{} \PlusBreak{\null \bigglP}
    \biggl( 2 X x - 3 x Y + {1 \over 2} + x \biggr) \li3(-y)
  + \biggl( {5 \over 12} x Y - {1 \over 6} + {3 \over 8} x \biggr) X^3
  + {1 \over 16} x Y^4
{} \PlusBreak{\null \bigglP}
    \biggl( {1 \over 4} x X^2
           -\biggl( x + {1 \over 2} - {1 \over 2} x Y \biggr) X
           -{3 \over 4} x Y^2 + \biggl( {1 \over 2} + x \biggr) Y
           -{5 \over 12} \pi^2 x
    \biggr) \li2(-x)
{} \PlusBreak{\null \bigglP}
   \biggl( 2 x - {5 \over 4} x Y + {7 \over 4}
          + {1 \over 4} x Y^2
   \biggr) X^2
  -{5 \over 6} x Y^3 X
  -\biggl( {15 \over 4} + \biggl( {\pi^2 \over 3} + 4 \biggr) x \biggr) Y X
{} \PlusBreak{\null \bigglP}
    \biggl( {5 \over 24} \pi^2 x - {\pi^2 \over 12} + 3 \biggr) X
  + \biggl( 2 + \biggl( 2 + {\pi^2 \over 8} \biggr) x \biggr) Y^2
-\biggl( 3 + {\pi^2 \over 3} + {\pi^2 \over 2} x \biggr) Y
{} \PlusBreak{\null \bigglP}
    \biggl( 2 \pi^2 - {59 \over 720} \pi^4 \biggr) x
  + {3 \over 2} \pi^2 + {5 \over 2} \zeta_3
\biggr] {x \over y}
-{1 \over 8} (11 x - 7) Y^2 X
+ {1 \over 6} x Y^3
%%%%%%%%%%%%%%% 1/y %%%%%%%%%%%%%
{} \PlusBreak{}
\biggl[
  -{7 \over 2} \li4 \biggl( -{x \over y} \biggr)
  -2 \li4(-x)
   + \li4(-y)
  -{1 \over 48} X^4
  -\biggl( {1 \over 2} - 2 Y + X \biggr) \li3(-x)
{} \MinusBreak{\null + \bigglP}
     \biggl( 3 X + {1 \over 2} - Y \biggr) \li3(-y)
   -{5 \over 48} Y^4
   + \biggl( {\pi^2 \over 6} - {5 \over 4} Y^2
            + {5 \over 8}-{1 \over 4} Y
     \biggr) X^2
{} \PlusBreak{\null + \bigglP}
       \biggl( {1 \over 4} X^2
      -{1 \over 2} (3 Y-1) X
      -{1 \over 4} Y^2-{1 \over 2} Y
     + {19 \over 12} \pi^2 \biggr) \li2(-x)
   + \biggl( {1 \over 12} Y + {5 \over 24} \biggr) X^3
{} \PlusBreak{\null + \bigglP}
     \biggl( {1 \over 3} Y^3
            + \biggl( \pi^2 - {15 \over 4} \biggr) Y
            + \zeta_3 - 3 + {11 \over 24} \pi^2
     \biggr) X
  -\biggl( {1 \over 2} - {11 \over 24} \pi^2 \biggr) Y^2
{} \MinusBreak{\null + \bigglP}
     \biggl( 4 \zeta_3 + {51 \over 16}
    -{23 \over 12} \pi^2 \biggr) Y
   + {25 \over 4} \zeta_3 + {97 \over 48} \pi^2
  -{163 \over 720} \pi^4 - {511 \over 64}
\biggr]{1 \over y}
-{1 \over 2 x y} Y^2
%
%%%%%%%%%%% Imaginary part %%%%%%%%%%%%
%
{} \PlusBreak{}
  i \pi {} \Biggl\{
\biggl[
  - x \li3(-x)
  - x \li3(-y)
  + x {}(X-Y) \li2(-x)
  + \biggl( {3 \over 8} x + {1 \over 2} x Y - {1 \over 4} \biggr) X^2
{} \MinusBreak{\null + i \pi \BigglP \bigglP \null }
   \biggl( x Y^2
           - \biggl( {1 \over 2} - {3 \over 4} x \biggr) Y
           + {1 \over 4} - {\pi ^2 \over 6}  x
   \biggr) X
  + \biggl( {1 \over 4} + {\pi^2 \over 2} x \biggr) Y
-{\pi^2 \over 12} + {3 \over 8} \pi^2 x
\biggr] {x \over y}
{} \PlusBreak{\null + i \pi \BigglP }
\biggl[
   \li3(-x)
  -2 \li3(-y)
  -(X + 2 Y) \li2(-x)
   + {1 \over 8} X^2
{} \MinusBreak{\null + i \pi \BigglP  + \, \bigglP \null }
   \biggl( {5 \over 2}-{\pi^2 \over 2}
          + {7 \over 4} Y + {3 \over 2} Y^2
   \biggr) X
  -\biggl( {19 \over 4} - {\pi^2 \over 2} \biggr) Y
  -{99 \over 16} - 3 \zeta_3
   + {13 \over 24} \pi^2
\biggr] {1 \over y}
{} \PlusBreak{\null + i \pi \BigglP }
{1 \over 6} (1 - x) Y^3
-{1 \over 8} (3 x - 5) Y^2
-{1 \over x y} Y
\Biggr\}
%%%%% end : II52Paper
\,, \label{II52} \\[1pt plus 4pt]
%_____________________________________________________________________________
%
J^{[2]}_5 & = &
%%%%% begin : JJ52Paper
\biggl[
-{1 \over 3} (\li3(-y) + Y \li2(-x))
-{1 \over 6} Y^2 X - {5 \over 18} Y^3
+{37 \over 36} Y^2
+\biggl( {11 \over 18} \pi^2 - {145 \over 54} \biggr) Y
{} \MinusBreak{\null \bigglP}
  {41 \over 72} \pi^2 + {49 \over 36} \zeta_3
+{455 \over 108}
\biggr] {1 \over y}
%%%%%%%%%%%%%%%%%% Imaginary part %%%%%%%%%%%%%
+i {\pi \over 3 y} {} \biggl\{
{\pi^2 \over 6}
-{5 \over 2} Y^2 + {37 \over 6} Y
- \li2(-x)
-{145 \over 18}
\biggr\}
%%%%% end : JJ52Paper
\,, \label{JJ52}\\[1pt plus 4pt]
%___________________________________________________________________________
%
K^{[2]}_5 & = &
%%%%% begin : KK52Paper
%%%%%%%%%%%% x/y %%%%%%%%%%%%
\biggl[
  {1 \over 18} x X^3
-\biggl( {1 \over 6} + {4 \over 9} x \biggr) X^2
-\biggl(  {1 \over 6} x Y^2
          -\biggl( {2 \over 3} + {8 \over 9} x \biggr) Y
          -{\pi^2 \over 18} x
          +{8 \over 9} \biggr) X
{} \PlusBreak{\null \bigglP}
  {1 \over 9} (8 + \pi^2 x) Y
-{5 \over 18} \pi^2
-{4 \over 9} \pi^2 x
\biggl] {x \over y}
%%%%%%%%%%% 1/y %%%%%%%%%%%%%%
{} \PlusBreak{}
\biggl[
  -{1 \over 3} \li3(-x)
  -{2 \over 3} \li3(-y)
  -{1 \over 3} (2 Y - X) \li2(-x)
  +{1 \over 18} X^3
  -\biggl( {19 \over 36} - {1 \over 6} Y \biggr) X^2
{} \PlusBreak{\null + \bigglP}
   \biggl(  {7 \over 27} + {\pi^2 \over 8}
           +{7 \over 18} Y - {2 \over 3} Y^2 \biggr) X
  +\biggl( {19 \over 24} \pi^2 - {35 \over 27} \biggr) Y
  +{11 \over 24} \pi^2 + {47 \over 36} \zeta_3
{} \PlusBreak{\null + \bigglP}
   {685 \over 162}
\biggr] {1 \over y}
%%%%%%%%%%%%%% others %%%%%%%%%%%
+{1 \over 9} (1 - x) Y^3
+\biggl( {1 \over 18} + {4 \over 9} x \biggr) Y^2
%
%%%%%%%%%%%% Imaginary part %%%%%%%%%
%
{} \PlusBreak{}
i \pi {} \Biggl\{
%%%%%%%%%%% x/y %%%%%%%%%%%
\biggl[
  {1 \over 6} x X^2
-{1 \over 3} (x Y - 1) X
+{\pi^2 \over 6} x + {1 \over 6} x Y^2
-{1 \over 3} Y
\biggr] {x \over y}
%%%%%%%%%%%% 1/y %%%%%%%%%%%%%
{} \MinusBreak{\null + i \pi \BigglP }
\biggl[
   \li2(-x)
  -{1 \over 2} X^2
  +(Y + 2) X
  +2 Y^2 + {28 \over 9}
  -{5 \over 12} \pi^2 - {5 \over 6} Y
\biggr] {1 \over 3 y}
%%%%%%%%%%%%% others %%%%%%%%%%
\Biggr\}
%%%%% end : KK52Paper
\,, \label{KK52}\\[1pt plus 4pt]
%___________________________________________________________________________
%
L^{[2]}_5 & = &
%%%%% begin : LL52Paper
\biggl[
  {\pi^2 \over 9}
-{1 \over 9} Y^2
+{10 \over 27} Y
-{25 \over 81}
\biggr] {1 \over y}
%%%%%%%% Imaginary part %%%%%%%%%5
+i \pi {} \biggl[
-{2 \over 9} Y + {10 \over 27}
\biggr] {1 \over y}
%%%%% end : LL52Paper
\,, \label{LL52} %\\[1pt plus 4pt]
%___________________________________________________________________________
%
\end{eqnarray}

For $h=6$ in \eqn{h6} and color factor $\trc^{[1]}$ in \eqn{basis56}:
\begin{eqnarray}
A^{[1]}_6 & = &
%%%%% begin : AA61Paper
%%%%%%%%%%% x/y %%%%%%%%%%%%
\biggl[
  \li4 \biggl( - {x \over y} \biggr)
-\biggl( Y - X - {29 \over 6} \biggr) (\li3(-x) + \li3(-y))
+{1 \over 12} X^4 + {1 \over 6} Y^4 - Y^3 X
{} \MinusBreak{\null \bigglP}
  \biggl(  {1 \over 2} X^2 + \biggl( {29 \over 6} - Y \biggr) X
          +{1 \over 2} Y^2 + {\pi^2 \over 6}
          -{29 \over 6} Y \biggr) \li2(-x)
-\biggl( {1 \over 2} Y + {49 \over 36} \biggr) X^3
{} \MinusBreak{\null \bigglP}
  \biggl(  {1 \over 6} Y - {5 \over 4} Y^2
          -{\pi^2 \over 12} - {56 \over 9} - {3 \over 2} x \biggr) X^2
+{53 \over 12} Y^2 X
-\biggl( {125 \over 18} + {\pi^2 \over 2} + 3 x \biggr) Y X
{} \MinusBreak{\null \bigglP}
  \biggl(  {79 \over 27}
          +{287 \over 144} \pi^2
          +{3 \over 2} \zeta_3 \biggr) X
- {23 \over 18} Y^3
+\biggl( {3 \over 2} x + {\pi^2 \over 3} + {49 \over 12} \biggr) Y^2
{} \MinusBreak{\null \bigglP}
  \biggl( {17 \over 6} \pi^2 + {1513 \over 432} + 2 \zeta_3 \biggr) Y
+{23 \over 36} \pi^2 + {23213 \over 5184}
+{113 \over 1440} \pi^4 + {65 \over 72} \zeta_3
+{3 \over 2} \pi^2 x
\biggr] {x \over y}
%%%%%%%%%%% 1/y %%%%%%%%%%%%%
{} \PlusBreak{}
\biggl[
  -3 \biggl( \li4 \biggl( - {x \over y} \biggr)
            + \li4(-x) + \li4(-y) - \li3(-x) + (X - Y - 1) \li3(-y) \biggr)
{} \PlusBreak{\null + \bigglP}
   \biggl( 3 Y - 3 X + {\pi^2 \over 2} \biggr) \li2(-x)
  -{3 \over 4} (2 Y + Y^2) X^2
  -{1 \over 2} Y^3 + {\pi^2 \over 4} Y^2
{} \PlusBreak{\null + \bigglP}
   \biggl( 3 Y^2 + 3 \zeta_3
          +\pi^2 Y - {\pi^2 \over 2}
          +{1 \over 2} Y^3 \biggr) X
  -\biggl( {3 \over 2} \pi^2 + 3 \zeta_3 \biggr) Y
  +{\pi^4 \over 30} + {\pi^2 \over 2}
\biggr] {1 \over y}
%%%%%%%%%%%%%%%%%% 1/x/y %%%%%%%%%%%%%%
{} \PlusBreak{}
\biggl[
  -3 \li4 \biggl( - {x \over y} \biggr)
  -3 \li4(-x)
  -3 \li4(-y)
  -3 (X - Y) \li3(-y)
  +{\pi^2 \over 2} \li2(-x)
{} \MinusBreak{\null + \bigglP}
   {3 \over 4} Y^2 X^2
  +\biggl( \pi^2 Y + {1 \over 2} Y^3 + 3 \zeta_3 \biggr) X
  +{\pi^2 \over 4} Y^2 - 3 Y \zeta_3
  +{\pi^4 \over 30}
\biggr] {1 \over x y}
%
%%%%%%%%%%%%% Imaginary part %%%%%%%%%%%%
%
{} \PlusBreak{}
i \pi {} \Biggl\{
-{11 \over 6} (X^2 + Y^2)
+\biggl( {11 \over 2} + {11 \over 3} Y \biggr) X
+{11 \over 9} Y - {7 \over 2} \zeta_3
-{2777 \over 432}
-{77 \over 48} \pi^2
\Biggr\} {x \over y}
%%%%% end : AA61Paper
\,, \label{AA61}\\[1pt plus 4pt]
%____________________________________________________________________________
%
B^{[1]}_6 & = &
%%%%% begin : BB61Paper
%%%%%%%%%%%%% x/y %%%%%%%%%%%%%%%
\biggl[
-\li4 \biggl( - {x \over y} \biggr)
+2 \li4(-x)
+4 \li4(-y)
-{1 \over 4} X^4
-{1 \over 8} Y^4
-\biggl( 3 X + {29 \over 3} - Y \biggr) \li3(-x)
{} \MinusBreak{\null \bigglP}
  \biggl( 2 X + Y + {37 \over 3} \biggr) \li3(-y)
+\biggl( {3 \over 2} Y + {125 \over 36} \biggr) X^3
+\biggl(  {4 \over 3} Y^3 - {229 \over 24} Y^2 \biggr) X
{} \PlusBreak{\null \bigglP}
  \biggl(  {3 \over 2} X^2
          -\biggl( 2 Y - {29 \over 3} \biggr) X
          -{\pi^2 \over 6} - {37 \over 3} Y
          +{1 \over 2} Y^2 \biggr) \li2(-x)
{} \MinusBreak{\null \bigglP}
  \biggl( 3 Y^2 + {7 \over 6} Y + {7 \over 2} x + {869 \over 72} \biggr) X^2
-\biggl( {19 \over 18} + {7 \over 2} x + {25 \over 24} \pi^2 \biggr) Y^2
{} \PlusBreak{\null \bigglP}
  \biggl( \biggl( {4 \over 3} \pi^2 + 7 x + {383 \over 36} \biggr) Y
          -{4 \over 27} + {329 \over 72} \pi^2 + 3 \zeta_3 \biggr) X
{} \MinusBreak{\null \bigglP}
  \biggl(  {811 \over 54} - {1267 \over 144} \pi^2
          -{3 \over 2} \zeta_3 \biggr) Y
+{30659 \over 1296} - {83 \over 48} \pi^2
+{37 \over 72} \zeta_3 - {7 \over 2} \pi^2 x
-{13 \over 32} \pi^4
\biggr] {x \over y}
%%%%%%%%%%%%%% 1/y %%%%%%%%%%%%%%
{} \PlusBreak{}
\biggl[
   6 \biggl( \li4 \biggl( - {x \over y} \biggr)
            + \li4(-x) + \li4(-y) - \li3(-x) + (X - 1 - Y) \li3(-y) \biggr)
{} \PlusBreak{\null + \bigglP}
   (6 X - \pi^2 - 6 Y) \li2(-x)
  +\biggl( {3 \over 2} Y^2 + 3 Y \biggr) X^2
  -Y^3 X - {23 \over 4} Y^2 X
{} \MinusBreak{\null + \bigglP}
   \biggl( {1 \over 4} + 2 \pi^2 \biggr) Y X
  -{1 \over 2} (\pi^2 - 9) Y^2
  -\biggl( {64 \over 9} - 6 \zeta_3 - {15 \over 4} \pi^2 \biggr) Y
{} \PlusBreak{\null + \bigglP}
   \biggl( {5 \over 6} \pi^2 - 6 \zeta_3 \biggr) X
  -{91 \over 36} \pi^2 + 4 \zeta_3
  -{\pi^4 \over 15}
\biggr] {1 \over y}
%%%%%%%%%%%%%%%% others %%%%%%%%%%%%%
+{1 \over 36} \biggl( {29 \over x} - 47 \biggr) Y^3
{} \PlusBreak{}
\biggl[
     4 \li4 \biggl( - {x \over y} \biggr)
    +4 \li4(-x)
    +2 \li4(-y)
    +2 Y \li3(-x)
    -(2 Y + 1 - 5 X) \li3(-y)
{} \MinusBreak{\null + \bigglP}
     \biggl( Y - Y^2 + X Y + {\pi^2 \over 3} \biggr) \li2(-x)
    + Y^2 X^2
    -{5 \over 3} \pi^2 Y X - 5 \zeta_3 X
    +{1 \over 6} Y^3 X
{} \MinusBreak{\null + \bigglP}
     {7 \over 8} Y^2 X
    +\biggl( {73 \over 18} - {23 \over 24} \pi^2 \biggr) Y^2
    +\biggl( 2 \zeta_3 + {85 \over 36} \pi^2 \biggr) Y
    -{\pi^4 \over 45} + \zeta_3
\biggr] {1 \over x y}
%
%%%%%%%%%%% Imaginary part %%%%%%%%%%%%%%
%
{} \PlusBreak{}
i \pi {} \Biggl\{
%%%%%%%%%%% x/y %%%%%%%%%%%%
\biggl[
-2 \li3(-x)
-3 \li3(-y)
-\biggl( {8 \over 3} + Y - X \biggr) \li2(-x)
+\biggl( {1 \over 2} Y + {53 \over 12} \biggr) X^2
{} \MinusBreak{\null + i \pi \BigglP \bigglP \null }
  \biggl( {27 \over 2} + 2 Y^2 + {109 \over 12} Y
         -{2 \over 3} \pi^2 \biggr) X
+{5 \over 6} Y^3 + {13 \over 24} Y^2
+\biggl( {3 \over 4} \pi^2 + {307 \over 36} \biggr) Y
{} \MinusBreak{\null + i \pi \BigglP \bigglP \null }
  {91 \over 6} + {9 \over 2} \zeta_3 + {601 \over 144} \pi^2
\biggr] {x \over y}
%%%%%%%%%%%%%%% 1/y %%%%%%%%%%%%
+\biggl[
  \biggl( {1 \over 2} Y - {1 \over 4} \biggr) X
  -{5 \over 4} Y^2 - {64 \over 9}
  +{35 \over 4} Y + {\pi^2 \over 12}
\biggr] {1 \over y}
%%%%%%%%%%%%% 1/x/y %%%%%%%%%%%%%%
{} \PlusBreak{\null + i \pi \BigglP }
\biggl[
   2 \li3(-x)
  +3 \li3(-y)
  -(1 + X - Y) \li2(-x)
  +\biggl( Y^2 - {3 \over 4} Y \biggr) X
{} \PlusBreak{\null + i \pi \BigglP  + \, \bigglP \null }
   {1 \over 6} Y^3 - {67 \over 24} Y^2
  +\biggl( {73 \over 9} - {7 \over 12} \pi^2 \biggr) Y
  -3 \zeta_3
\biggr] {1 \over x y}
\Biggr\}
%%%%% end : BB61Paper
\,, \label{BB61}\\[1pt plus 4pt]
%_____________________________________________________________________________
%
C^{[1]}_6 & = &
%%%%% begin : CC61Paper
%%%%%%%%%%% x/y %%%%%%%%%%%%%
\biggl[
-2 \li4 \biggl( - {x \over y} \biggr)
-6 \li4(-x)
-8 \li4(-y)
+(4 X + 2 Y) \li3(-x)
+{5 \over 24} X^4
-{3 \over 2} Y X^3
{} \PlusBreak{\null \bigglP}
  (1 + X + 8 Y) \li3(-y)
+\biggl( Y + X Y - {4 \over 3} \pi^2
         +3 Y^2 - 2 X^2 \biggr) \li2(-x)
-{9 \over 4} X^3
{} \MinusBreak{\null \bigglP}
  \biggl( {7 \over 12} \pi^2 - {3 \over 2} x
         -{9 \over 2} Y - 2 Y^2 - {7 \over 8} \biggr) X^2
+{13 \over 6} Y^3 X
-\biggl( 3 x - {23 \over 4} + {7 \over 6} \pi^2 \biggr) Y X
{} \PlusBreak{\null \bigglP}
  \biggl( 12 - \zeta_3 - {7 \over 4} \pi^2 \biggr) X
-\biggl( {11 \over 8} \pi^2 - {3 \over 2} x - {3 \over 2} \biggr) Y^2
-\biggl( {93 \over 16} + {17 \over 12} \pi^2 - 3 \zeta_3 \biggr) Y
{} \PlusBreak{\null \bigglP}
  {3 \over 2} \pi^2 x - {99 \over 16} \pi^2
-{19 \over 4} \zeta_3 + {44 \over 45} \pi^4
+{511 \over 64}
\biggr] {x \over y}
%%%%%%%%%%%% 1/y %%%%%%%%%%%%%
{} \PlusBreak{}
\biggl[
   4 \li3(-y)
  +4 Y \li2(-x)
  +\biggl( {3 \over 4} Y + {\pi^2 \over 2} \biggr) X
  +{3 \over 2} Y^2
  -\biggl( 6 - {7 \over 4} \pi^2 \biggr) Y
  -{11 \over 12} \pi^2
{} \MinusBreak{\null + \bigglP}
   4 \zeta_3
\biggr] {1 \over y}
%%%%%%%%%% others %%%%%%%%%%%%
+{11 \over 8} Y^2 X
-\biggl( {21 \over 8} Y^2 X - {1 \over 3} Y^3 \biggr) {1 \over x}
%%%%%%%%%%%%% 1/x/y %%%%%%%%%%%%
{} \PlusBreak{}
\biggl[
     6 \biggl( \li4 \biggl( - {x \over y} \biggr)
              + \li4(-x) - Y \li3(-x) \biggr)
    +(3 X + 3 - 4 Y) \li3(-y)
    +{1 \over 6} Y^4
{} \PlusBreak{\null + \bigglP}
     {3 \over 2} Y^2 X^2
    +(3 Y + 3 X Y - Y^2 - 2 \pi^2) \li2(-x)
    -\biggl( \pi^2 Y + 3 \zeta_3 + {3 \over 2} Y^3 \biggr) X
{} \MinusBreak{\null + \bigglP}
     \biggl( {\pi^2 \over 8} - {3 \over 2} \biggr) Y^2
    +\biggl( 4 \zeta_3 - {19 \over 12} \pi^2 \biggr) Y
    -3 \zeta_3
\biggr] {1 \over x y}
%
%%%%%%%%%%%%% Imaginary part %%%%%%%%%%%%%
%
{} \PlusBreak{}
i \pi {} \Biggl\{
%%%%%%%%%%%% x/y %%%%%%%%%%%%%%%
\biggl[
  6 \li3(-x)
+9 \li3(-y)
+(1 - 3 X + 7 Y) \li2(-x)
-\biggl( {3 \over 2} Y + {9 \over 4} \biggr) X^2
{} \PlusBreak{\null + i \pi \BigglP }
  \biggl( {15 \over 2} + 6 Y^2 - 2 \pi^2 + {21 \over 4} Y \biggr) X
+\biggl( {35 \over 4} - {35 \over 12} \pi^2 \biggr) Y
+{99 \over 16} - {29 \over 12} \pi^2
+2 \zeta_3
\biggr] {x \over y}
%%%%%%%%%%%%% 1/y %%%%%%%%%%%%%%%
{} \PlusBreak{\null + i \pi \BigglP }
\biggl[
   4 \li2(-x)
  +{3 \over 4} (1 - 2 Y) X
  +{\pi^2 \over 12} - 6
  +{15 \over 4} Y
\biggr] {1 \over y}
%%%%%%%%%%%%% 1/x %%%%%%%%%%
-\biggl( {1 \over 8} Y^2 + {1 \over 6} Y^3 \biggr) {1 \over x}
%%%%%%%%%%%%%% 1/x/y %%%%%%%%%%%
{} \PlusBreak{\null + i \pi \BigglP }
\biggl[
  -6 \li3(-x)
  -\li3(-y)
  +(Y + 3 + 3 X) \li2(-x)
  +\biggl( Y^2 + {9 \over 4} Y \biggr) X
{} \PlusBreak{\null + i \pi \BigglP  + \, \bigglP \null }
   \biggl( 3 + {\pi^2 \over 12} \biggr) Y
  +\zeta_3
\biggr] {1 \over x y}
%%%%%%%% others %%%%%%%%%%%%
+{1 \over 6} Y^3 + {15 \over 8} Y^2
\Biggr\}
%%%%% end : CC61Paper
\,, \label{CC61}\\[1pt plus 4pt]
%____________________________________________________________________________
%
D^{[1]}_6 & = &
%%%%% begin : DD61Paper
\biggl[
-{1 \over 3} (\li3(-x) + \li3(-y) - (X - Y) \li2(-x))
+{1 \over 9} X^3 + {5 \over 18} Y^3
-\biggl( {29 \over 36} - {1 \over 6} Y \biggr) X^2
{} \PlusBreak{\null \bigglP}
  \biggl(  {13 \over 72} \pi^2 + {31 \over 27}
          +{11 \over 18} Y - {2 \over 3} Y^2 \biggr) X
- {37 \over 36} Y^2
+\biggl( {\pi^2 \over 3} + {145 \over 54} \biggr) Y
+{55 \over 72} \pi^2 - {455 \over 108}
{} \MinusBreak{\null \bigglP}
  {37 \over 36} \zeta_3
\biggr] {x \over y}
%%%%%%%%%%% Imaginary part %%%%%%%%%%%%%
+i \pi {} \biggl\{
{1 \over 3} X^2
-\biggl( {2 \over 3} Y + 1 \biggr) X
+{7 \over 24} \pi^2 + {23 \over 6}
-{13 \over 9} Y + {1 \over 3} Y^2
\biggr\} {x \over y}
%%%%% end : DD61Paper
\,, \label{DD61}\\[1pt plus 4pt]
%____________________________________________________________________________
%
E^{[1]}_6 & = &
%%%%% begin : EE61Paper
%%%%%%%%%%%%% x/y %%%%%%%%%%%%%
\biggl[
  {2 \over 3} (\li3(-x) + 2 \li3(-y) - (X - 2 Y) \li2(-x))
-{2 \over 9} X^3
-\biggl( {1 \over 3} Y - {29 \over 18} \biggr) X^2
{} \PlusBreak{\null \bigglP}
  \biggl(  {5 \over 3} Y^2 - {13 \over 36} \pi^2
          -{62 \over 27} - {11 \over 9} Y \biggr) X
-\biggl( {101 \over 72} \pi^2 - {107 \over 27} \biggr) Y
-{\pi^2 \over 24} - {59 \over 36} \zeta_3
-{685 \over 162}
\biggr] {x \over y}
%%%%%%%%%%%%%%% others %%%%%%%%%%%%
{} \PlusBreak{}
\biggl( {16 \over 9} Y + {4 \over 9} \pi^2 \biggr) {1 \over y}
+{2 \over 9} Y^3 + {1 \over 9} Y^2
+(8 Y^2 - 2 Y^3) {1 \over 9 x}
  -{4 \over 9 x y} \pi^2 Y
%%%%%%%%%%%% Imaginary part %%%%%%%%%%%%%%
{} \PlusBreak{}
i \pi {} \Biggl\{
\biggl[
  {2 \over 3} \li2(-x)
-{2 \over 3} X^2 + \biggl( {4 \over 3} Y + 2 \biggr) X
+{1 \over 3} Y^2 - {13 \over 9} Y + {5 \over 3}
-{47 \over 72} \pi^2
\biggr] {x \over y}
{} \PlusBreak{\null + i \pi \BigglP }
\biggl( {16 \over 9} - 2 Y \biggr) {1 \over y}
-\biggl( {16 \over 9} Y - {2 \over 3} Y^2 \biggr) {1 \over x y}
\Biggr\}
%%%%% end : EE61Paper
\,, \label{EE61}\\[1pt plus 4pt]
%___________________________________________________________________________
%
F^{[1]}_6 & = &
%%%%% begin : FF61Paper
\biggl[
-{1 \over 9} \pi^2
+{1 \over 9} Y^2
-{10 \over 27} Y
+{25 \over 81}
\biggr] {x \over y}
%%%%%%%%%% Imaginary part %%%%%%%%%%
+i \pi {} \biggl[
{2 \over 9} Y - {10 \over 27}
\biggr] {x \over y}
%%%%% end : FF61Paper
\,, \label{FF61} %\\[1pt plus 4pt]
%___________________________________________________________________________
%
\end{eqnarray}

For $h=6$ in \eqn{h6} and color factor $\trc^{[2]}$ in \eqn{basis56}:
\begin{eqnarray}
G^{[2]}_6 & = &
%%%%% begin : GG62Paper
%%%%%%%%%% x/y %%%%%%%%%%%%
\biggl[
  2 \li4 \biggl(- {x \over y} \biggr)
-4 \li4(-x)
-{5 \over 2} \li4(-y)
+\biggl( {9 \over 2} + 4 X - Y \biggr) \li3(-x)
-{1 \over 12} Y X^3
{} \PlusBreak{\null \bigglP}
  \biggl( Y + {29 \over 6} + X \biggr) \li3(-y)
+\biggl(  {3 \over 4} Y^2 - {\pi^2 \over 3}
          +{29 \over 6} Y - {9 \over 2} X - X^2 \biggr) \li2(-x)
{} \PlusBreak{\null \bigglP}
  \biggl(  {3 \over 4} - {9 \over 4} Y + {1 \over 2} x
          +{5 \over 8} Y^2 - {3 \over 8} \pi^2 \biggr) X^2
-{1 \over 6} Y^3 X + {25 \over 6} Y^2 X
+{1 \over 24} Y^4 + {1 \over 3} Y^3
{} \MinusBreak{\null \bigglP}
  \biggl( x + {11 \over 8} + {5 \over 6} \pi^2 \biggr) Y X
-\biggl( {5 \over 12} \pi^2 + \zeta_3 \biggr) X
-\biggl( {29 \over 36} - {2 \over 3} \pi^2 - {1 \over 2} x \biggr) Y^2
{} \MinusBreak{\null \bigglP}
  \biggl( {67 \over 18} \pi^2 - 2 \zeta_3 - {3049 \over 432} \biggr) Y
+{115 \over 72} \pi^2 - {23213 \over 5184}
+{\pi^2 \over 2} x - {341 \over 72} \zeta_3
+{59 \over 160} \pi^4
\biggr] {x \over y}
%%%%%%%%%%%% 1/y %%%%%%%%%%%%
{} \PlusBreak{}
\biggl[
  -3 \biggl( \li4 \biggl(- {x \over y} \biggr)
            + \li4(-x) + \li4(-y) - \li3(-x)
            + \biggl( X - Y - {1 \over 2} \biggr) \li3(-y) \biggr)
{} \MinusBreak{\null + \bigglP}
   \biggl( 3 X - {3 \over 2} Y - {\pi^2 \over 2} \biggr) \li2(-x)
  -{3 \over 4} \biggl( Y^2 + 2 Y \biggr) X^2
  +{1 \over 2} Y^3 X + {7 \over 4} Y^2 X
{} \MinusBreak{\null + \bigglP}
   {1 \over 3} Y^3
  +\biggl( \pi^2 - {3 \over 2} \biggr) Y X
  -\biggl( {\pi^2 \over 6} - 3 \zeta_3 \biggr) X
  +{1 \over 4} (\pi^2 + 1) Y^2
{} \MinusBreak{\null + \bigglP}
   \biggl( 3 \zeta_3 + {13 \over 12} \pi^2 - {32 \over 9} \biggr) Y
  +{59 \over 36} \pi^2 + {\pi^4 \over 30}
  -2 \zeta_3
\biggr] {1 \over y}
%%%%%%%%%%%%%%% 1/x/y %%%%%%%%%%%%
{} \PlusBreak{}
\biggl[
  -3 \li4 \biggl(- {x \over y} \biggr)
  -3 \li4(-x)
  +{1 \over 2} \li4(-y)
  +\biggl( {3 \over 2} - 3 X + Y \biggr) \li3(-y)
  -{3 \over 4} Y^2 X^2
{} \MinusBreak{\null + \bigglP}
   {1 \over 12} Y^4
  -\biggl(  {1 \over 4} Y^2 - {\pi^2 \over 6}
           -{3 \over 2} Y \biggr) \li2(-x)
  +\biggl( {3 \over 2} Y^2 + {2 \over 3} \pi^2 Y
          +3 \zeta_3 + {1 \over 4} Y^3 \biggr) X
{} \PlusBreak{\null + \bigglP}
   {1 \over 9} Y^3
  -\biggl( {37 \over 36} - {3 \over 8} \pi^2 \biggr) Y^2
  -\biggl( \zeta_3 + {31 \over 18} \pi^2 \biggr) Y
  -{\pi^4 \over 180} - {3 \over 2} \zeta_3
\biggr] {1 \over x y}
%
%%%%%%%%%%%% Imaginary part %%%%%%%%%%
%
{} \PlusBreak{}
i \pi {} \Biggl\{
%%%%%%%%%%%% x/y %%%%%%%%%%%%%%
\biggl[
  3 \li3(-x)
+2 \li3(-y)
+\biggl( {1 \over 3} + {3 \over 2} Y - 2 X \biggr) \li2(-x)
-{1 \over 12} X^3 + {1 \over 2} X^2 Y
{} \MinusBreak{\null + i \pi \BigglP \bigglP \null }
  \biggl( Y + {13 \over 12} \pi^2 - {1 \over 8} \biggr) X
-{5 \over 12} Y^3 + {11 \over 4} Y^2
-\biggl( {215 \over 72} + {\pi^2 \over 3} \biggr) Y
{} \PlusBreak{\null + i \pi \BigglP \bigglP \null }
  {3049 \over 432} + \zeta_3
+{\pi^2 \over 36}
\biggr] {x \over y}
%%%%%%%%%%%%% others %%%%%%%%%%%%%%
-{1 \over 2} Y^2 X
+{1 \over 2 x} X Y^2
%%%%%%%%%%%%%%%% 1/y %%%%%%%%%%%%
{} \PlusBreak{\null + i \pi \BigglP }
\biggl[
  -{3 \over 2} \li2(-x)
  -\biggl( {3 \over 2} + Y \biggr) X + {\pi^2 \over 12}
  +{32 \over 9} - Y
\biggr] {1 \over y}
%%%%%%%%%%%%%%% 1/x/y %%%%%%%%%%%%%
{} \PlusBreak{\null + i \pi \BigglP }
\biggl[
    -2 \li3(-y)
    -{1 \over 2} (Y - 3) \li2(-x)
    +{3 \over 2} X Y - {1 \over 6} Y^3
    +{13 \over 12} Y^2
{} \MinusBreak{\null + i \pi \BigglP  + \, \bigglP \null }
     \biggl( {37 \over 18} - {\pi^2 \over 4} \biggr) Y
    +2 \zeta_3
\biggr] {1 \over x y}
\Biggr\}
%%%%% end : GG62Paper
\,, \label{GG62}\\[1pt plus 4pt]
%____________________________________________________________________________
%
H^{[2]}_6 & = &
%%%%% begin : HH62Paper
%%%%%%%%%%%% x/y %%%%%%%%%%%
\biggl[
-\li4 \biggl(- {x \over y} \biggr)
+6 \li4(-x)
+4 \li4(-y)
-\biggl( 5 X - {1 \over 3} \biggr) \li3(-x)
+{1 \over 24} X^4
-{1 \over 12} Y^4
{} \MinusBreak{\null \bigglP}
  \biggl( X - {13 \over 6} + 4 Y \biggr) \li3(-y)
+\biggl(  {13 \over 6} Y + {3 \over 2} \pi^2
          -{1 \over 3} X - {5 \over 2} Y^2
          +{3 \over 2} X^2 \biggr) \li2(-x)
{} \PlusBreak{\null \bigglP}
  \biggl( {1 \over 12} Y - {11 \over 18} \biggr) X^3
+\biggl(  {1 \over 2} x + {17 \over 24} \pi^2
          -{3 \over 8} Y^2 + {7 \over 12} Y + {277 \over 72} \biggr) X^2
-{17 \over 12} Y^3 X
{} \PlusBreak{\null \bigglP}
  {47 \over 24} Y^2 X
+\biggl( {5 \over 6} \pi^2 - {347 \over 72} - x \biggr) Y X
-\biggl(  {79 \over 27} + {143 \over 144} \pi^2
          -{1 \over 2} \zeta_3 \biggr) X
{} \PlusBreak{\null \bigglP}
  \biggl( {1 \over 2} x + {5 \over 6} \pi^2 - {29 \over 9} \biggr) Y^2
-\biggl(  {3 \over 2} \zeta_3 + {215 \over 144} \pi^2
          -{781 \over 54} \biggr) Y
-{1031 \over 1440} \pi^4 + {467 \over 144} \pi^2
{} \PlusBreak{\null \bigglP}
  {\pi^2 \over 2} x + {311 \over 72} \zeta_3
-{30659 \over 1296}
\biggr] {x \over y}
%%%%%%%%%%%%% 1/y %%%%%%%%%%%%%%%
{} \PlusBreak{}
\biggl[
  -\li3(-y)
  -Y \li2(-x)
  +\biggl( {5 \over 4} Y - {\pi^2 \over 2} + {1 \over 4} Y^2 \biggr) X
  -{21 \over 4} Y^2
  -\biggl( {9 \over 4} \pi^2 - {59 \over 9} \biggr) Y
{} \PlusBreak{\null + \bigglP}
   \zeta_3 + {8 \over 9} \pi^2
\biggr] {1 \over y}
%%%%%%%%%%%%% others %%%%%%%%%%%%%
+{13 \over 36} Y^3
-{31 \over 36 x} Y^3
%%%%%%%%%%% 1/x/y %%%%%%%%%%%
{} \PlusBreak{}
\biggl[
    -2 \biggl( \li4 \biggl( - {x \over y} \biggr)
              + \li4(-x) - Y \li3(-x) \biggr)
    -\biggl( X + {5 \over 2} - Y \biggr) \li3(-y)
    -{1 \over 24} Y^4
{} \MinusBreak{\null + \bigglP}
     \biggl( {5 \over 2} Y + X Y - \pi^2 \biggr) \li2(-x)
    -{1 \over 2} Y^2 X^2
    +\biggl(  {2 \over 3} \pi^2 Y + \zeta_3
             +{1 \over 6} Y^3 - {19 \over 8} Y^2 \biggr) X
{} \PlusBreak{\null + \bigglP}
     \biggl( {11 \over 24} \pi^2 - {127 \over 36} \biggr) Y^2
    -\biggl( \zeta_3 - {19 \over 36} \pi^2 \biggr) Y
    +{5 \over 2} \zeta_3
\biggr] {1 \over x y}
%
%%%%%%%%%%% imaginarey part %%%%%%%%%%%%%
%
{} \PlusBreak{}
i \pi {} \Biggl\{
%%%%%%%%%%%% x/y %%%%%%%%%%%%%
\biggl[
-5 (\li3(-x) + \li3(-y))
-\biggl( 5 Y - {11 \over 6} - 3 X \biggr) \li2(-x)
+{1 \over 12} X^3
-{1 \over 4} Y^3
{} \MinusBreak{\null + i \pi \BigglP \bigglP \null }
  {13 \over 12} X^2
+\biggl(  {35 \over 12} Y + {23 \over 8}
          -{5 \over 2} Y^2 + {7 \over 4} \pi^2 \biggr) X
- {5 \over 24} Y^2
-\biggl( {811 \over 72} - {3 \over 2} \pi^2 \biggr) Y
{} \PlusBreak{\null + i \pi \BigglP \bigglP \null }
  {623 \over 54} - \zeta_3
-{85 \over 72} \pi^2
\biggr] {x \over y}
%%%%%%%%%%%% 1/y %%%%%%%%%%%%%%
{} \PlusBreak{\null + i \pi \BigglP }
\biggl[
  -\li2(-x)
  +\biggl( {5 \over 4} + {3 \over 2} Y \biggr) X
  -{\pi^2 \over 4} + {9 \over 4} Y^2
  -{37 \over 4} Y + {59 \over 9}
\biggr] {1 \over y}
%%%%%%%%%%%%% 1/x/y %%%%%%%%%%%%
{} \PlusBreak{\null + i \pi \BigglP }
\biggl[
   2 \li3(-x)
  -\biggl( Y + X + {5 \over 2} \biggr) \li2(-x)
  -\biggl( Y^2 + {9 \over 4} Y \biggr) X
  -{1 \over 6} Y^3 + {35 \over 24} Y^2
{} \MinusBreak{\null + i \pi \BigglP  + \, \bigglP \null }
   \biggl( {127 \over 18} - {\pi^2 \over 4} \biggr) Y
\biggr] {1 \over x y}
\Biggr\}
%%%%% end : HH62Paper
\,, \label{HH62}\\[1pt plus 4pt]
%____________________________________________________________________________
%
I^{[2]}_6 &=&
%%%%% begin : II62Paper
%%%%%%%%%%% x/y %%%%%%%%%%%%%%
\biggl[
   \li4 \biggl( -{x \over y} \biggr)
  + 2 \li4(-x)
  + {5 \over 2} \li4(-y)
  -(Y + X) \li3(-x)
  -\biggl( 3 Y + {1 \over 2} \biggr) \li3(-y)
{} \PlusBreak{\null \bigglP}
   \biggl( -{5 \over 4} Y^2 - {1 \over 2} Y
           + {1 \over 2} X^2 + {\pi^2 \over 2}
   \biggr) \li2(-x)
-{1 \over 12} X^4
-{11 \over 12} Y^3 X
  + \biggl( {3 \over 4} + {1 \over 2} Y \biggr) X^3
{} \MinusBreak{\null \bigglP}
    \biggl( {3 \over 2} Y + {1 \over 2} Y^2
            - {\pi^2 \over 6} + {1 \over 2} x - {3 \over 8}
    \biggr) X^2
   + \biggl( x -{13 \over 4} + {\pi^2 \over 3} \biggr) Y X
   - \biggl( 6 - {7 \over 12}  \pi^ 2 \biggr) X
{} \PlusBreak{\null \bigglP}
    {7 \over 12} \pi^2 Y^2
  + \biggl( {2 \over 3} \pi^2-3 \zeta_3 + {45 \over 16} \biggr) Y
-{511 \over 64} - {\pi^2 \over 2} x
-{109 \over 360} \pi^4
  + {15 \over 4} \zeta_3
  + {39 \over 16} \pi^2
\biggr] {x \over y}
%%%%%%%%%%%%%%%%%%% 1/y %%%%%%%%%%%%%%
{} \PlusBreak{}
\biggl[
  -{3 \over 2} \li3(-y)
  -{3 \over 2} Y \li2(-x)
   - \biggl( {1 \over 4} Y + {\pi^2 \over 6} \biggr) X
  -Y^2 + \biggl( 3 - {2 \over 3} \pi^2 \biggr) Y
   + \zeta_3
{} \PlusBreak{\null + \bigglP}
     {5 \over 12} \pi^2
\biggr] {1 \over y}
+ {1 \over 2} x Y^2
- {3 \over 8} Y^2 X
+ \biggl( {7 \over 8} Y^2 X-{1 \over 6} Y^3 \biggr) {1 \over x}
%%%%%%%%%%%% others %%%%%%%%%%%%%%%%%%
{} \PlusBreak{}
\biggl[
    -2 \li4 \biggl( -{x \over y} \biggr)
    -2 \li4(-x)
     + {1 \over 2} \li4(-y)
     + 2 Y \li3(-x)
     + (Y-1-X) \li3(-y)
{} \PlusBreak{\null + \bigglP}
       \biggl( -Y + {2 \over 3} \pi^2
            -Y X + {1 \over 4} Y^2 \biggr) \li2(-x)
    -{1 \over 2} Y^2 X^2
     + \biggl( {\pi^2 \over 3} Y + {5 \over 12} Y^3
         + \zeta_3 \biggr) X
{} \MinusBreak{\null + \bigglP}
     {1 \over 24} Y^4 - Y^2
     + \biggl( {5 \over 12} \pi^2 - \zeta_3 \biggr) Y
    -{\pi^4 \over 180} + \zeta_3
\biggr]{1 \over x y}
%
%%%%%%%%%%%% Imaginary part %%%%%%%%%%%%%%
%
{} \PlusBreak{}
  i \pi {} \Biggl\{
\biggl[
-2 \li3(-x)
-3 \li3(-y)
-\biggl( {5 \over 2} Y + {1 \over 2}-X \biggr) \li2(-x)
  + \biggl( {3 \over 4} + {1 \over 2} Y \biggr) X^2
{} \PlusBreak{\null + i \pi \BigglP \bigglP \null }
    \biggl( {2 \over 3} \pi^2-{5 \over 2}
       -2 Y^2-{7 \over 4} Y \biggr) X
  + \biggl( \pi^2 - {17 \over 4} \biggr) Y
-3 \zeta_3 + \pi^2 - {51 \over 16}
\biggr] {x \over y}
{} \PlusBreak{\null + i \pi \BigglP }
\biggl[
   - {3 \over 2} \li2(-x)
   + \biggl( {1 \over 2} Y - {1 \over 4} \biggr) X + 3 - {9 \over 4} Y
\biggr] {1 \over y}
-{5 \over 8} Y^2
   -{1 \over 8 x} Y^2
{} \PlusBreak{\null + i \pi \BigglP }
\biggl[
     2 \li3(-x)
    -\biggl( X + 1 + {1 \over 2} Y \biggr) \li2(-x)
    -\biggl( {1 \over 2} Y + {3 \over 4} \biggr) Y X - 2 Y
  \biggr]{1 \over x y}
\Biggr\}
%%%%% end : II62Paper
\,, \label{II62} \\[1pt plus 4pt]
%_____________________________________________________________________________
%
J^{[2]}_6 & = &
%%%%% begin : JJ62Paper
\biggl[
-{1 \over 3} \li3(-y)
-{1 \over 3} Y \li2(-x)
-{1 \over 6} Y^2 X - {1 \over 6} Y^3
+{13 \over 12} Y^2
-\biggl( {193 \over 54} - {7 \over 18} \pi^2 \biggr) Y
+{455 \over 108}
{} \PlusBreak{\null \bigglP}
  {49 \over 36} \zeta_3
-{19 \over 24} \pi^2
\biggr] {x \over y}
-\biggl( {8 \over 9} Y + {2 \over 9} \pi^2
        -{1 \over 2} Y^2 \biggr) {1 \over y}
-(Y^3 - 4 Y^2 - 2 \pi^2 Y) {1 \over 9 x y}
%%%%%%%%%%% Imaginary part %%%%%%%%%%%%%%
{} \PlusBreak{}
i \pi {} \Biggl\{
\biggl[
-{1 \over 2} Y^2 + {13 \over 6} Y
-{1 \over 3} \li2(-x)
-{193 \over 54} + {\pi^2 \over 18}
\biggr] {x \over y}
-\biggl( {8 \over 9} - Y \biggr) {1 \over y}
{} \PlusBreak{\null + i \pi \BigglP }
\biggl( {8 \over 9} Y - {1 \over 3} Y^2 \biggr) {1 \over x y}
\Biggr\}
%%%%% end : JJ62Paper
\,, \label{JJ62}\\[1pt plus 4pt]
%_____________________________________________________________________________
%
K^{[2]}_6 & = &
%%%%% begin : KK62Paper
\biggl[
-{1 \over 3} \li3(-x)
-{2 \over 3} \li3(-y)
+{1 \over 3} (X - 2 Y) \li2(-x)
+{1 \over 9} X^3
-\biggl( {29 \over 36} - {1 \over 6} Y \biggr) X^2
{} \MinusBreak{\null \bigglP}
  \biggl(  {5 \over 6} Y^2 - {13 \over 72} \pi^2
          -{31 \over 27} - {11 \over 18} Y \biggr) X
+\biggl( {49 \over 72} \pi^2 - {83 \over 27} \biggr) Y
+{5 \over 72} \pi^2 + {47 \over 36} \zeta_3
+{685 \over 162}
\biggr] {x \over y}
{} \MinusBreak{}
(8 Y + 2 \pi^2) {1 \over 9 y}
-{1 \over 9} Y^3 - {1 \over 18} Y^2
-(4 Y^2 - Y^3) {1 \over 9 x}
+{2 \pi^2 \over 9 x y} Y
%%%%%%%%%% Imaginary part %%%%%%%%%%
{} \PlusBreak{}
i \pi {} \Biggl\{
\biggl[
-{1 \over 3} \li2(-x)
+{1 \over 3} X^2 - \biggl( {2 \over 3} Y + 1 \biggr) X
-{1 \over 6} Y^2 + {13 \over 18} Y
-{52 \over 27} + {11 \over 36} \pi^2
\biggr] {x \over y}
{} \MinusBreak{\null + i \pi \BigglP \bigglP \null }
\biggl( {8 \over 9} - Y \biggr) {1 \over y}
+\biggl( {8 \over 9} Y - {1 \over 3} Y^2 \biggr) {1 \over x y}
\Biggr\}
%%%%% end : KK62Paper
\,, \label{KK62}\\[1pt plus 4pt]
%_____________________________________________________________________________
%
L^{[2]}_6 & = &
%%%%% begin : LL62Paper
\biggl[
  {\pi^2 \over 9}
-{1 \over 9} Y^2
+{10 \over 27} Y
-{25 \over 81}
\biggr] {x \over y}
%%%%%%%%%% Imaginary part %%%%%%%%%
+i \pi {} \biggl[
-{2 \over 9} Y + {10 \over 27}
\biggr] {x \over y}
%%%%% end : LL62Paper
\,, \label{LL62} %\\[1pt plus 4pt]
%_____________________________________________________________________________
%
\end{eqnarray}

\section{Auxiliary functions for two-loop scheme shifts}
\label{OrdepsdeltaRemainderAppendix}

In this appendix we present auxiliary functions appearing in
\eqn{Twoloopc12} for the shift in the two-loop amplitudes
under scheme changes.  These functions correspond to the
$\delta_R$-dependent parts of the $\Ord(\e)$ terms in the one-loop
amplitude remainders.  They are given by,
\begin{eqnarray}
M^{(1),[1]\epsilon,\delta_R}_1 &=&
%%%%% begin : M11aux
- {19 \over 18} y N - \biggl( Y - {y \over 2} \biggr) {1 \over N}
- i \pi {} \biggl[ {y \over 3} N + (x + 3) {1 \over 2 N} \biggr]
%%%%% end : M11aux
\,, \\
M^{(1),[2]\epsilon,\delta_R}_1 &=&
%%%%% begin : M12aux
{1 \over 2} Y + {19 \over 18} y + (Y - y) {1 \over 2 N^2} 
+i \pi {} \biggl[ - {x \over 3} + {1 \over 6}
                  + \biggl( 1 + {1 \over 2} x \biggr) 
{1 \over N^2} \biggr]
%%%%% end : M12aux
\,, \\
M^{(1),[1]\epsilon,\delta_R}_2 &=&
%%%%% begin : M21aux
-{19 \over 18} x N + \biggl( - {1 \over 2 x} Y^2 + Y + {x \over 2} \biggr) 
{1
\over N}
{} \PlusBreak{}
i \pi {}\biggl[ - {x \over 3} N
                 + \biggl( -{1 \over x} Y + {x \over 2} + 1 \biggr) {1 \over 
N}
\biggr]
%%%%% end: M21aux
\,, \\
M^{(1),[2]\epsilon,\delta_R}_2 &=&
%%%%% begin : M22aux
{1 \over 4 x} Y^2 - {1 \over 2} Y + {19 \over 18} x 
+ \biggl( {1 \over 4 x} Y^2 - {1 \over 2} Y - {x \over 2} \biggr)
{1 \over N^2}
{} \PlusBreak{}
i \pi {} \biggl[ {1 \over 2 x} Y + {x \over 3} - {1 \over 2} 
+ \biggl( {1 \over x} Y + y \biggr) {1 \over 2 N^2}
\biggr]
%%%%% end : M22aux
\,, \\
M^{(1),[1]\epsilon,\delta_R}_3 &=&
%%%%% begin : M31aux
\biggl[ -{\pi^2 \over 2} y + X - {y \over 2} (X-Y)^2
         + {1 \over 2 y}- \biggl( {1 \over 2 y} + 1 \biggr) Y \biggr] 
{1 \over N}
{} \PlusBreak{}
\biggl( {1 \over 3} Y - {19 \over 18} \biggr) {N \over y}
%%%%% end : M31aux
\,, \\
M^{(1),[2]\epsilon,\delta_R}_3 &=&
%%%%% begin : M32aux
\biggl[ {\pi^2 \over 4} y - {1 \over 2} X + 1
       + {y \over 4} (X-Y)^2 \biggr] \biggl( 1 + {1 \over N^2} \biggr)
+\biggl( x + {37 \over 18} \biggr) {1 \over y}
{} \PlusBreak{}
\biggl( x + {1 \over 2}  - {x \over 2} Y \biggr) {1 \over y N^2}
+\biggl( {1 \over 2} - {1 \over 3 y} \biggr) Y
%%%%% end : M32aux
\,, \\
M^{(1),[1]\epsilon,\delta_R}_4 &=&
%%%%% begin : M41aux
\biggl[ -X + \biggl( {3 \over 2} + {1 \over 2 y} \biggr) Y + {x \over 2 y}
\biggr] {1 \over N}
+ \biggl( {1 \over 3} Y - {19 \over 18} \biggr) {x \over y} N
%%%%% end : M41aux
\,, \\
M^{(1),[2]\epsilon,\delta_R}_4 &=&
%%%%% begin : M42aux
\biggl[ {1 \over 2} X
       + {1 \over y} \biggl( x + {1 \over 2} \biggr) Y - {x \over 2 y} 
\biggr]
{1 \over N^2}
+{1 \over 2} X + {1 \over 6 y} (x+3) Y + {19 \over 18} {x \over y}
%%%%% end : M42aux
\,, \hskip0.5cm \\
M^{(1),[1]\epsilon,\delta_R}_5 &=&
%%%%% begin : M51aux
\biggl[ \biggl( 1 - {1 \over 2 y} \biggr) Y + {1 \over 2 y} \biggr] {1 \over 
N}
+\biggl( {1 \over 3} Y - {19 \over 18} \biggr) {N \over y} + i \pi {} {1 
\over
N}
%%%%% end : M51aux
\,, \\
M^{(1),[2]\epsilon,\delta_R}_5 &=&
%%%%% begin : M52aux
\biggl[ \biggl( {x \over 2} + 1 \biggr) Y - {1 \over 2} \biggr] {1 \over y 
N^2}
-\biggl( {1 \over 3 y} + {1 \over 2} \biggr) Y + {19 \over 18 y} 
- i {} {\pi \over 2} \biggl( 1 + {1 \over N^2} \biggr)
%%%%% end : M52aux
\,, \\
M^{(1),[1]\epsilon,\delta_R}_6 &=&
%%%%% begin : M61aux
\biggl[ -{y \over 2 x} Y^2 + {1 \over y} \biggl( {x \over 2} + 1 \biggr) Y
        + {x \over 2 y} \biggr] {1 \over N}
+\biggl( {1 \over 3} Y - {19 \over 18} \biggr) {x \over y} N
{} \MinusBreak{}
i \pi {} \biggl( {y \over x} Y + 1 \biggr) {1 \over N}
%%%%% end : M61aux
\,, \\
M^{(1),[2]\epsilon,\delta_R}_6 &=&
%%%%% begin : M62aux
\biggl[ {y \over 4 x} Y^2 - {1 \over 2 y} Y - {x \over 2 y} \biggr] {1 \over
N^2}
+{y \over 4 x} Y^2 + \bigg( {1 \over 2} - {x \over 3 y} \biggl) Y + {19 
\over
18} {x \over y}
{} \PlusBreak{}
i {} {\pi \over 2} \biggl( {y \over x} Y + 1 \biggr) 
\biggl( 1 + {1 \over N^2} \biggr)
%%%%% end : M62aux
\,.
\end{eqnarray}

%%%%%%%%%%%%%%%%%%%%%%%%%%%%%%%%%%%%%%%%%%%%%%%%

\end{document}